\documentclass[reprint,amsmath,amssymb,aps,prb,superscriptaddress,floatfix,nofootinbib,nobibnotes]{revtex4-2}

\pdfoutput=1
\usepackage{graphicx}% Include figure files
\usepackage{dcolumn}% Align table columns on a decimal point
\usepackage{bm}% bold math
\usepackage{braket}
\usepackage[mathlines]{lineno}% Enable numbering of text and display math
\usepackage{stackengine}
\usepackage{verbatim}
\usepackage{graphics}
\usepackage{hyperref}
\usepackage{xcolor}
\usepackage{appendix}
\usepackage{soul}
\usepackage{ulem} % needed for \sout-command
\usepackage{multirow}

\begin{document}
\preprint{APS/123-QED}
%\title{Phase classification in quasiperiodic systems using machine learning}
\title{Supervised and unsupervised learning of the many-body critical phase, phase transitions,
and critical exponents in disordered quantum systems}

\author{Aamna Ahmed}
\email{aamna.ahmed@uni-a.de}
\affiliation{Institute of Physics, University of Augsburg, Augsburg, Germany, 86159\\}
\author{Nilanjan Roy}%
\email{nilanjan.roy@ntu.edu.sg}
\affiliation{Division of Physics and Applied Physics, Nanyang Technological University, Singapore 637371}

\date{\today}% It is always \today, today,

%%%%%%%%%%%%%%%%%%%%%%%%%%%%%%%%%%%%%%%%%%%%%%%%%%
%%%%%%%%%%%%%%%%%%%%%%%%%%%%%%%%%%%%%%%%%%%%%%%%%%
\begin{abstract}
Supervised and unsupervised machine-learning-based frameworks have been implemented in the past to
classify various dynamical phases of disordered quantum systems using rather sophisticated (highly preprocessed
or difficult to generate) input data. In this paper, we begin by questioning the existence of a nonergodic extended
phase, namely, the many-body critical (MBC) phase in finite systems of an interacting quasiperiodic system.
We find that this phase can be separately detected from the other phases, such as the many-body ergodic
(ME) and many-body localized (MBL) phases, in the model through supervised neural networks made for both
binary and multiclass classification tasks, utilizing, rather unpreprocessed, eigenvalue spacings and eigenvector
probability densities as input features. Moreover, the output of our trained neural networks can also indicate
the critical points separating ME, MBC, and MBL phases, which are consistent with the same obtained from
other conventional methods. We also employ unsupervised learning techniques, particularly principal component
analysis (PCA) of eigenvector probability densities, to investigate how this framework, without any training,
captures the, rather unknown, many-body phases (ME, MBL, and MBC) and single-particle phases (delocalized,
localized, and critical) of the interacting and noninteracting systems, respectively. Our findings reveal that PCA
entropy serves as an effective indicator (order parameter) for detecting phase transitions in the single-particle
systems. Moreover, this method proves applicable to many-body systems when the data undergoes a suitable
preprocessing. Interestingly, when it comes to the extraction of critical (correlation length) exponents through a
finite-size scaling, we find that for single-particle systems, scaling collapse of neural network outputs is obtained
using components of inverse participation ratio (IPR) as input data. Remarkably, we observe identical critical
exponents as obtained from the scaling collapse of the IPR directly for different single-particle phase transitions.
However, the current finite-size scaling approach to the network output cannot reliably extract critical exponents
for the many-body system that needs further investigation.
\end{abstract}

\maketitle
%\keywords{Suggested keywords}%Use the show keys class option if a keyword
                              %display desired \maketitle

%\tableofcontents

%%%%%%%%%%%%%%%%%%%%%%%%%%%%%%%%%%%%%%%%%%%%%%%%%%
%%%%%%%%%%%%%%%%%%%%%%%%%%%%%%%%%%%%%%%%%%%%%%%%%%
\section{INTRODUCTION}
Machine learning techniques have emerged as transformative tools in the study of quantum systems, offering innovative approaches to tackle problems that are computationally intensive for traditional methods~\cite{PhysRevB.95.245134, PhysRevE.96.022140, PhysRevE.95.062122, PhysRevB.108.184202, PhysRevB.99.155136, Carleo2018Constructing, Torlai2018Neural, Cai2018Approximating,Gardas2018Quantum, Carrasquilla2021How,Schmitt2020Quantum, PhysRevLett.120.257204, Cao_2024}. These techniques excel in capturing complex correlations and high-dimensional features from raw data. Among various strategies, supervised learning has gained prominence in phase classification tasks by utilizing labelled data to learn distinct phase boundaries within the parameter space~\cite{PhysRevX.7.031038, Broecker2017,doi:10.1126/science.aag2302,PhysRevLett.118.216401,PhysRevB.108.155128}. This approach has demonstrated significant success in identifying phase transitions across a range of systems, from disordered to strongly correlated models. In parallel, unsupervised techniques like the principal component analysis (PCA) have offered a complementary perspective, capable of detecting phase transitions without requiring prior knowledge of the phases~\cite{PhysRevE.110.064121, PhysRevB.110.024204,10.21468/SciPostPhysCore.6.4.086,PhysRevB.111.205116}. PCA effectively reduces the dimensionality of large datasets, extracting dominant features that encode critical behaviours, making it a compelling alternative to traditional analytical tools. 

The study of quantum dynamics of isolated many-body systems has been pivotal in understanding phase transitions in out-of-equilibrium quantum matter, especially in the presence of disorder~\cite{abanin2019colloquium,alet2018many,sierant2403many}. These systems in 1D exhibit two well-established phases: the ergodic phase and the many-body localized (MBL) phase. In the ergodic phase, the system adheres to the Eigenstate Thermalization Hypothesis (ETH)~\cite{Deutsch_2018}, achieving thermal equilibrium with subsystem properties aligning with statistical mechanics predictions. Conversely, the MBL phase avoids thermalization under strong disorder, retaining memory of its initial state over a long period.

Recent studies have highlighted the existence of a third phase, namely, the nonergodic extended (NEE) phase which is an intermediate one between the ergodic and MBL phases~\cite{li2015many, PhysRevLett.126.080602, Rispoli2019,roy2023diagnostics,roy2025manybodycriticalphasequasiperiodic}. It was first proposed in the interacting generalized Aubry-Andr\'e-Harper (GAAH) model where NEE states were found in a small fraction of the many-body spectrum~\cite{li2015many}. Very recently, a more robust presence of NEE states has been found throughout the spectrum, dubbed as the many-body critical (MBC) phase, in an interacting extended AAH (EAAH) model~\cite{PhysRevLett.126.080602}. While the MBC phase (NEE states) violates ETH, its entanglement entropy still follows the volume law like the ergodic phase. To distinguish nonergodic phases, methods such as the one-particle density matrix and entanglement spectrum have been widely applied~\cite{bera2017one, PhysRevLett.117.160601,ahmed2023interplay}. However, these conventional techniques often entail significant computational overhead.
Notably, in this context earlier studies have explored various types of sophisticated input data for machine learning including entanglement spectra, intrinsic dimensions, and others~\cite{PhysRevB.110.024204,10.21468/SciPostPhysCore.6.4.086, PhysRevB.95.245134, PhysRevLett.121.245701, Rao_2018,PhysRevLett.120.257204}.  

While conventional measures are effective for well-understood phase transitions, they rely on predefined observables tailored to specific cases, limiting their applicability to more intricate regimes. For example, metrics such as the gap ratio and many-body inverse participation ratio (MIPR) successfully distinguish thermal and MBL phases but face significant challenges in identifying complex phases such as many-body critical (MBC) and other non-ergodic phases (with a lot of spectral gaps or non-monotonic density of states). In these cases, the standard metrics struggle and mere deviations from standard indicators are often interpreted as signatures of criticality rather than definitive phase boundaries (see Appendix~\ref{appA}). These limitations underscore the need for a more flexible and data-driven framework—one that can systematically classify phases without relying on prior assumptions about order parameters or correlation functions. Machine learning provides such a framework, enabling unbiased phase identification and uncovering subtle structures in the spectrum that conventional measures may overlook because of the lack of complex regression analysis unlike the supervised machine learning which performs a complex non-linear regression per se before it generates a stable output~\cite{PhysRevLett.120.257204}. This advantage is particularly evident in cases where traditional measures, such as level statistics, fail due to spectral complexities (see Appendix~\ref{appC} for details). 

In this work, we introduce a machine learning-based framework that leverages simple yet highly informative input features—eigenvalue spacings and eigenvector probability densities (PDs)—to uncover and analyze phase transitions in the interacting EAAH model mentioned earlier. Building on the findings of Ref.~\cite{PhysRevLett.126.080602}, which claimed the existence of a many-body critical phase alongside the ergodic and many-body localized phases, we provide a detailed study of searching the MBC phase alongside ergodic and MBL phases, albeit in finite systems, using supervised learning. Fully connected neural networks trained on eigenvalue spacings and eigenvector PDs not only classify these phases with remarkable accuracy but also reveal the tantalizing possibility of distinct sub-phases within the many-body critical regime. Furthermore, our application of PCA to eigenvector PDs highlights its potential as a tool for unsupervised phase identification, with the first principal component capturing dominant phase-transition-related information after suitable preprocessing in the many-body systems. 

For single-particle systems, we demonstrate the effectiveness of PCA entropy~\cite{WOLD198737} and its numerical derivative in systematically identifying phase transitions. Another significant outcome is the validation of critical exponents through scaling collapse analyses. The agreement between exponents derived from neural network outputs and those obtained via direct scaling of input data highlights the reliability of this approach.  We further demonstrate that ML output surpasses conventional measures in phase classification, particularly when level statistics prove inconclusive. Applied to a model with ambiguous spectral signatures, our neural network accurately discerns phases by identifying subtle patterns in eigenvalue spacings(see Appendix~\ref{appC}). By integrating data-driven techniques with traditional scaling analyses, our work establishes a versatile framework for exploring (unknown) complex phase transitions in disordered quantum systems.

This paper is organized as follows: Section~\ref{sec:level2} describes the Hamiltonian model and input data. Section~\ref{sec:level3} outlines the neural network architecture for supervised learning and presents binary, three-class, and four-class classification results. In Section~\ref{sec:level4}, we apply PCA to eigenvector probability densities to explore phase transitions. Section~\ref{sec:level5} discusses scaling collapse and extraction of critical exponents. Finally, we summarize our findings and discuss the outlook of our work in Section~\ref{sec:level6}. %followed by additional analyses of gap ratio, many-body participation ratio, and multi-class classifiers for single-particle systems provided in the Appendix.

%%%%%%%%%%%%%%%%%%%%%%%%%%%%%%%%%%%%%%%%%%%%%%%%%%%
%%%%%%%%%%%%%%%%%%%%%%%%%%%%%%%%%%%%%%%%%%%%%%%%%%%
%%%%%%%%%%%%%%%%%%%%%%%%%%%%%%%%%%%%%%%%%%%%%%%%%%%

\section{Model and Methods}\label{sec:level2}
\subsection{Hamiltonian}
The interacting EAAH model is described by the Hamiltonian~\cite{PhysRevB.42.8282, PhysRevB.50.11365, PhysRevE.70.066203, PhysRevB.91.014108}:
\begin{equation}
\hat{H}= \hat{H}_{\text{hop}}+ \hat{H}_{\text{os}}+ \hat{H}_{\text{int}},
\label{eq1}
\end{equation}
where the components are defined as follows:
\begin{align}
\nonumber \hat{H}_{\text{hop}}=&\sum_{i=1}^{N}  \left\{ 1+\mu\cos\left[2\pi \left(i+ \frac{1}{2}\right)\alpha +\theta_p\right] \right\}\hat{c}_i^{\dagger}\hat{c}_{i+1}+\text{H.c.},\\
\nonumber \hat{H}_{\text{os}}=&\lambda \sum_{i=1}^{N} \cos(2\pi \alpha i+\theta_p)\hat{c}_i^{\dagger}\hat{c}_{i}, \\
\nonumber \hat{H}_{\text{int}}=&U\sum_{i=1}^{N-1}\hat{c}_i^{\dagger}\hat{c}_{i}\hat{c}_{i+1}^{\dagger}\hat{c}_{i+1}.
\end{align}

Here, $\hat{c}_i^{\dagger}$ ($\hat{c}_i$) denotes the fermionic creation (annihilation) operator at site $i$. The parameter $\mu$ represents the modulation strength of the incommensurate nearest-neighbor hopping, while $\lambda$ characterizes the strength of the quasiperiodic on-site potential. The quasiperiodicity parameter $\alpha$ is taken to be the golden mean, $(\sqrt{5}-1)/2$~\cite{Modugno_2009}, ensuring an irrational value. The interaction strength is denoted by $U$, and $\theta_p$, the global phase, is randomly chosen from a uniform distribution within $[0, 2\pi]$. The system consists of $N$ lattice sites under periodic boundary conditions.

At half-filling, corresponding to a particle number $N_p$ with a filling fraction $N_p/N = 0.5$, the model has been claimed to exhibit distinct many-body phases~\cite{PhysRevLett.126.080602}. For $\mu < 1$ and $0<\lambda \lesssim 3$, the system resides in the ergodic phase. When $\mu > 1$, roughly for $\lambda<3$ and $3<\lambda \lesssim 2\mu$, the system is in the MBC phase which is a more robust example of nonergodic extended phase~\cite{PhysRevLett.126.080602} than the one found earlier in the generalized Aubry-Andre-Harper model~\cite{PhysRevLett.121.245701}. The remaining region of the phase diagram corresponds to the MBL phase. Unless stated otherwise, the interaction strength $U$ is fixed at $U = 1$.

In the non-interacting limit ($U = 0$), the model reduces to the well-known Aubry-André-Harper (AAH) model when $\mu = 0$~\cite{aubry1980analyticity, PhysRevB.28.4272}. For the AAH model, a critical point occurs at $\lambda = 2$, where all single-particle eigenstates exhibit multifractal behavior~\cite{RevModPhys.80.1355}. At this critical point, the model maps onto itself in both position and momentum space. The non-interacting phase diagram is similarly well-established~\cite{PhysRevB.50.11365,chang1997multifractal, PhysRevB.109.035164}: for $\mu < 1$ and $\lambda < 2$, the system lies in the delocalized phase, whereas for $\mu > 1$ and below the line $\lambda = 2\mu$, it transitions to the critical (all states multifractal) phase. The remaining region corresponds to the localized phase.

We refrain from showing schematics of phase diagrams of both the interacting and non-interacting models which can be found, along with a detailed study, in the existing literature~\cite{PhysRevLett.126.080602, PhysRevB.42.8282, PhysRevB.50.11365, PhysRevE.70.066203, PhysRevB.91.014108}.

\subsection{Input data}
We consider two types of input datasets for supervised learning. For a system size $N$ with $N_p$ fermions, the Hilbert space dimension is given by $D = {N \choose N_p}$. We perform exact diagonalization of the many-body Hamiltonian across multiple disorder realizations. The first dataset comprises the spacings of the eigenvalues arranged in ascending order, $\{E_{i+1}-E_{i}\}$, where $E_i$ denotes the energy levels. The input layer dimension for the neural networks using eigenvalue spacings is $D-1$. To normalize the energy levels, they are rescaled to lie within $0 < \epsilon_i < 1$, where $\epsilon_i = \frac{E_i-E_1}{E_D-E_1}$, with $E_1$ and $E_D$ representing the minimum (ground state) and the maximum energy levels, respectively. The second dataset consists of the probability densities (PDs) corresponding to the eigenvectors in the energy window $\epsilon_i \in [0.53, 0.55]$. The probability densities (PDs) are defined as the squared amplitudes of the components of the eigenvector. The resulting vectors of PDs are not normalized. For this dataset, the input layer dimension for the neural network is $D$. Notably, no preprocessing is applied to either dataset, as we aim to uncover the intrinsic features of the system directly through the energy level spacings and eigenstates. 

For the numerical simulations, we consider a system size of $N = 14$ with half filling, resulting in a Hilbert space dimension of $D = 3432$. For eigenvalue spacings and eigenvector PDs, the training datasets consist of $1250$ and $10000$ samples per class, respectively. For testing, we average over $300$ samples for eigenvalue spacings and $500$ samples for eigenvector PDs. Note that for eigenvalue spacings, the number of training and testing samples corresponds directly to the number of disorder realizations. However, in the case of eigenvector probability densities (PDs), the samples are drawn from the energy window $\epsilon_i \in [0.53, 0.55]$, leading to multiple samples being obtained from a single disorder realization. Hence, the number of samples differs from the number of disorder realizations. Notably, for the eigenvalue spacing dataset, the network achieved high accuracy even with a relatively smaller dataset. 

The trained networks are utilized to predict phase transitions along three distinct paths in the parameter space: (i) $\lambda = 0.5$, $0<\mu<2.5$ to explore the ergodic to MBC transition, (ii) $\mu = 0.5$, $0<\lambda<5$ to study the ergodic to MBL transition, and (iii) $\mu = 1.8$, $0<\lambda<5$ to investigate the MBC to MBL transition. A detailed discussion of the network architectures and methodologies is presented in the following section.

For the PCA analysis, we employ the same $500$ eigenvector PDs from the test set used in supervised learning for the many-body case. In the single-particle system, we generate $500$ eigenvector PDs for various system sizes to study transitions along specific parameter paths: (i) $\lambda = 0.5$, $0<\mu<2.5$ for the delocalized to critical phase transition, (ii) $\mu = 0$, $0<\lambda<5$ for the delocalized to localized transition, and (iii) $\mu = 1.5$, $0<\lambda<5$ for the critical to localized phase transition. These datasets also serve as testing data for the supervised learning analysis described in Appendix~\ref{appB}.
%%%%%%%%%%%%%%%%%%%%%%%%%%%%%%%%%%%%%%%%%%%%%%%%%%%%%%%%
\begin{figure*}
\centering
\stackunder{\hspace{-5cm}(a)}{\includegraphics[width=4.5cm]{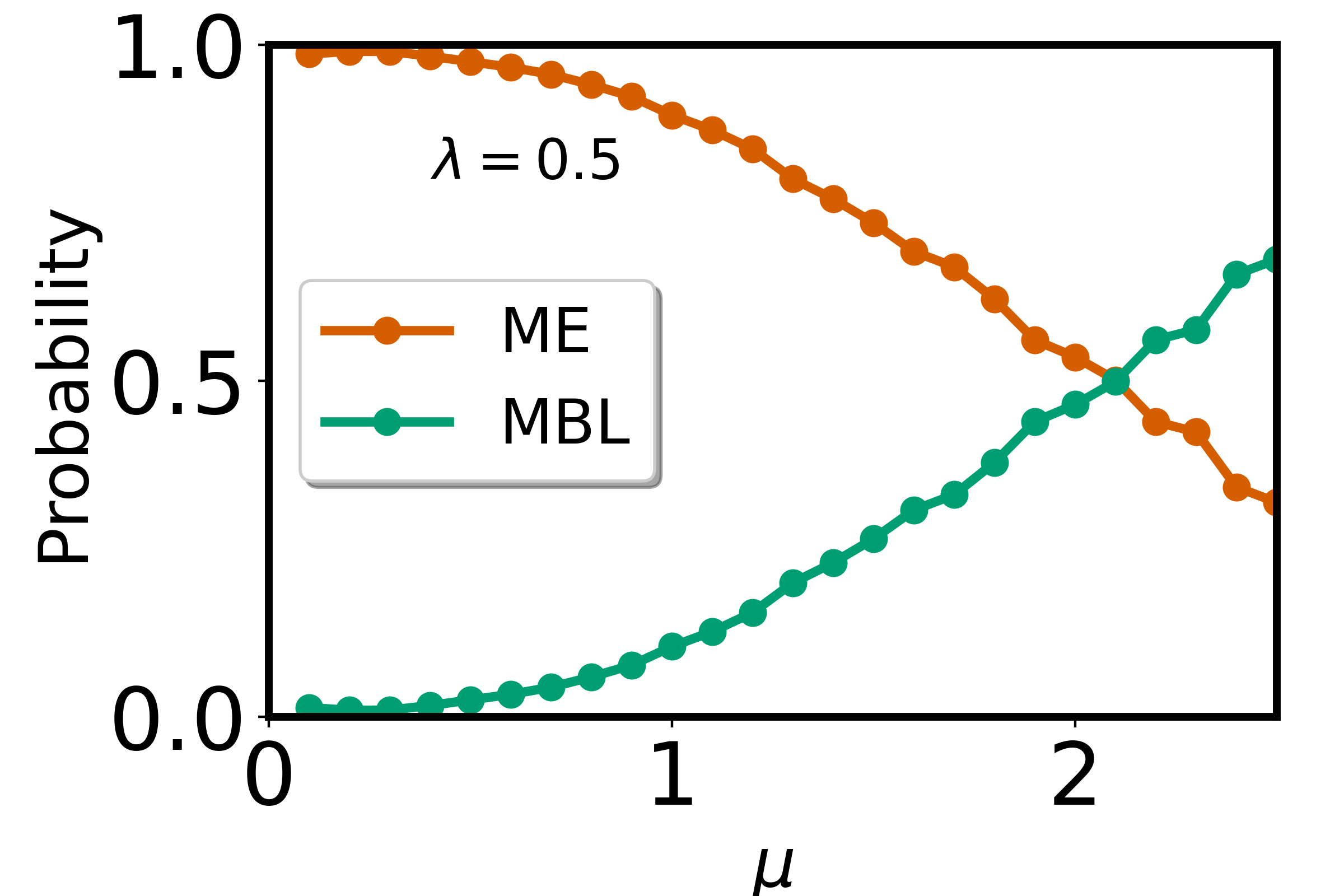}}
\stackunder{\hspace{-4.5cm}(b)}{\includegraphics[width=4.5cm]{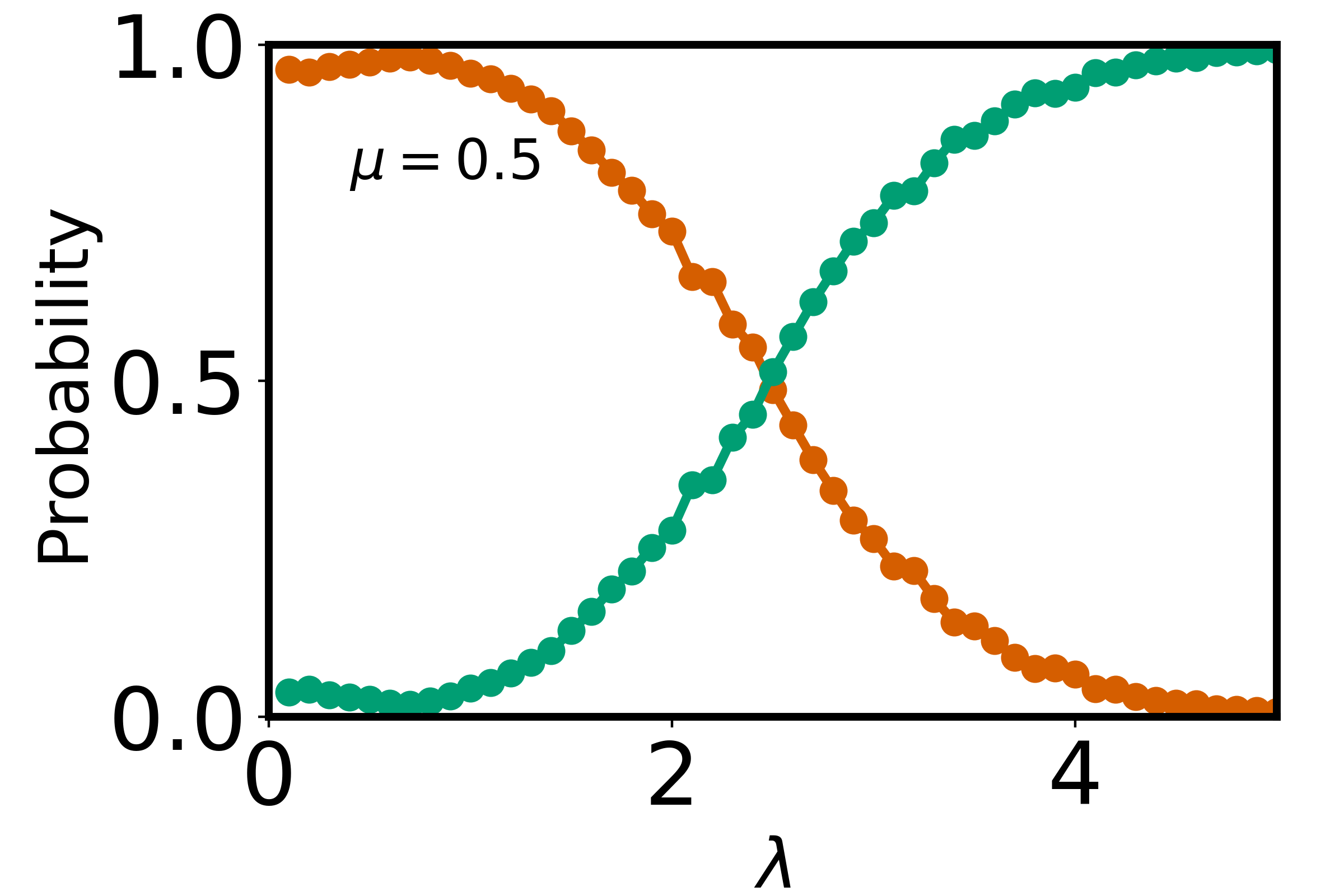}}
\stackunder{\hspace{-4.5cm}(c)}{\includegraphics[width=4.5cm]{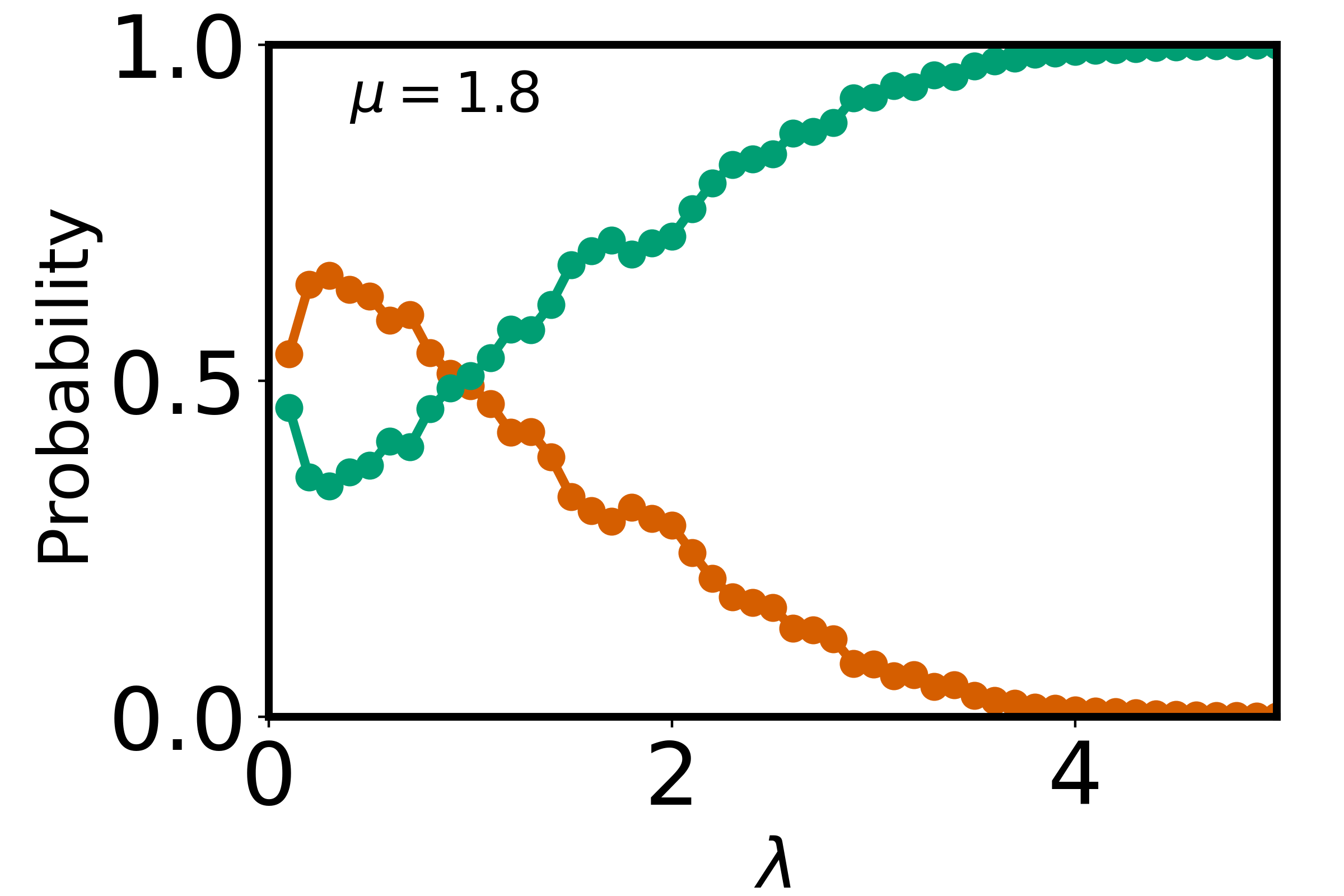}}
\vspace{-0.4cm}

\stackunder{\hspace{-5cm}(d)}{\includegraphics[width=4.5cm]{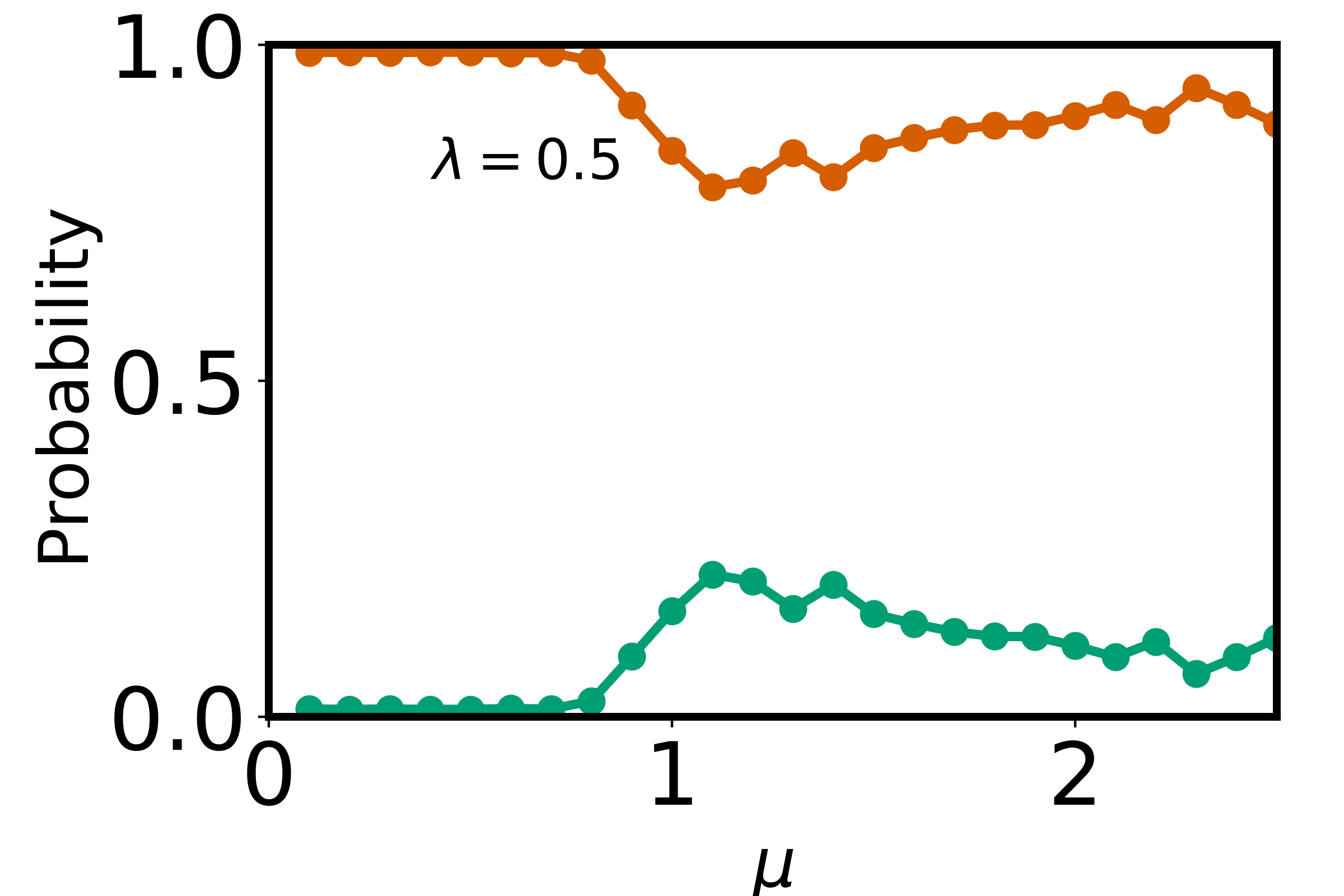}}
\stackunder{\hspace{-4.5cm}(e)}{\includegraphics[width=4.5cm]{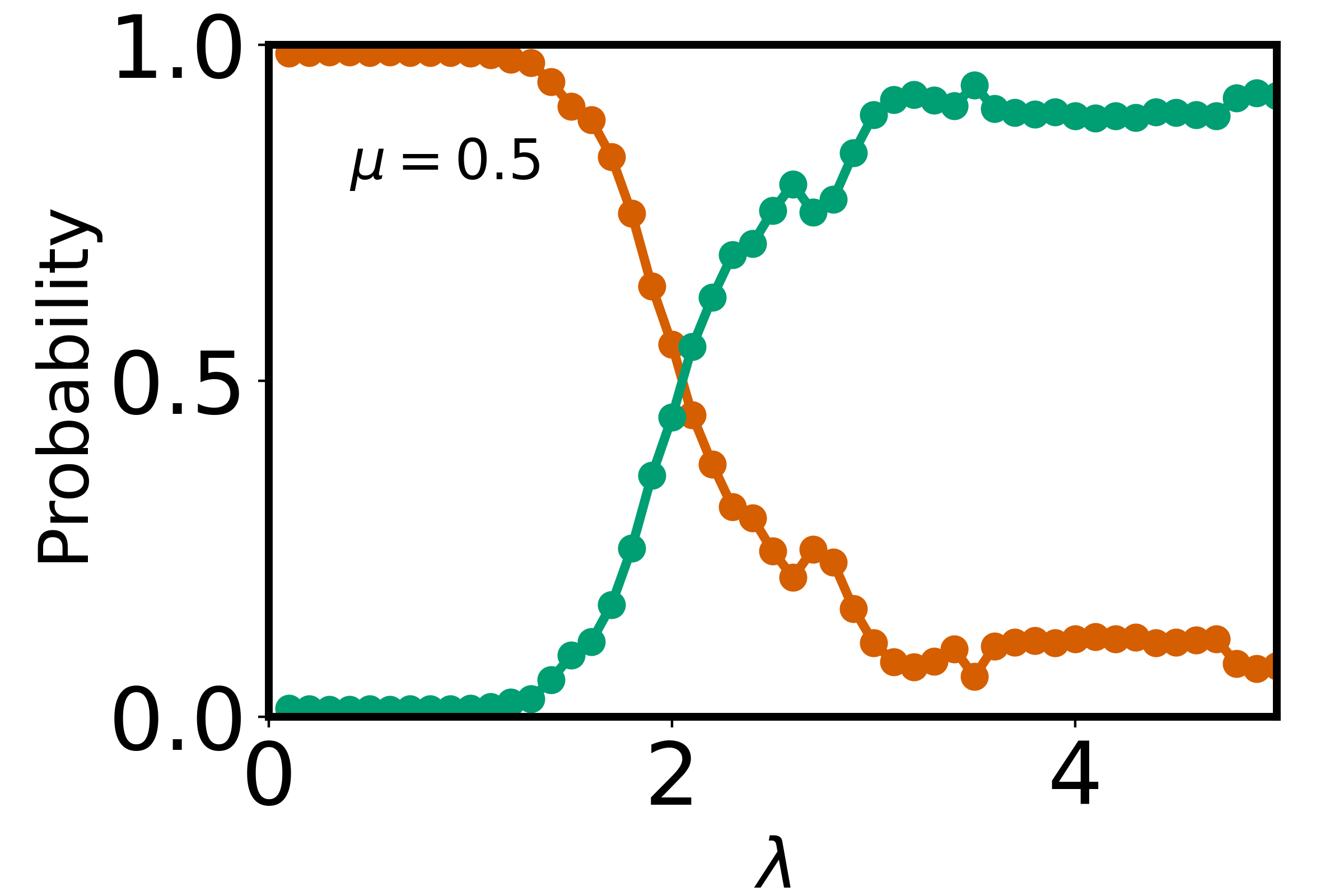}}
\stackunder{\hspace{-4.5cm}(f)}{\includegraphics[width=4.5cm]{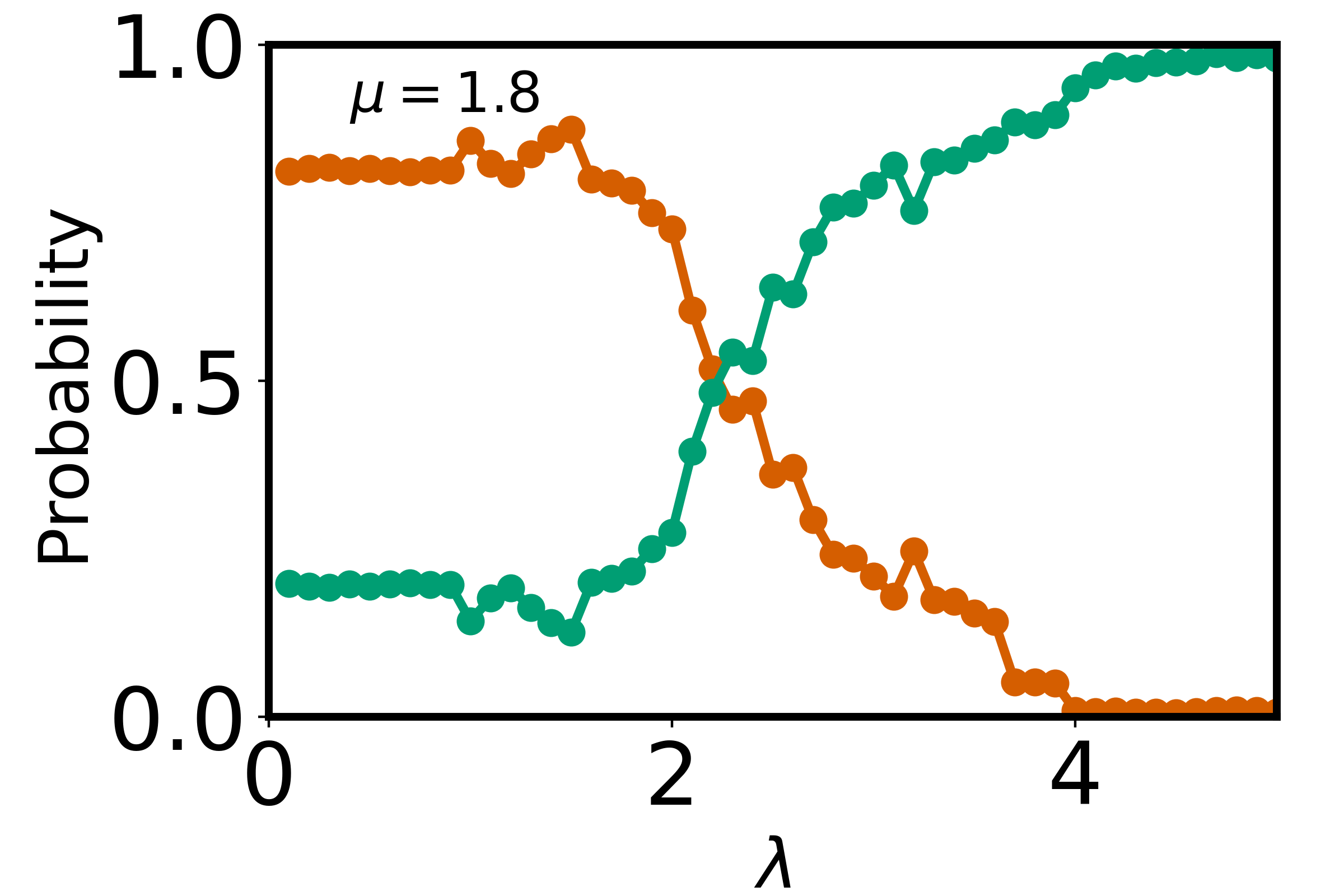}}
\caption{\label{fig1}Neural network predictions for the binary classifier are shown for (a, d) increasing hopping amplitude $\mu$ at $\lambda = 0.5$, and for increasing disorder strength $\lambda$ at fixed (b, e) $\mu = 0.5$  and (c, f) $\mu = 1.8$. Panels (a–c) utilize eigenvalue spacings, with averages computed over $300$ test datasets, while panels (d–f) are based on the probability density corresponding to eigenstates, averaged over $500$ test datasets. The system size is $N = 14$ and half-filling is considered.}
\end{figure*}

%%%%%%%%%%%%%%%%%%%%%%%%%%%%%%%%%%%%%%%%%%%%%%%%%%%%%%%%
\section{Supervised learning}\label{sec:level3}

\subsection{Neural Network Framework}
In this section, we introduce the general terminology and concepts underlying the neural network-based approach employed in this study. We implement a fully connected artificial neural network (ANN) that applies linear transformations and non-linear activation functions~\cite{https://doi.org/10.48550/arxiv.2109.14545} to input data, propagating it through successive layers of neurons. Each layer is characterized by weights and biases that map vectors from one layer to the next. The output layer comprises nodes corresponding to the desired outputs. ANNs with multiple hidden layers constitute what is commonly referred to as deep learning~\cite{GoodBengCour16}. For supervised learning tasks, the network is trained on a labelled dataset, enabling it to classify unseen data from a test set with high accuracy.
For the hidden layers in our implementation, we employ two widely used activation functions~\cite{Fukushima1975Cognitron}:

1. The Rectified Linear Unit (ReLU)~\cite{agarap2019deep}:
\begin{equation}
  \text{ReLU: } f(x_i) =
    \begin{cases}
      0, & x_i < 0, \\
      x_i, & x_i \geq 0,
    \end{cases}       
\end{equation}

2. The Exponential Linear Unit (ELU)~\cite{clevert2016fastaccuratedeepnetwork}:
\begin{equation}
  \text{ELU: } f(x_i) =
    \begin{cases}
      \alpha(e^{x_i} - 1), & x_i \leq 0, \\
      x_i, & x_i > 0,
    \end{cases}       
\end{equation}

Although the initial inputs to the network are strictly positive, the use of activation functions in the hidden layers is essential due to the intermediate computations involving weights and biases. These computations can result in both positive and negative values propagating through the network. For ReLU, negative values are set to zero, introducing sparsity in the activations. In contrast, ELU smooths the output for negative values, with the parameter \(\alpha > 0\) controlling the saturation value. This smoothing property of ELU can improve gradient flow and lead to faster convergence compared to ReLU in deeper networks. Here $\alpha$ is set to the default value $1$.

In the output layer, the choice of activation function depends on the classification task. For binary classification, where the output is a single probability, we use the Sigmoid activation function~\cite{Feng2019Performance}:
\begin{equation}
  \text{Sigmoid: } f(x_i) = \frac{1}{1 + e^{-x_i}}.
\end{equation}
For multi-class classification, the Softmax activation function~\cite{Feng2019Performance} is employed:
\begin{equation}
  \text{Softmax: } f(x_i) = \frac{e^{x_i}}{\sum_j e^{x_j}}.
\end{equation}
The Softmax function outputs probabilities for each class, ensuring that the sum of probabilities across all classes is equal to $1$.

To address computational efficiency and memory constraints, the training data is divided into smaller batches, which are fed sequentially into the network. The process of iterating through the entire training dataset in these batches constitutes an epoch. After each epoch, the cross-entropy loss function~\cite{Goodfellow-et-al-2016} is computed, which measures the difference between the predicted and actual outputs. The optimizer updates the model parameters to minimize the loss function. In this study, we use the Stochastic Gradient Descent (SGD) optimizer~\cite{doi:10.1137/16M1080173} with its default learning rate unless stated otherwise.

The dataset is split into 80\% training and 20\% validation subsets. At every epoch, the model trains on the 80\% training dataset and is evaluated on the 20\% validation dataset. The use of a validation set—comprising previously unseen data—helps mitigate overfitting or underfitting. For a well-trained network, the training and validation losses should gradually decrease and converge toward zero as the number of epochs increases. Additionally, dropout layers are incorporated into the neural network to further mitigate overfitting. A dropout layer randomly sets a fraction of the neurons to zero during each training step, preventing the network from becoming overly reliant on specific neurons and promoting better generalization to unseen data.

%%%%%%%%%%%%%%%%%%%%%%%%%%%%%%%%%%%%%%%%%%%%%%%%%%%%%%%%

\begin{table}[b]
\centering
\begin{tabular}{ | m{8em} | m{8em} | m{8em} | }
  \hline
  \textbf{Input Data} & \textbf{Eigenvalue Spacings} & \textbf{Eigenstate Probability Density} \\
  \hline
  & &\\
  Neurons in Input Layer & $D-1$ & $D$ \\
  \hline
  & &\\
  Neurons in Hidden Layer,\hspace{1cm} with dropout & $\text{int}[(D\times 2)/10] + 1$,\hspace{1cm} None & 512,\hspace{2cm} 0.4 \\
  \hline
    & &\\
  Hidden Layer Activation Function & ELU & ReLU \\
  \hline
    & &\\
  Epochs & 50 & 800 \\
  \hline
   & &\\
  Batch Size &  30 & 50 \\
  \hline
\end{tabular}
\caption{\label{tab1} Neural network architecture for the binary classifier with a single hidden layer, utilizing eigenvalue spacings and eigenstate probability densities.}
\end{table}
%%%%%%%%%%%%%%%%%%%%%%%%%%%%%%%%%%%%%%%%%%%%%%%%%%%%%%%%
\begin{figure*}
\centering
\stackunder{\hspace{-5cm}(a)}{\includegraphics[width=4.5cm]{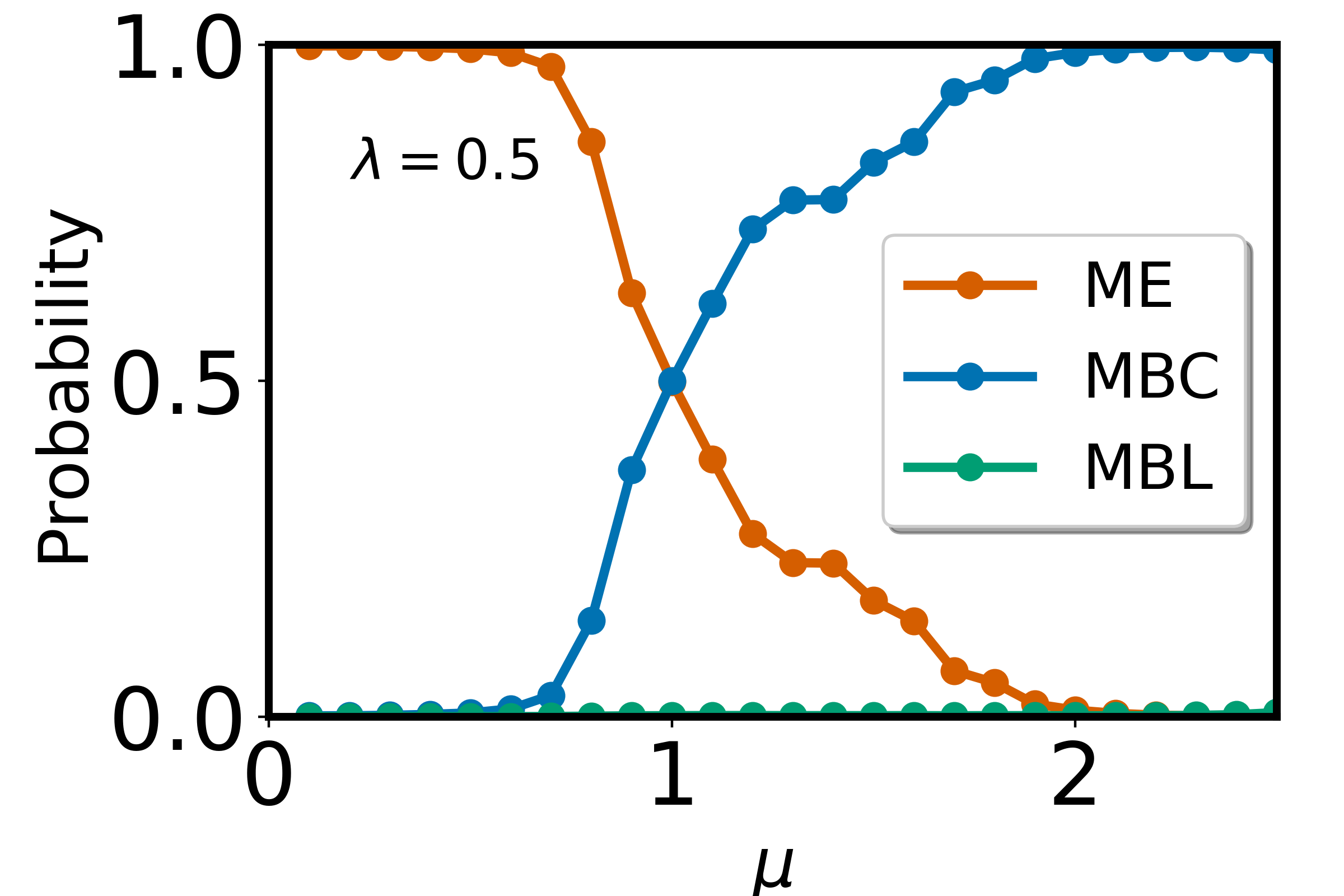}}
\stackunder{\hspace{-4.5cm}(b)}{\includegraphics[width=4.5cm]{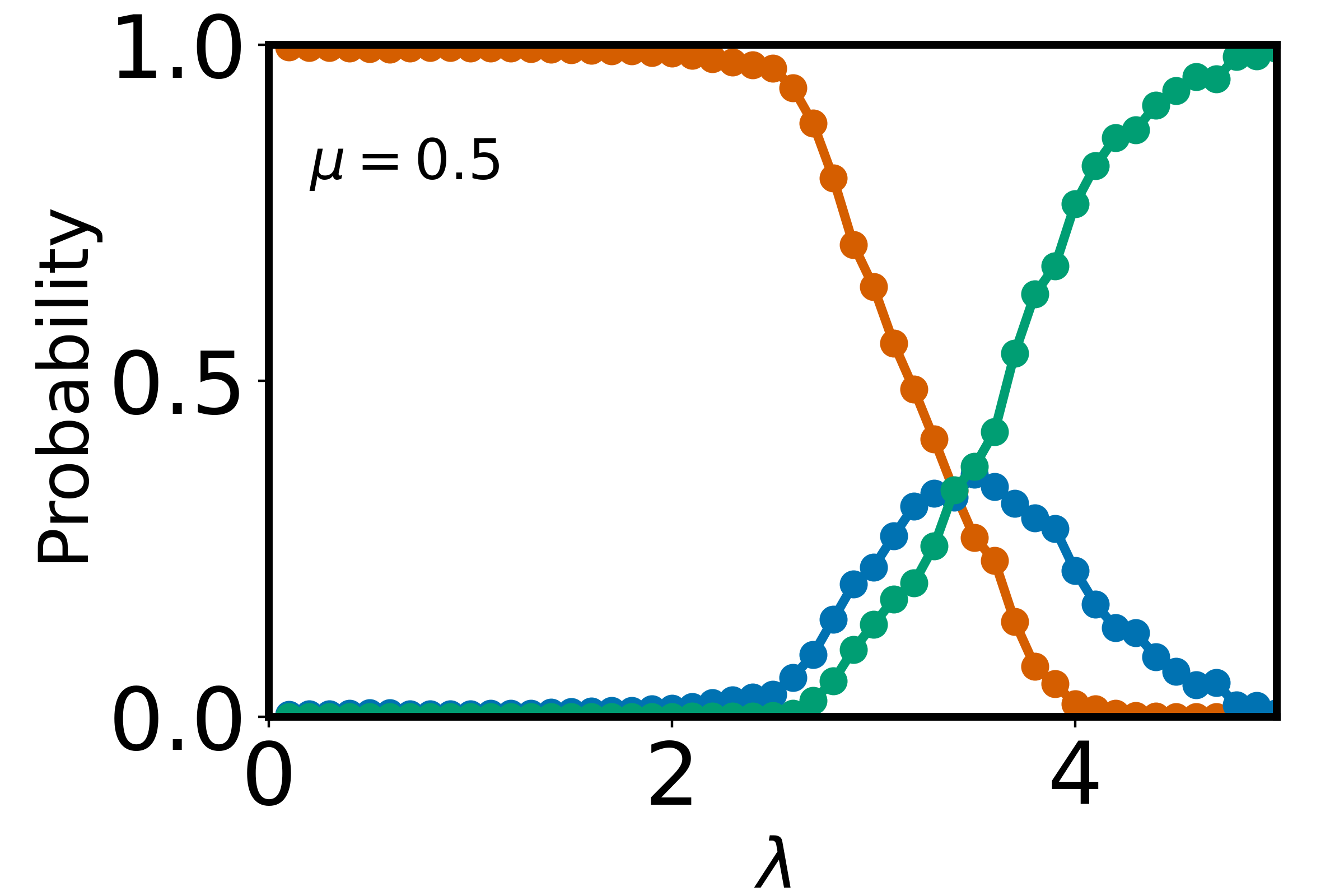}}
\stackunder{\hspace{-4.5cm}(c)}{\includegraphics[width=4.5cm]{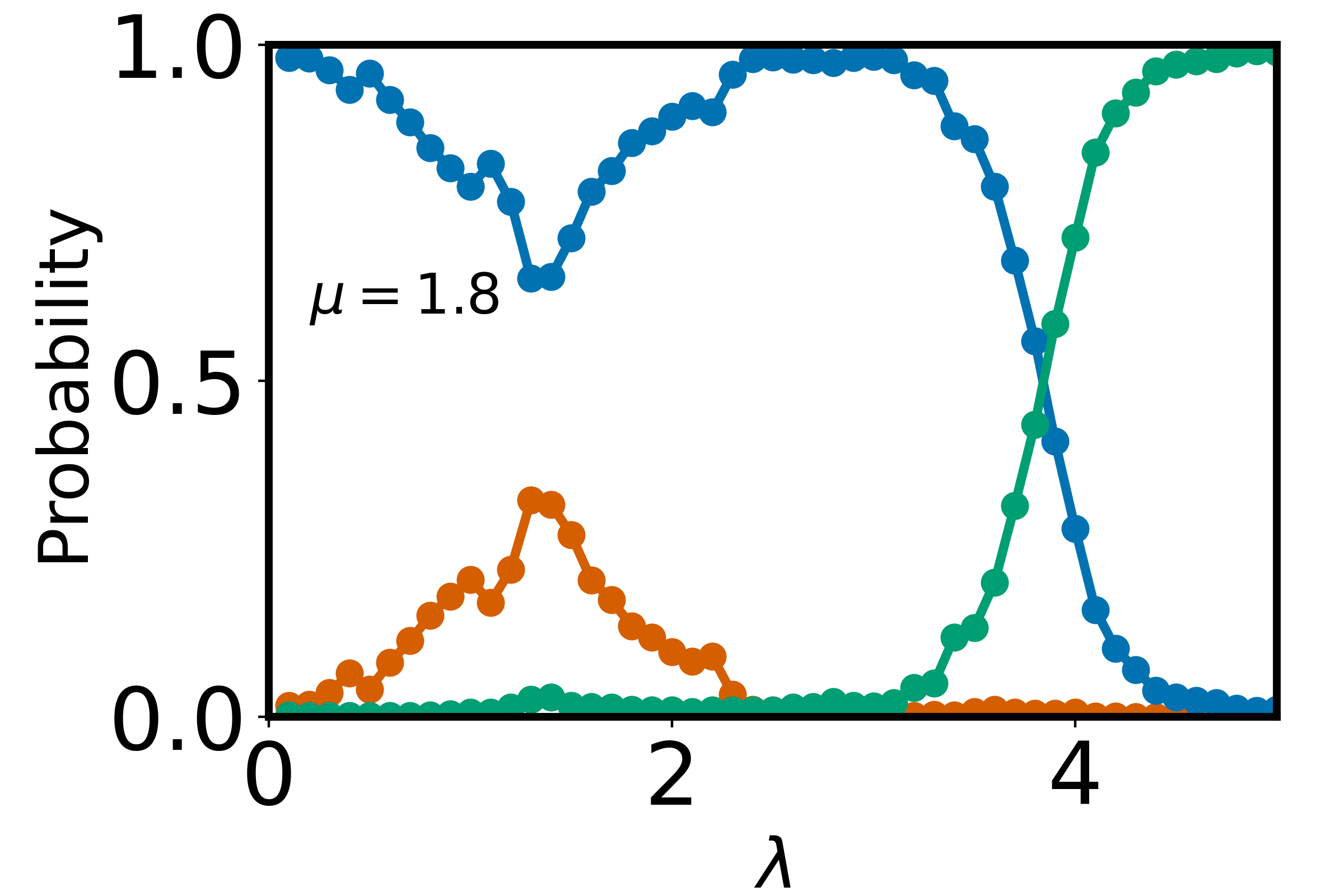}}
\vspace{-0.4cm}

\stackunder{\hspace{-4.5cm}(d)}{\includegraphics[width=4.5cm]{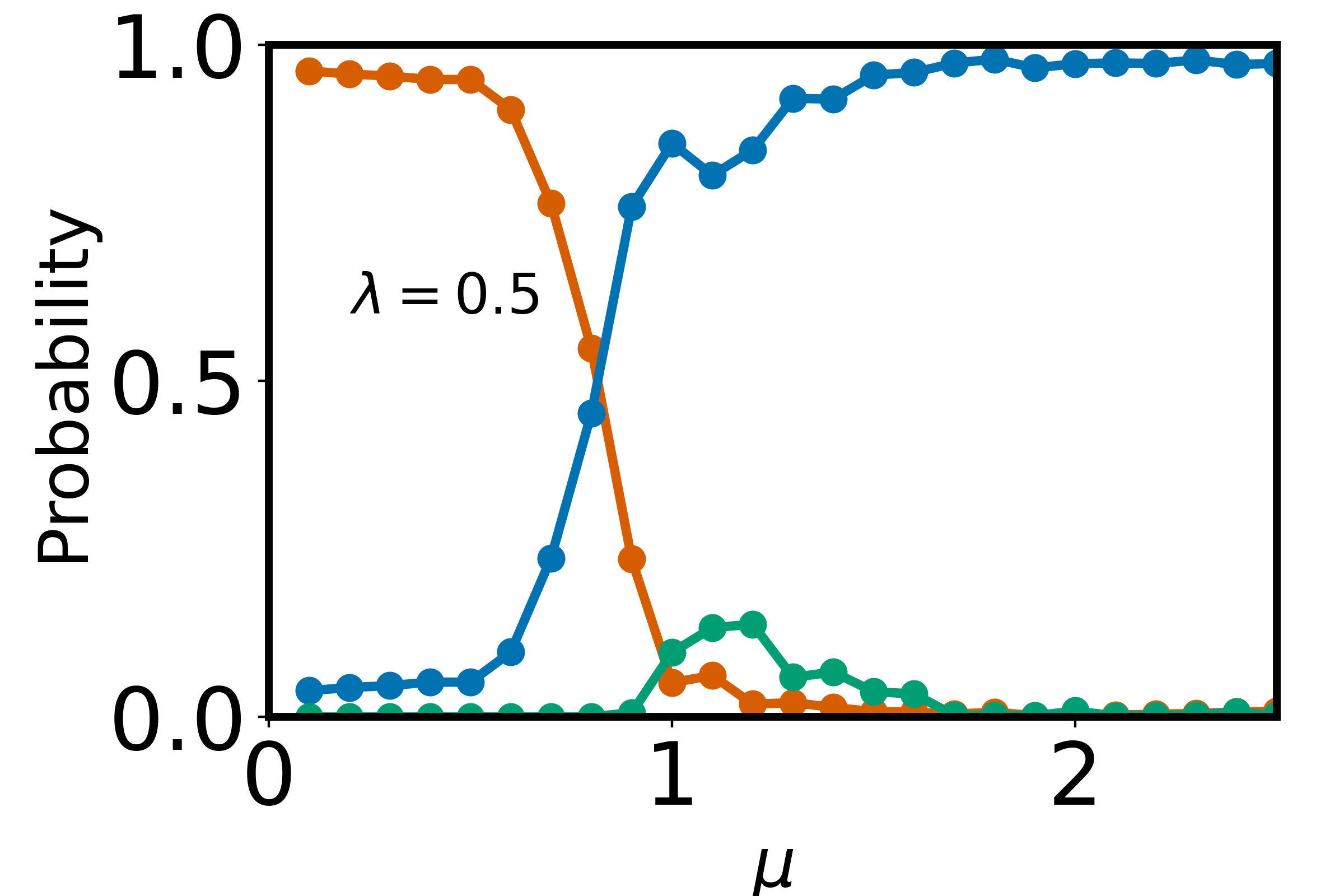}}
\stackunder{\hspace{-4.5cm}(e)}{\includegraphics[width=4.5cm]{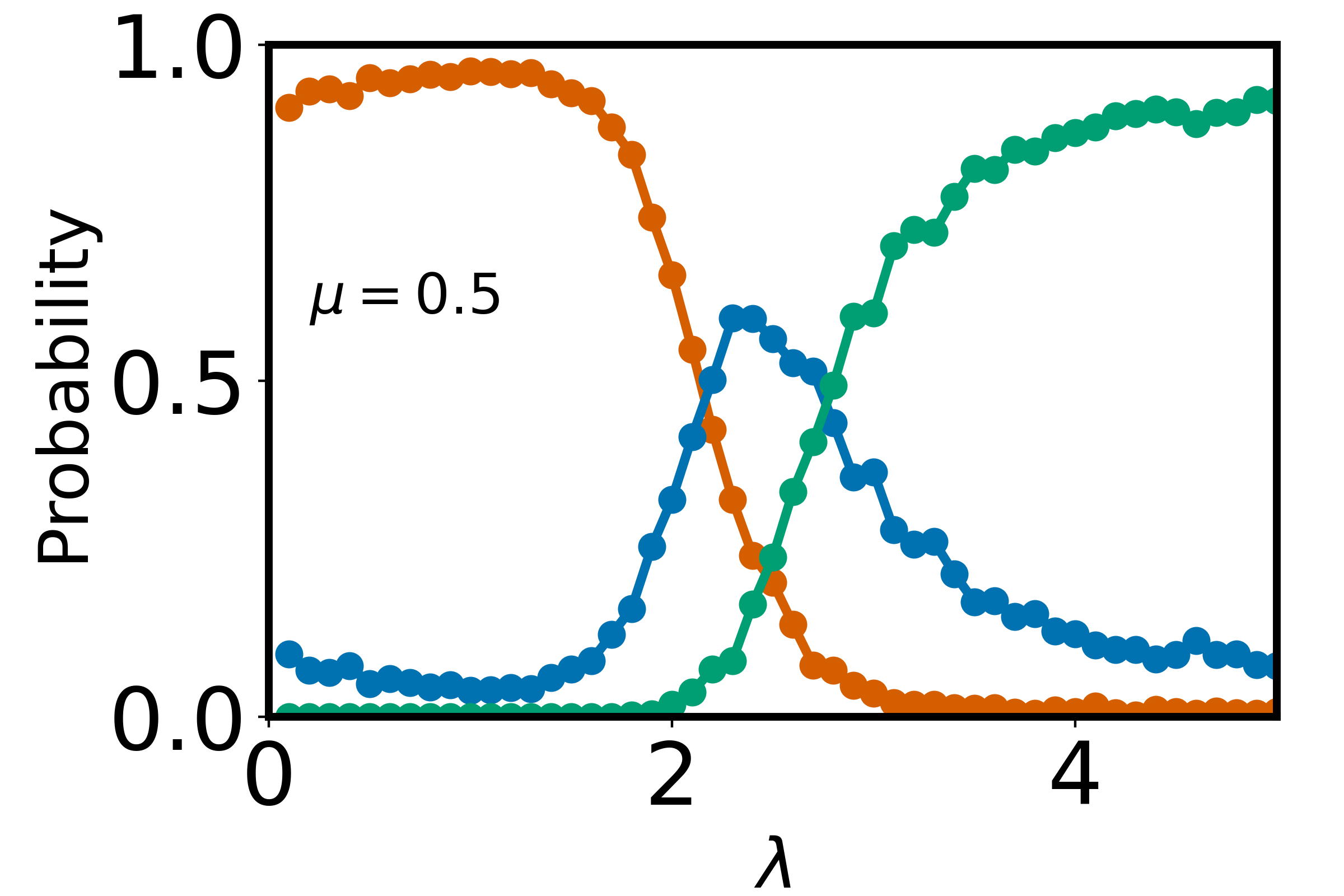}}
\stackunder{\hspace{-4.5cm}(f)}{\includegraphics[width=4.5cm]{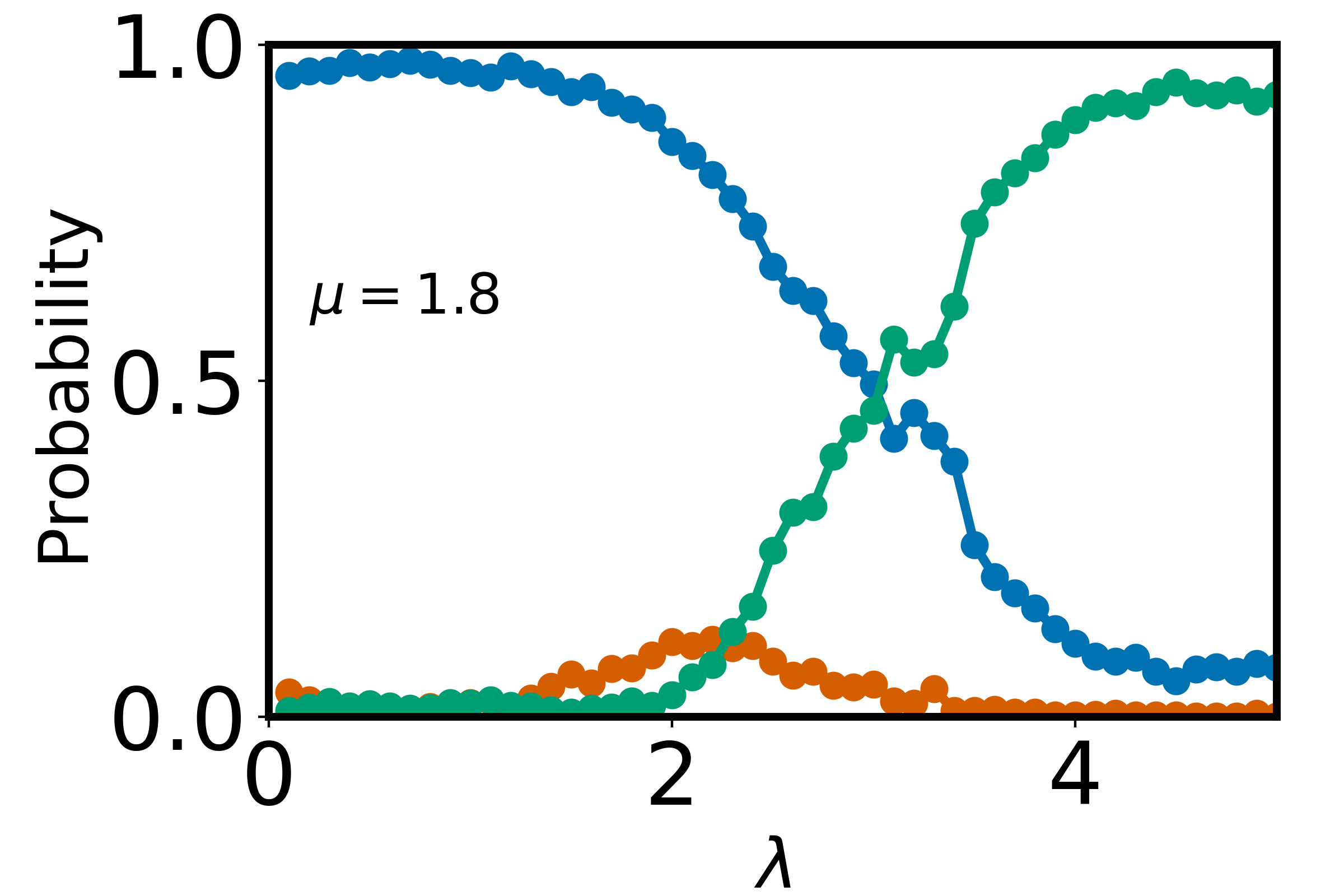}}
\caption{\label{fig2}Neural network predictions for the $3-$ class classifier are shown for varying hopping amplitude $\mu$ at (a, d) $\lambda = 0.5$, and for increasing disorder strength $\lambda$ at fixed (b, e) $\mu = 0.5$ and (c, f) $\mu = 1.8$. Panels (a–c) utilize eigenvalue spacings, with averages computed over $300$ test datasets, while panels (d–f) are based on the probability density corresponding to eigenstates, averaged over $500$ test datasets. The system size is $N = 14$ and half-filling is considered.}
\end{figure*} 
%%%%%%%%%%%%%%%%%%%%%%%%%%%%%%%%%%%%%%%%%%%%%%%%%%%%%%%%

\subsection{Binary classification}
We begin our analysis by training a binary classifier to differentiate between data from the ergodic (ME) and the MBL phase. We focus on how the network behaves when tasked with identifying the MBC phase, as it was not trained exclusively for this phase. For training, we generate datasets from points deep within the ME phase, with parameters $\mu=0.3$ and $\lambda=1$, and from points deep within the MBL phase, with $\mu=0.5, 2$ and $\lambda=4.5$. As previously mentioned, we generate $1250$ and $10000$ disorder realizations ($\theta_i$) per class for the eigenvalue spacings and probability densities (PDs) corresponding to the eigenstates, respectively.

In Table~\ref{tab1}, we summarize the architecture and details of the neural networks. The network consists of an input layer, a single hidden layer, and an output layer connected to a Sigmoid activation function. When the dropout layer is also considered, it is mentioned corresponding to that layer in the table. The cost function employed is binary cross-entropy, and the optimizer used is Stochastic Gradient Descent (SGD) with the default learning rate. Once trained, the neural networks are tested on a separate testing dataset. The binary classifier has a single neuron in the output layer, which represents the probability $P$ that a sample belongs to the MBL phase, while $1-P$ indicates the probability of the sample belonging to the ME phase. When the network encounters a class for which it was not trained, we expect the output probability $P$ to lie between $0$ and $1$.

To evaluate the classifier’s performance, we average the output probability $P$ over several test samples and plot the results in Fig.~\ref{fig1}. In Figs.~\ref{fig1}(a)-(c), we present the probabilities $P$ and $1-P$ obtained from the neural network trained on eigenvalue spacings. Notably, in both the ME and MBL phases, the respective probability approaches 1, indicating strong classification confidence. However, in the MBC phase, with $\mu>1$ in Fig.~\ref{fig1}(a) and $\lambda<4$ in Fig.~\ref{fig1}(c), the output probability tends to approximately 0.5, suggesting uncertainty in the network's classification.

A similar analysis is carried out using the probability densities of the eigenvectors, as shown in Figs.~\ref{fig1}(d)-(f). While the binary classifier trained on eigenvalue spacings clearly distinguishes between the ME and MBL phases, it also reveals a new phase (MBC). However, the classifier trained on eigenvector PDs seems to be confused in the MBC phase. The network classifies the ME and MBL phases with high confidence, but in the MBC phase, it assigns a higher probability to the ME phase unlike for eigenvalue spacings for which output shows no bias in the new phase. This discrepancy suggests that a different machine-learning algorithm for eigenvector PDs might be more suitable for classifying this new phase. Nonetheless, our primary focus here is the multi-class classification approach, which we will explore in the next section.

\subsection{Three-class classification}
In this section, we train a neural network for a three-class classification problem, distinguishing between the ME, MBC, and MBL phases. Utilizing the known phase diagram, we train the network on datasets deeply representative of each phase and then test it on the unseen data. The training data sets for each class are generated as follows: $1250$ disorder realizations for eigenvalue spacings and $10000$ for the probability densities corresponding to eigenstates. For the ME phase, we use $\mu=0.3$ and $\lambda=1, 2.5$; for the MBC phase, $\mu=2$ and $\lambda=0.3,1,2.5,3$; and for the MBL phase, $\mu=0.5, 2$ and $\lambda=4.5$ . The complete architecture of the neural network for the three-class classification is shown in Table~\ref{tab2}. The network comprises an input layer, several hidden layers, and an output layer, where the output is linked to the Softmax activation function. When the dropout layer is also considered, it is mentioned corresponding to that layer in the table. The cost function used is categorical cross-entropy, and the weights of the network are optimized using the SGD optimizer with a learning rate of $0.005$ for eigenvalue spacing data. The accuracy of the network, which measures how well the network has learned, is approximately $99\%$ for the eigenvalue spacings and $95\%$ for the eigenvector PDs. Once trained, the network outputs three values corresponding to the network’s confidence in classifying the input test data into each of the three phases.

\begin{table}
\centering
\begin{tabular}{ | m{8em} | m{8em} | m{8em} | }
  \hline

  \textbf{Input Data} & \textbf{Eigenvalue Spacings} & \textbf{Eigenstate Probability Density} \\
  \hline
      & &\\

  Neurons in Input Layer & $D-1$ & $D$ \\
  \hline
    & &\\

  Neurons in 1st Hidden Layer, with dropout & $\text{int}[(D\times 2)/3]+1$, $0.5$ & $512$, \hspace{2cm} $0.3$ \\
  \hline
    & &\\

  Neurons in 2nd Hidden Layer, with dropout & $512$, \hspace{2cm}$0.2$ & $32$, \hspace{2cm}$0.1$ \\
  \hline
    & &\\

  Neurons in 3rd Hidden Layer, with dropout & $32$, \hspace{2cm}$0.1$ & None \\
  \hline
    & &\\

  Hidden Layer Activation Function & ReLU & ReLU \\
  \hline
    & &\\

  Epochs & $40$ & $800$ \\
  \hline
   & &\\

  Batch Size &$30$ & $100$ \\
  \hline
\end{tabular}
\caption{\label{tab2} Neural network architecture for the three-class classifier utilizing eigenvalue spacings and eigenstate probability densities.}
\end{table}

The probabilities averaged over several datasets for the three-class classifier are shown in Fig.\ref{fig2}. Figs~\ref{fig2}(a-c) and Figs~\ref{fig2}(d-f) show the neural network probabilities corresponding to the three phases, obtained from the eigenvalue spacings and eigenvector PD classifiers, respectively. From Figs.\ref{fig2}(a) and (d), we observe a clear transition from the ME phase to the MBC phase as the hopping strength $\mu$ increases at a fixed disorder strength $\lambda=0.5$, with the probability in each phase approaching $1$. In Figs.\ref{fig2}(b) and (e), where $\mu=0.5$ is fixed and the disorder strength $\lambda$ increases, we observe the transition from the ME phase to the MBL phase. The finite low probability of the MBC phase at the transition point indicates that the MBC phase disappears with increasing system size, with $P_{\text{MBC}}$ approaching $0$. Finally, in Figs.\ref{fig2}(c) and (f), where $\mu=1.8$ is fixed and $\lambda$ increases, we observe the transition from the MBC phase to the MBL phase, with the probability in each phase again approaching $1$. This analysis supports the existence of the MBC phase and explains the ambiguity observed in the binary classifier, which was trained only on the ME and MBL phases.
%%%%%%%%%%%%%%%%%%%%%%%%%%%%%%%%%%%%%%%%%%%%%%%%%%%%%%%%

\begin{figure}[b]
\centering
\stackunder{\hspace{-4cm}(a)}{\includegraphics[width=4.2cm]{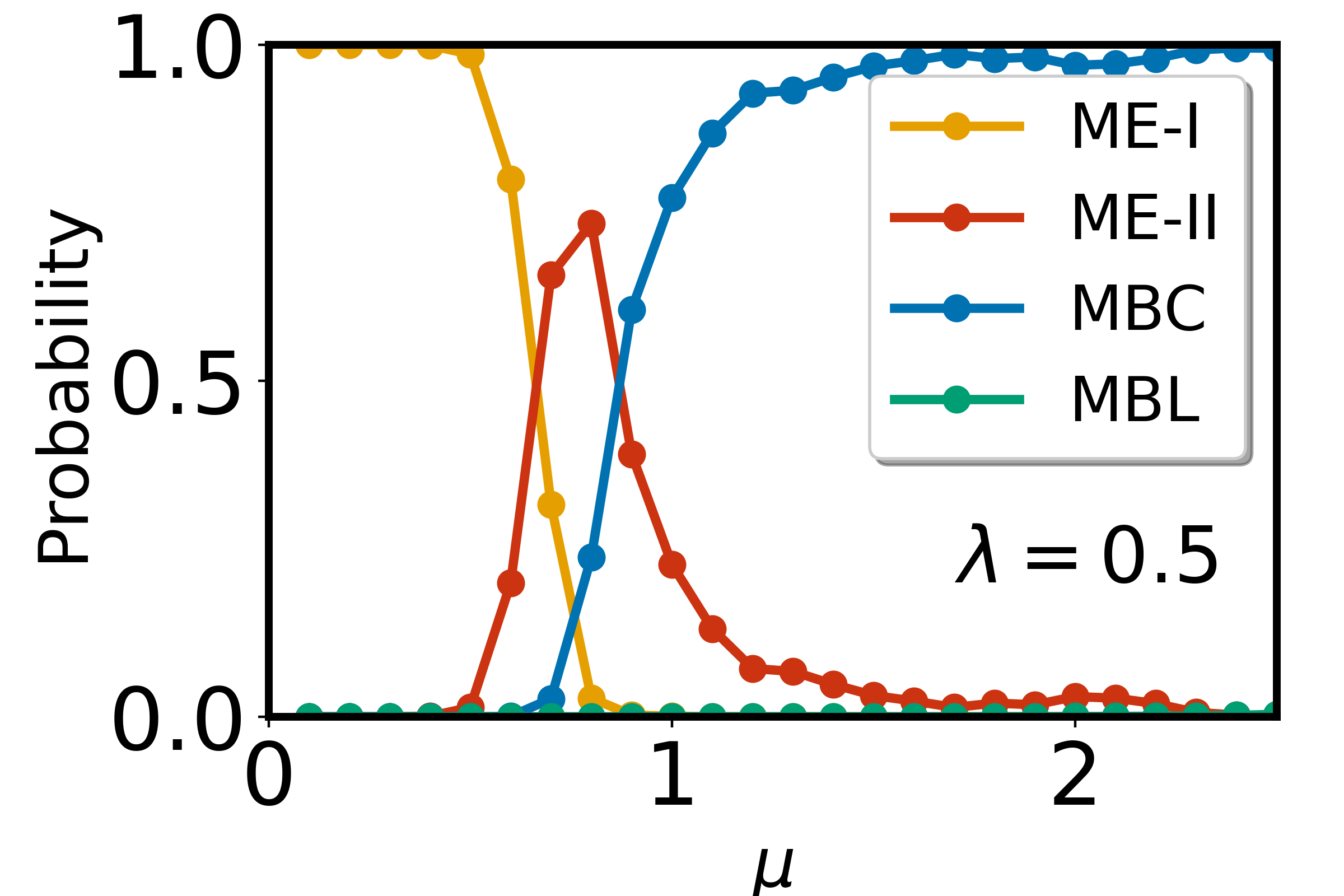}}
\stackunder{\hspace{-4cm}(b)}{\includegraphics[width=4.2cm]{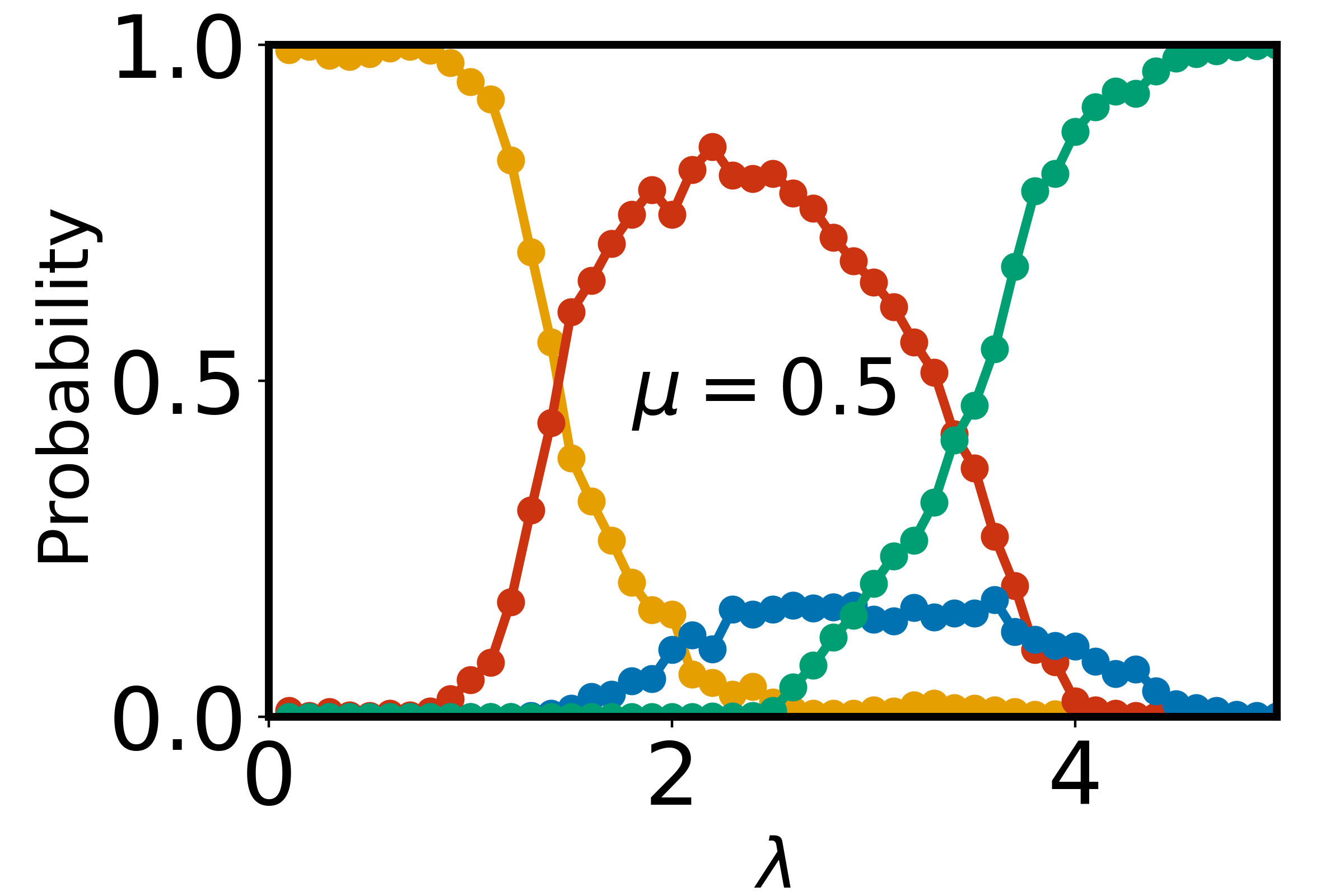}}
\vspace{-0.4cm}

\stackunder{\hspace{-4cm}(c)}{\includegraphics[width=4.2cm]{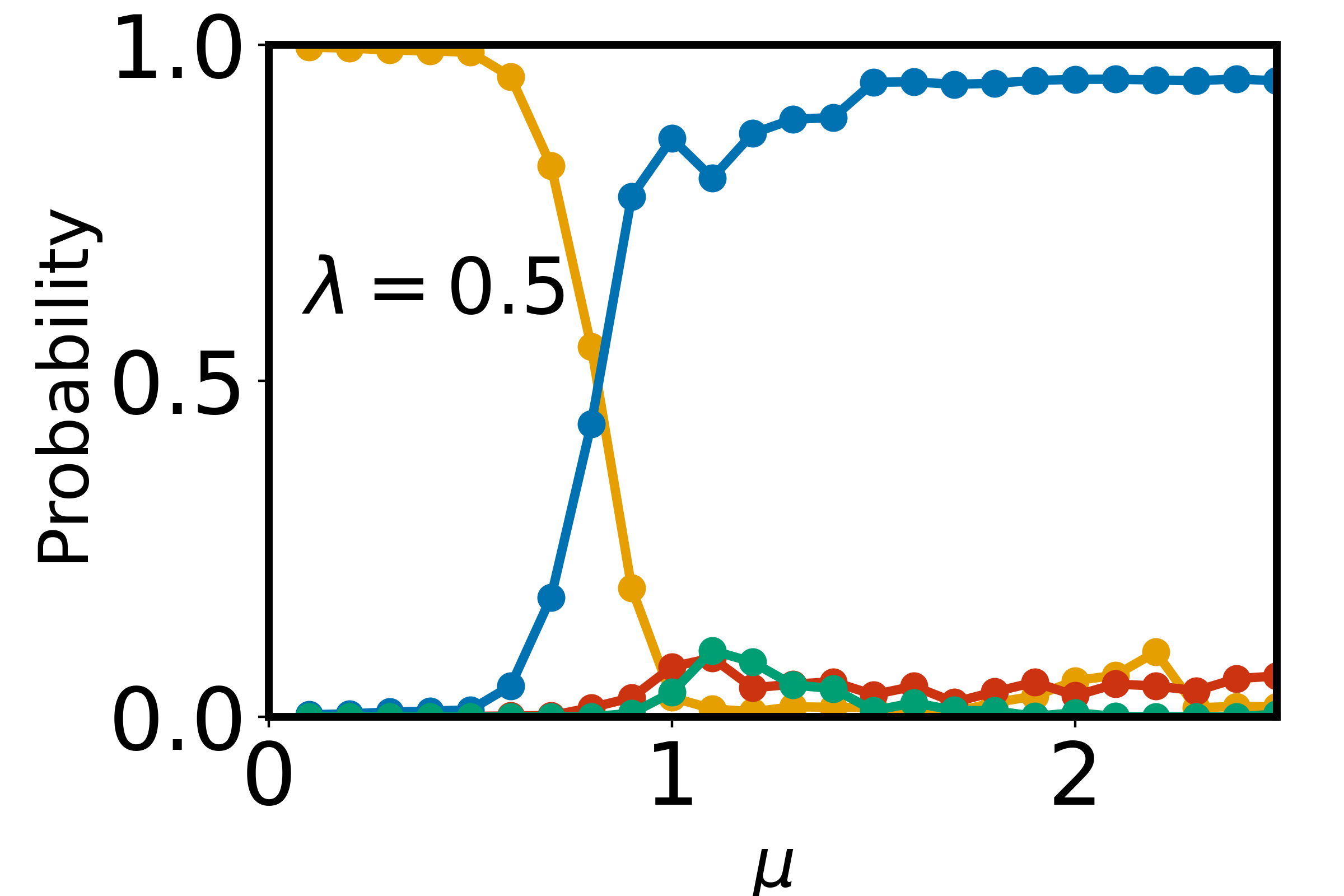}}
\stackunder{\hspace{-4cm}(d)}{\includegraphics[width=4.2cm]{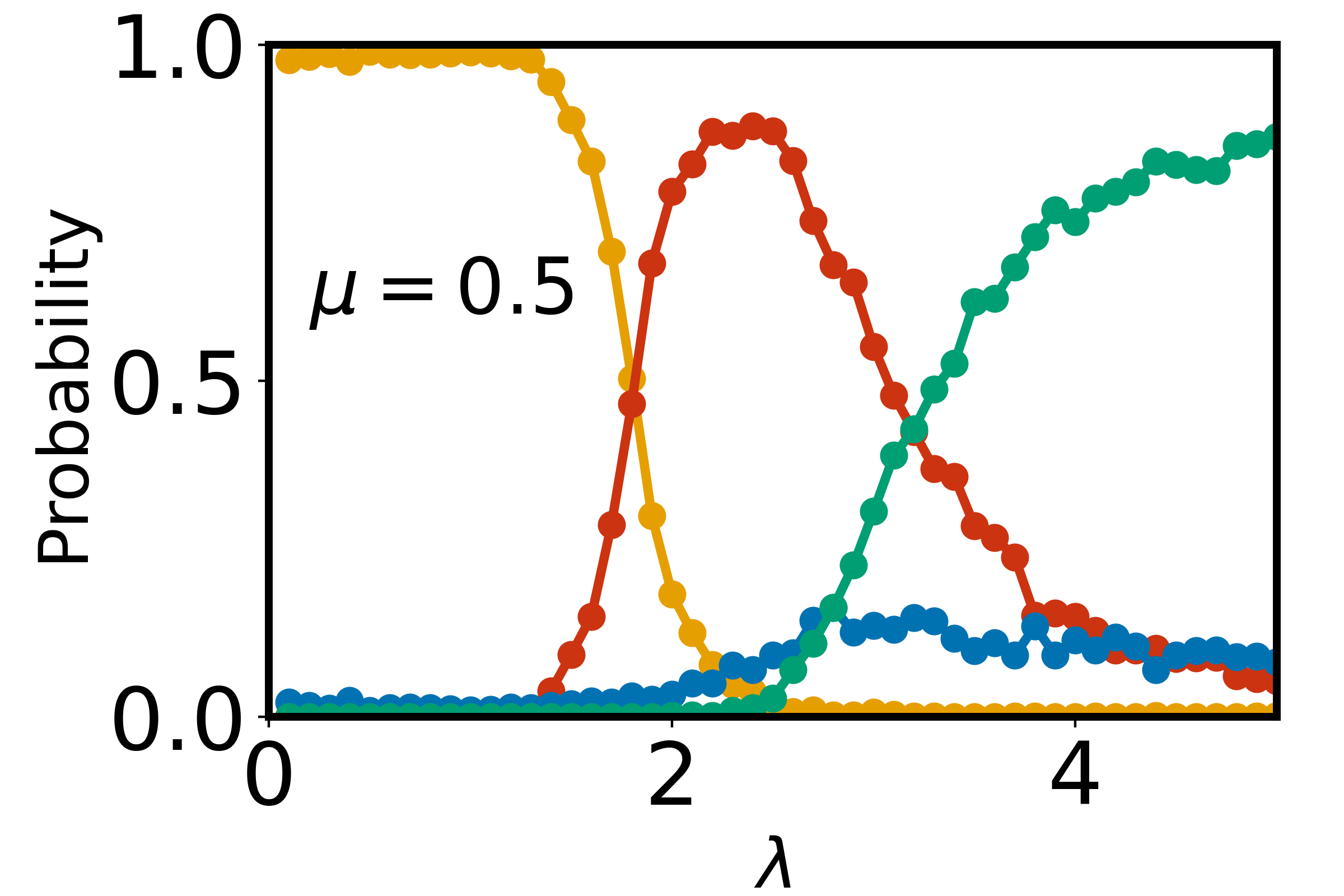}}
\caption{\label{fig3} Neural network probabilities for the $4$-class classifier, considering two ME phases, are shown for (a), (c) increasing hopping amplitude $\mu$ at fixed disorder strength $\lambda = 0.5$, and for (b), (d) increasing disorder strength $\lambda$ at fixed $\mu = 0.5$. Figures (a-b) are based on eigenvalue spacings, averaged over $300$ test datasets, while (c-d) are based on probability densities corresponding to eigenstates, averaged over $500$ test datasets. The system size is $N=14$, and half-filling is assumed.}
\end{figure}

%%%%%%%%%%%%%%%%%%%%%%%%%%%%%%%%%%%%%%%%%%%%%%%%%%%%%%%%
\subsection{Four-class classification}
\label{4-class}
In this section, we extend our analysis of the interacting EAAH model, which is known to exhibit only three distinct phases: ME, MBC, and MBL. To probe the possibility of further dividing these phases, we partition the phase diagram at $\lambda = 2$, creating four regions. This approach allows us to investigate whether the ME and MBC phases can be meaningfully split into subphases. The neural network is trained to classify these regions, providing four output probabilities corresponding to the confidence of the input data belonging to each phase. These probabilities sum to $1$. 

If a phase is artificially split, the probability of the non-existent subphase will remain high only near the training data points. Conversely, for a genuine subphase, the neural network output would exhibit a broad probability distribution across the phase. We use the same neural network architectures described in Table~\ref{tab2}, with the SGD optimizer and default learning rate. The network achieves an accuracy of $\approx 99\%$ in $30$ epochs for eigenvalue spacing data and $1200$ epochs for eigenvector probability densities (PDs).

We first divide the ME phase into two subphases: ME-I ($\lambda < 2$) and ME-II ($\lambda > 2$). The network is trained on data sampled deeply from these two regions, along with the MBC and MBL phases. The training datasets are as follows: $\mu=0.3$, $\lambda=1$ for ME-I; $\mu=0.3$, $\lambda=2.5$ for ME-II; $\mu=2$, $\lambda=0.3, 1, 2.5, 3$ for MBC; and $\mu=0.5, 2$, $\lambda=4.5$ for MBL. The results for the four-class classifier trained on eigenvalue spacings are shown in Figs.~\ref{fig3}(a-b), while Figs.~\ref{fig3}(c-d) present the corresponding results for eigenvector PDs.
%%%%%%%%%%%%%%%%%%%%%%%%%%%%%%%%%%%%%%%%%%%%%%%%%%%%%%%%
\begin{figure}
\centering
\stackunder{\hspace{-4cm}(a)}{\includegraphics[width=4.2cm]{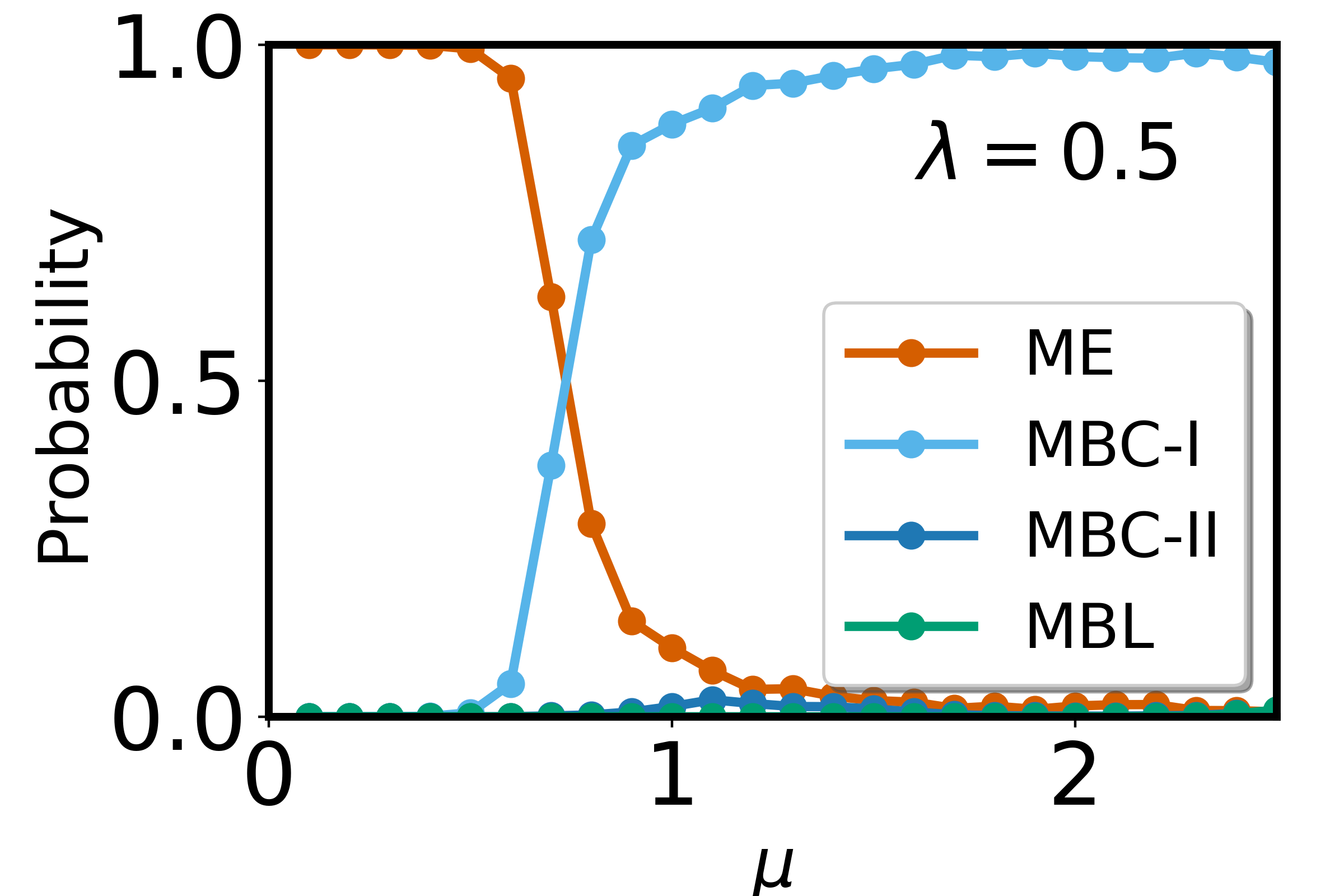}}
\stackunder{\hspace{-4cm}(b)}{\includegraphics[width=4.2cm]{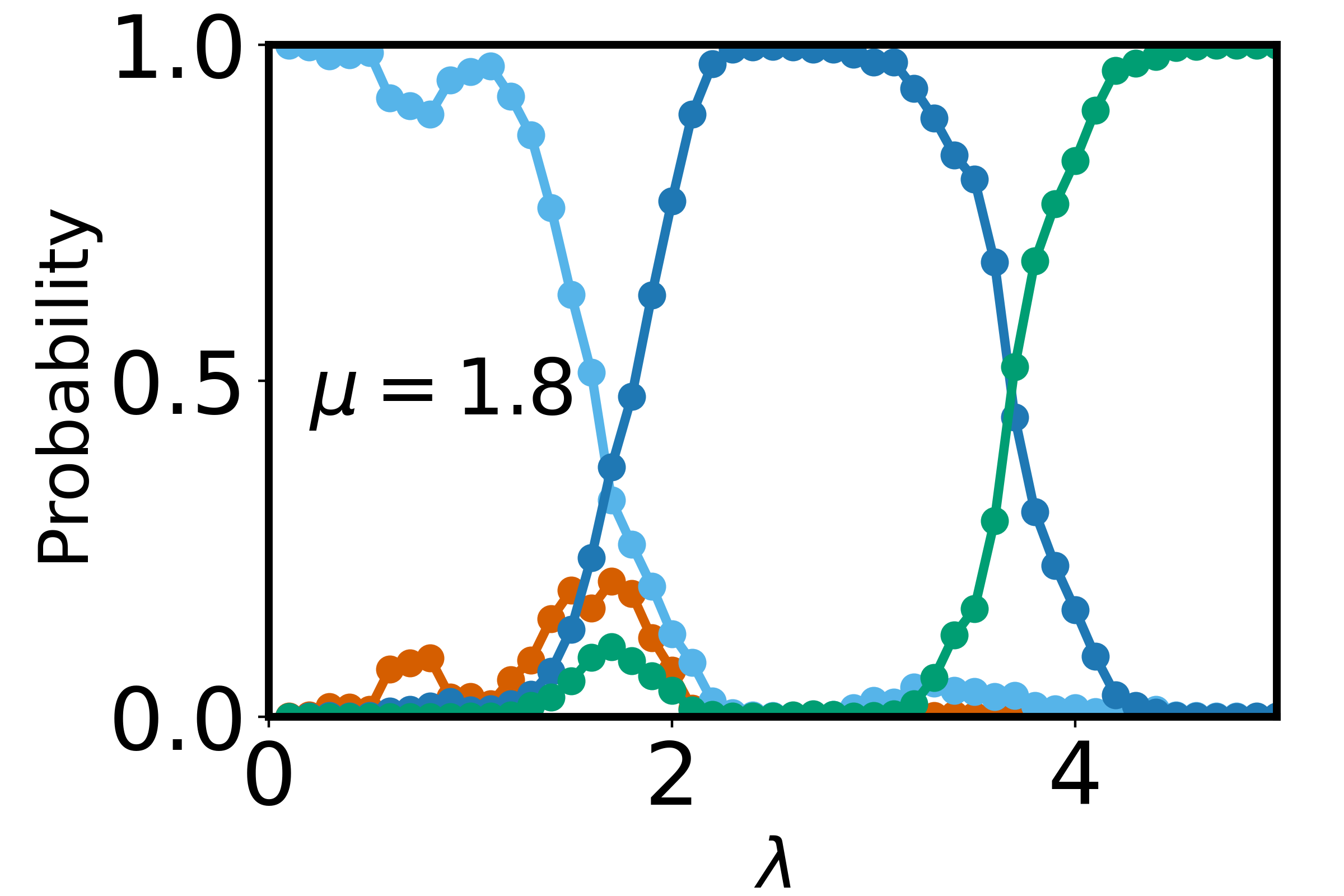}}
\vspace{-0.4cm}

\stackunder{\hspace{-4cm}(c)}{\includegraphics[width=4.2cm]{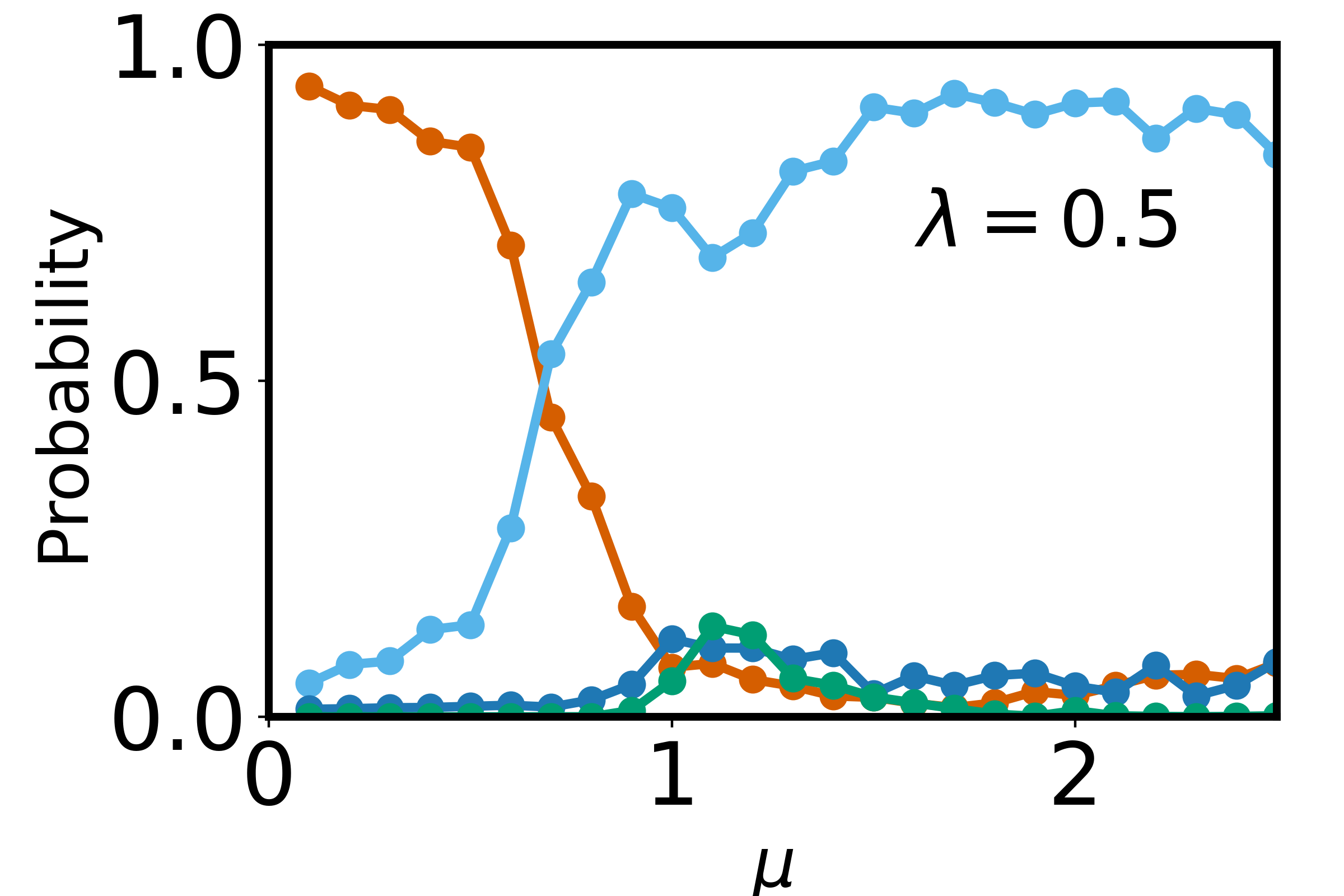}}
\stackunder{\hspace{-4cm}(d)}{\includegraphics[width=4.2cm]{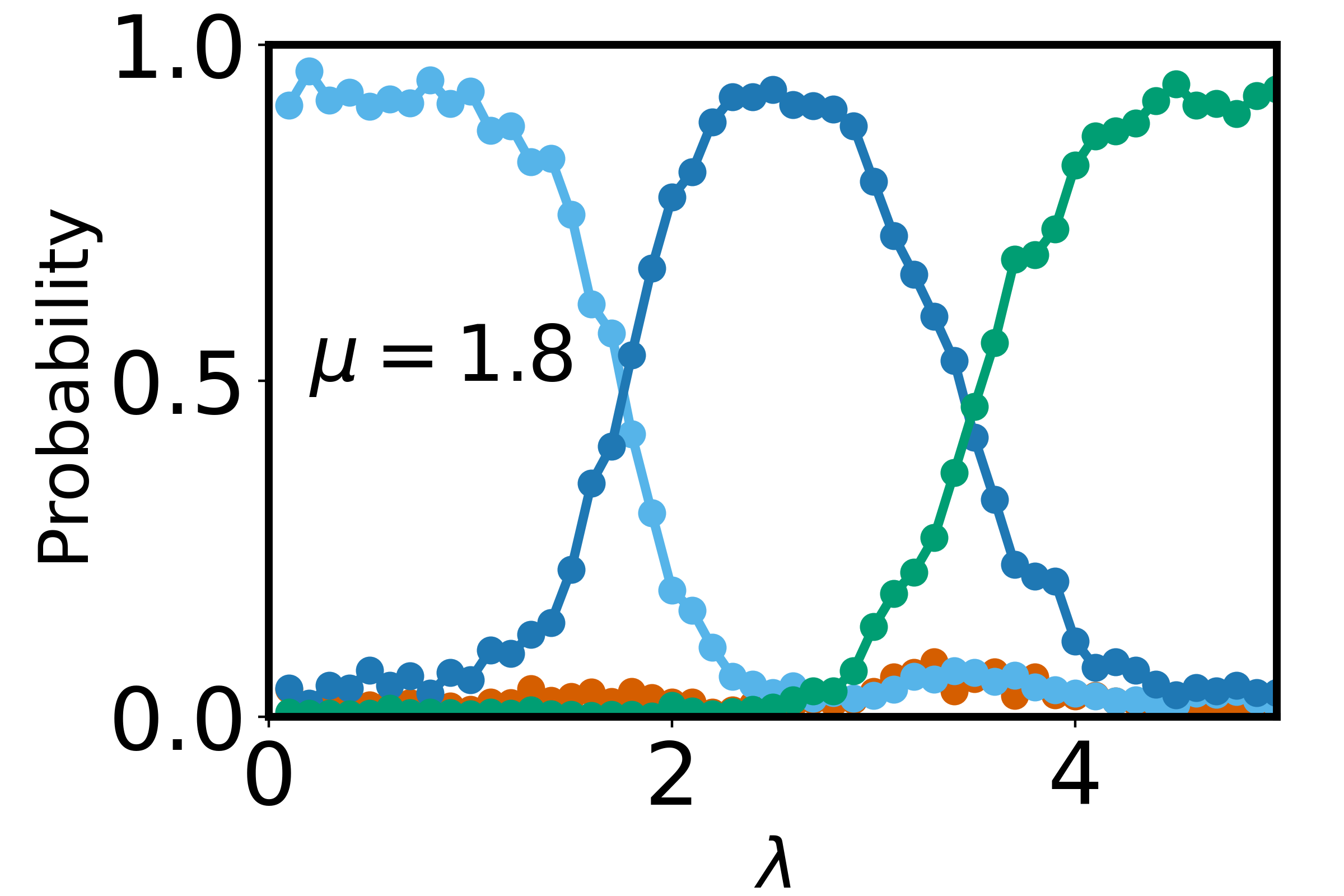}}
\caption{\label{fig4}Neural network probabilities for the $4$-class classifier, considering two MBC phases, are shown for (a), (c) increasing hopping amplitude $\mu$ at fixed disorder strength $\lambda = 0.5$, and for (b), (d) increasing disorder strength $\lambda$ at fixed $\mu = 1.8$. Figures (a-b) are based on eigenvalue spacings, averaged over $300$ test datasets, while (c-d) are based on probability densities corresponding to eigenstates, averaged over $500$ test datasets. The system size is $N=14$, and half-filling is assumed.
}
\end{figure}
%%%%%%%%%%%%%%%%%%%%%%%%%%%%%%%%%%%%%%%%%%%%%%%%%%%%%%%%
From Figs.~\ref{fig3}(a) and \ref{fig3}(c), we observe that the ME-I phase ($\lambda < 2$) exhibits a broad probability near $1$, confirming its existence. As the hopping amplitude increases, a transition from the ME phase to the MBC phase is observed. In contrast, Figs.~\ref{fig3}(b) and \ref{fig3}(d) show the probabilities with increasing $\lambda$, transitioning from ME-I to ME-II and finally to the MBL phase. While the ME-I phase probability remains $\approx 1$, the ME-II phase probability approaches $1$ only near the training points, suggesting that ME-II is not a distinct phase. This analysis confirms that the ME phase cannot be meaningfully divided at $\lambda=2$.

Next, we perform a similar analysis by dividing the MBC phase into two subphases: MBC-I and MBC-II. The training datasets are as follows: $\mu=0.3$, $\lambda=1, 2.5$ for ME; $\mu=2$, $\lambda=0.3, 1$ for MBC-I; $\mu=2$, $\lambda=2.5, 3$ for MBC-II; and $\mu=0.5, 2$, $\lambda=4.5$ for MBL. The results for the four-class classifier trained on eigenvalue spacings are shown in Figs.~\ref{fig4}(a-b), and the corresponding results for eigenvector PDs are shown in Figs.~\ref{fig4}(c-d).

From Figs.~\ref{fig4}(a) and \ref{fig4}(c), we observe that the MBC-I phase ($\lambda < 2$) exhibits a broad probability $\approx 1$, confirming its existence. As the hopping amplitude increases, a clear ME-to-MBC-I transition is observed. Similarly, Figs.~\ref{fig4}(b) and \ref{fig4}(d) reveal transitions from MBC-I to MBC-II and finally to the MBL phase with increasing $\lambda$. Both MBC-I and MBC-II phases show broad probabilities $\approx 1$, independent of the training points, indicating that the MBC phase can be meaningfully divided into MBC-I and MBC-II subphases. 

Thus, our machine learning analysis shows while the ME phase remains indivisible, the MBC phase is shown to consist of two distinct subphases, MBC-I and MBC-II, indicating the possibility of a, rather unusual, crossover or a transition in the multifractal nature of the many-body eigenstates at $\lambda\approx2$~\cite{roy2025manybodycriticalphasequasiperiodic}. 
This highlights the ability of ML techniques to uncover previously unexplored phases. A similar crossover around $\mu\approx 1$ has been predicted from the four class-classifier analysis within the MBL phase, which is discussed in Appendix~\ref{sec:MBL_subphases}.

\section{Principle component analysis}\label{sec:level4}
Principal Component Analysis (PCA) is a widely used statistical technique for dimensionality reduction, data visualization, and model training. By transforming the original features into a new set of uncorrelated variables, known as principal components, PCA simplifies the complexity of the data. This transformation is particularly useful for visualizing data in lower dimensions, aiding in the understanding of class separability and feature distributions.

\subsection{PCA Methodology}

For $m$ observations (vectors), each of dimension $n \times 1$ (i.e., $n$ features), the data matrix $X$ of dimension $m \times n$ is constructed, where each column represents a feature, and each row represents an observation. Here, the observations are represented by eigenvector probability densities (PDs), where the number of features \(n\) corresponds to the dimension of the Hilbert space.

PCA begins by centring the data by subtracting the mean of each feature. The elements of the centred data matrix $X'$ are computed as:  
\begin{equation}
    X'_{ij} = X_{ij} - \frac{1}{m}\sum_{i=1}^m X_{ij}.
\end{equation}

Next, the covariance matrix $C$ of dimension $n \times n$ is calculated as:
\begin{equation}
    C = \frac{1}{m-1} X'^T X',
\end{equation}
which is then decomposed using Singular Value Decomposition (SVD) into eigenvectors and eigenvalues. The eigenvectors, termed principal components, represent the directions in the feature space along which the data exhibits the highest variance. The eigenvalues correspond to the variance captured by each principal component.

\subsubsection{Explained Variance and Dimensionality}

The maximum number of principal components is determined by the rank of the data matrix $X$, given by $\text{min}(n, m)$. Importantly, the eigenvalues of the covariance matrix are non-negative and sum up to the total variance of the original dataset. The principal components are ordered by the magnitude of variance they capture, with the first few components typically accounting for most of the variability.

In cases where the input data consists of distinct classes, well-separated in the feature space, this separation is often reflected in the first few principal components. However, the first two components, while capturing the maximum variance, may not always distinguish between classes. In such cases, subsequent principal components may carry relevant discriminatory information.

\subsubsection{PCA-Entropy}

The eigenvalues $e_1, e_2, \ldots, e_k$ of the covariance matrix can be normalized to compute the explained variance ratios (EVR):
\begin{equation}
    p_i = \frac{e_i}{\sum_j e_j},
\end{equation}
where $p_i$ represents the fraction of variance captured by the $i$th principal component. The PCA entropy is then defined as~\cite{shannon1948mathematical}:
\begin{equation}
    S_{PCA} = -\sum_{i=1}^k p_i \log_2 p_i.
\end{equation}
This entropy is rescaled between $0$ and $1$ as:
\begin{equation}
    \Tilde{S}_{\text{PCA}} = \frac{-\sum_{i=1}^k p_i \log_2 p_i}{\log_2 k}.
\end{equation}
The PCA entropy measures the distribution of variance among the principal components. A low PCA entropy indicates that variance is concentrated in a few components, suggesting a simpler data structure with clear dominant variance directions. Conversely, high PCA entropy implies a more even distribution of variance across components, indicating a complex dataset with no strong underlying structure. It should be noted that this normalized entropy is particularly useful for comparing datasets with different numbers of principal components.

\subsubsection{Transition Region and Numerical Derivative}

To analyze the transition region of the PCA entropy, its numerical derivative is computed using the symmetric difference quotient:
\begin{equation}
    \frac{\delta \Tilde{S}_{\text{PCA}}}{\delta \lambda} = \frac{\Tilde{S}_{PCA}(\lambda + \Delta\lambda) - \Tilde{S}_{PCA}(\lambda - \Delta\lambda)}{2\Delta\lambda}.
    \label{eq2}
\end{equation}
Here, $\lambda$ represents any disorder parameter. This approach identifies regions of significant change in PCA entropy, providing insights into transitions in the dataset's structure. It should be noted that the symmetric approach provides a more accurate estimation of the derivative by considering changes on both sides of the point of interest. This ensures that the derivative captures transitions effectively without amplifying noise. In subsequent sections, we will discuss the results for single-particle systems and extend these observations to many-body cases.

\begin{figure}[b]
\centering
\stackunder{\hspace{-4cm}(a)}{\includegraphics[width=4.2cm,height=3.2cm]{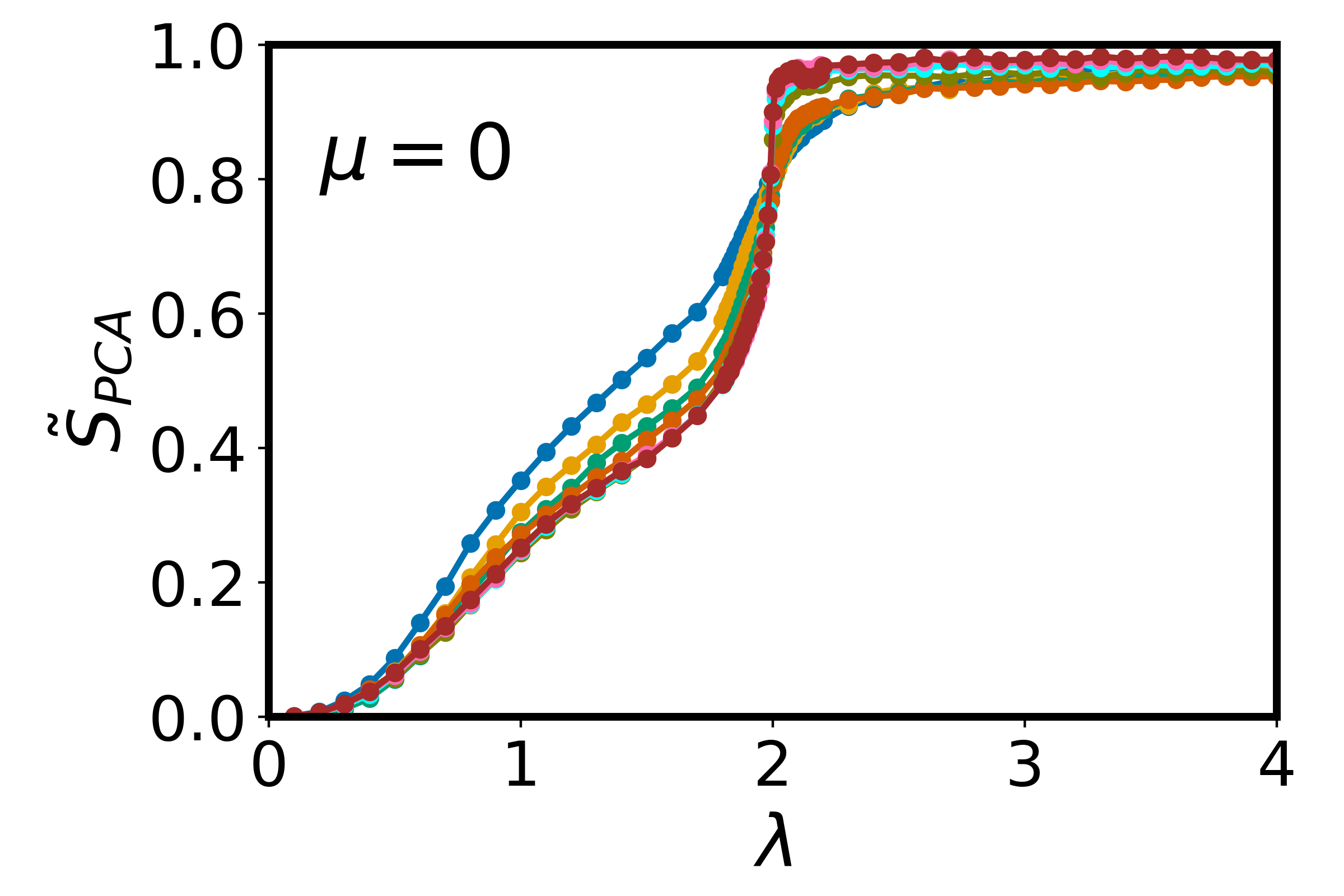}}
\stackunder{\hspace{-4cm}(b)}{\includegraphics[width=4.2cm,height=3.2cm]{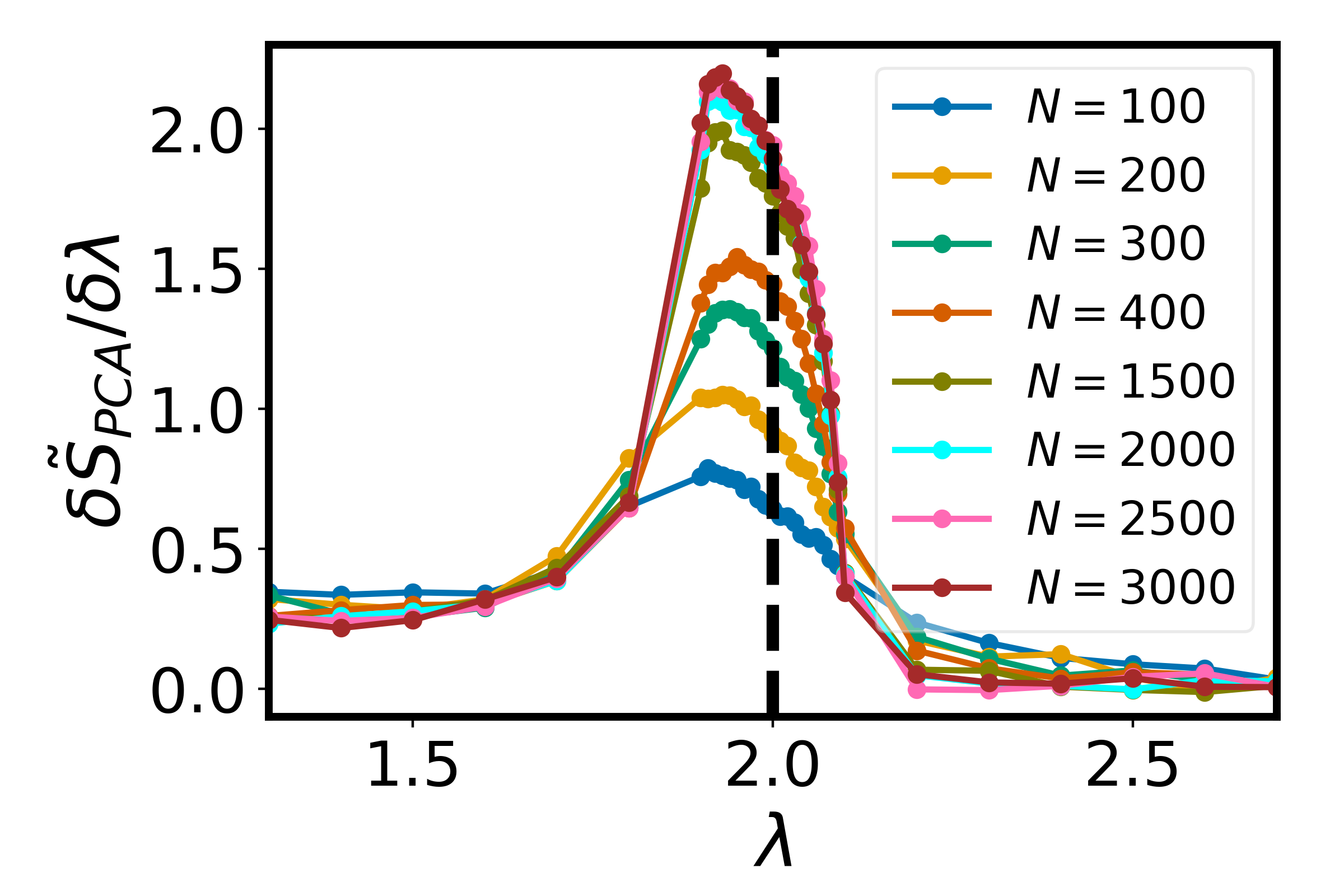}}
\caption{\label{fig5}(a) PCA-entropy $\Tilde{S}_{PCA}$ as a function of disorder strength $\lambda$ for the AAH model across various system sizes $N$. (b) Numerical derivative of the PCA-entropy, $\delta \Tilde{S}_{PCA}/\delta \lambda$, calculated using $\Delta \lambda = 0.01$ near the theoretically known transition point indicated by the black dashed line $\lambda_c = 2$ and $\Delta \lambda=0.1$ otherwise. The analysis is performed using $m = 500$ disorder samples for all cases. 
}
\end{figure}
\subsection{Single-particle system}
We begin our analysis with the Aubry-Andre-Harper (AAH) model, characterized by $U = 0$ and $\mu = 0$ in Eq.~\ref{eq1}. The behaviour of the PCA-entropy $\Tilde{S}_{PCA}$ is studied as a function of disorder strength $\lambda$ using $m = 500$ samples of eigenvector probability densities (PDs) for various system sizes. Two sets of system sizes are considered: ${R} = \{100, 200, 300, 400\}$, where $n < m$, and ${L} = \{1500, 2000, 2500, 3000\}$, where $n > m$. 

We use eigenvector PDs at infinite temperature (central eigenstate in single particle systems) generated for $500$ disorder realizations ($\theta$ values) corresponding to each system size to construct the covariance matrix. Ideally, the sample-to-feature ratio for PCA should be $10:1$ or higher to avoid overfitting. In our analysis, we observe similar PCA-entropy behaviour for both  $m>n$ (set ${R} ={100,200,300,400}$) and $n>m$ (set ${L} ={1500,2000,2500,3000}$). Despite $n>m$ leading to a rank-deficient covariance matrix, PCA remains effective as it captures the variance using the largest $m-1$ principal components. Moreover, consistency in results between these two regimes empirically validates the robustness of our findings, justifying the choice of $m=500$. To validate our analysis, we also checked the PCA-entropy $\Tilde{S}_{PCA}$ for members of ${R}$ with $m = 1000$, observing no qualitative differences from results obtained with $m = 500$. Thus, we fix $m = 500$ for all subsequent discussions.

In Fig.~\ref{fig5}(a), we plot the PCA-entropy $\Tilde{S}_{PCA}$ as a function of disorder strength $\lambda$. In the delocalized phase ($\lambda < 2$), $\Tilde{S}_{PCA}$ is independent of system size for $m > n$ (set ${L}$). However, for $m < n$ (set ${R}$), PCA-entropy shows system size dependence. In contrast, in the localized phase ($\lambda > 2$), $\Tilde{S}_{PCA}$ exhibits system size dependence for both $m > n$ and $m < n$. 

Drawing an analogy to thermodynamic entropy, it is expected that the PCA-entropy would be higher in the delocalized phase and approach $0$ in the localized phase. However, the actual behaviour is counterintuitive. In the delocalized phase, the probability density of each eigenvector is spread uniformly across multiple sites, resulting in a strong structural similarity among vectors. PCA detects these correlations and captures most of the variance in a few principal components ($O(10)$), reflecting the global structure of the delocalized states. This leads to a low PCA entropy due to the high concentration of variance in the leading components, akin to a low configurational entropy system.
\begin{figure}
\centering
\stackunder{\hspace{-4cm}(a)}{\includegraphics[width=4.2cm,height=3.2cm]{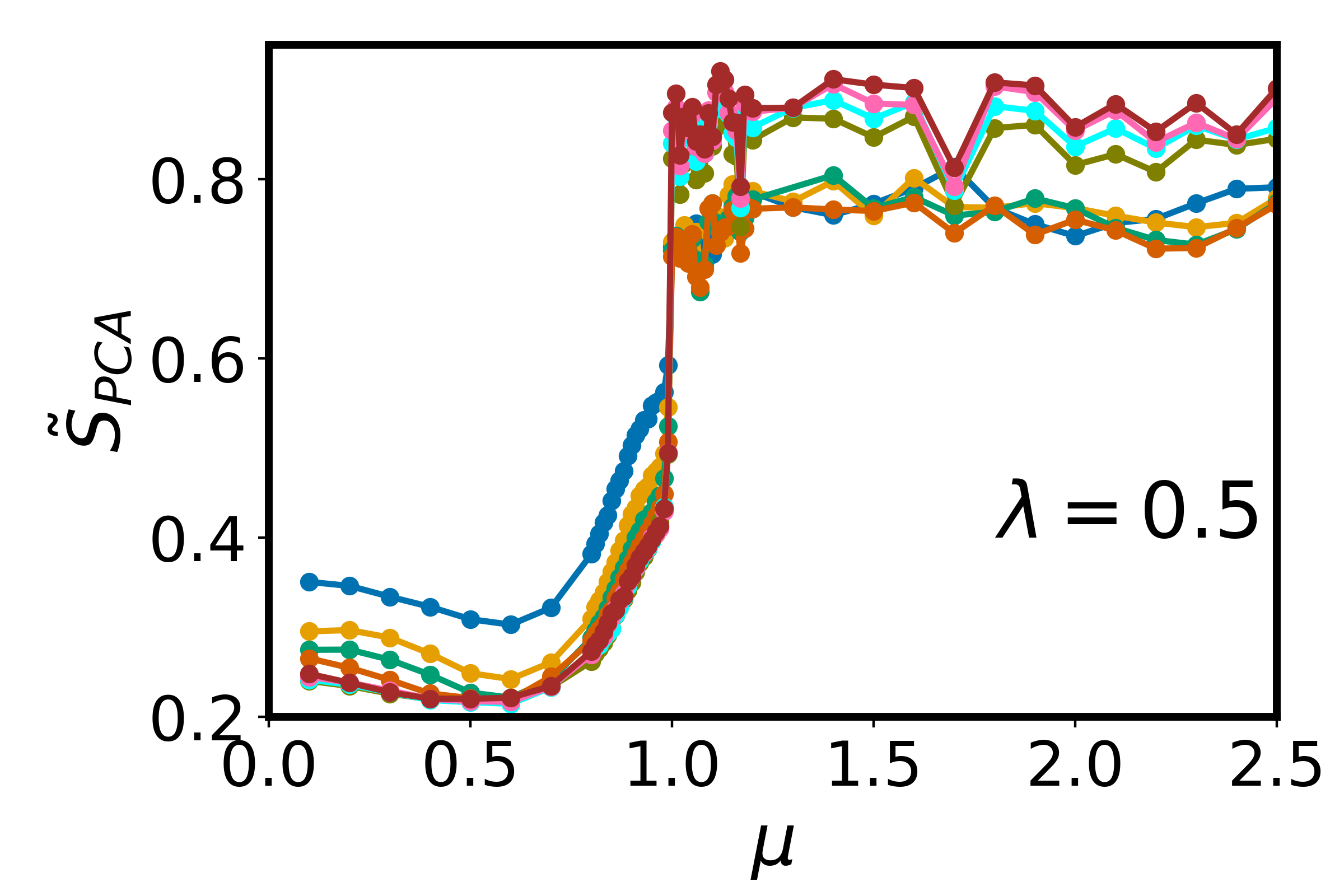}}
\stackunder{\hspace{-4cm}(b)}{\includegraphics[width=4.2cm,height=3.2cm]{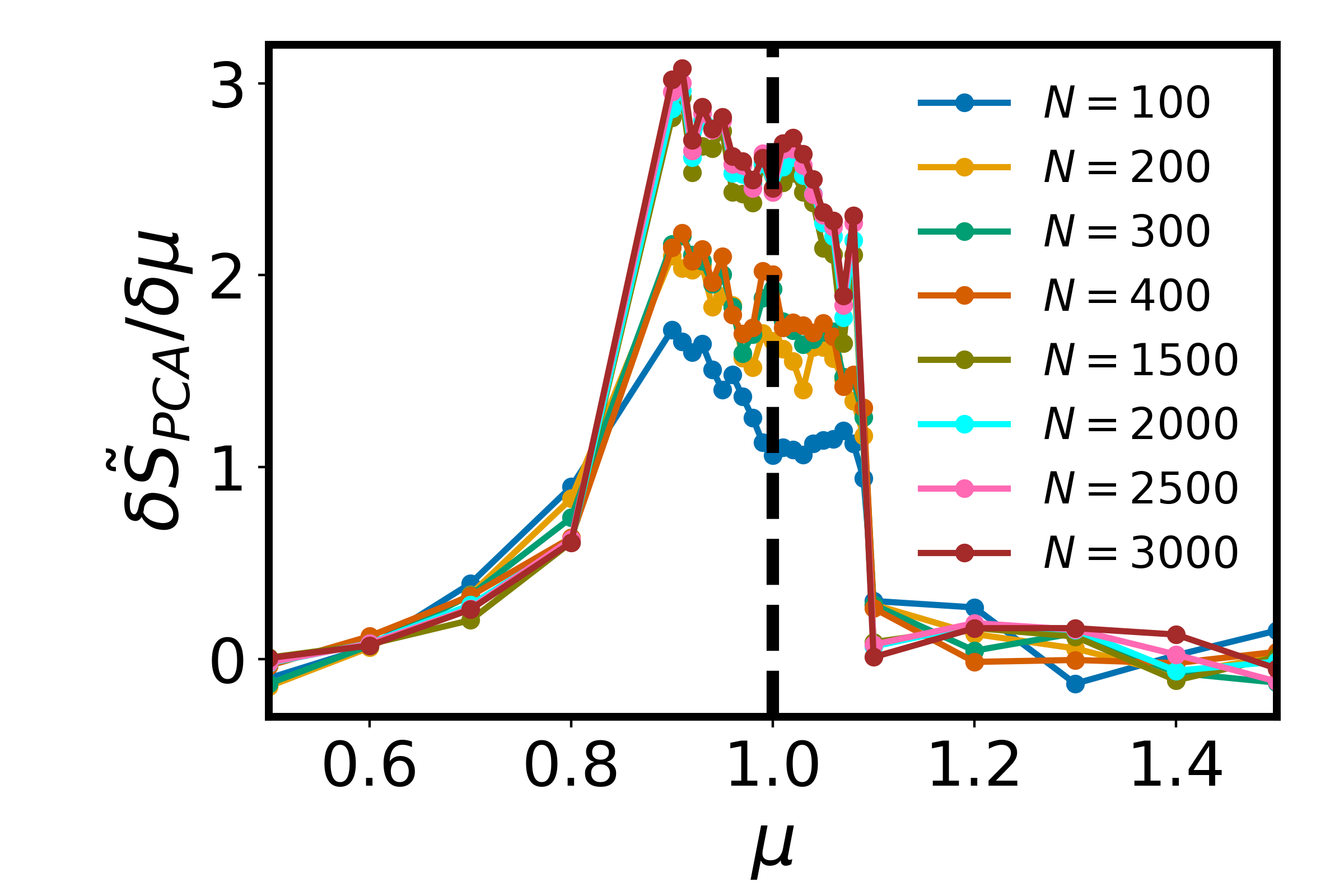}}
\vspace{-0.4cm}

\stackunder{\hspace{-4cm}(c)}{\includegraphics[width=4.2cm,height=3.2cm]{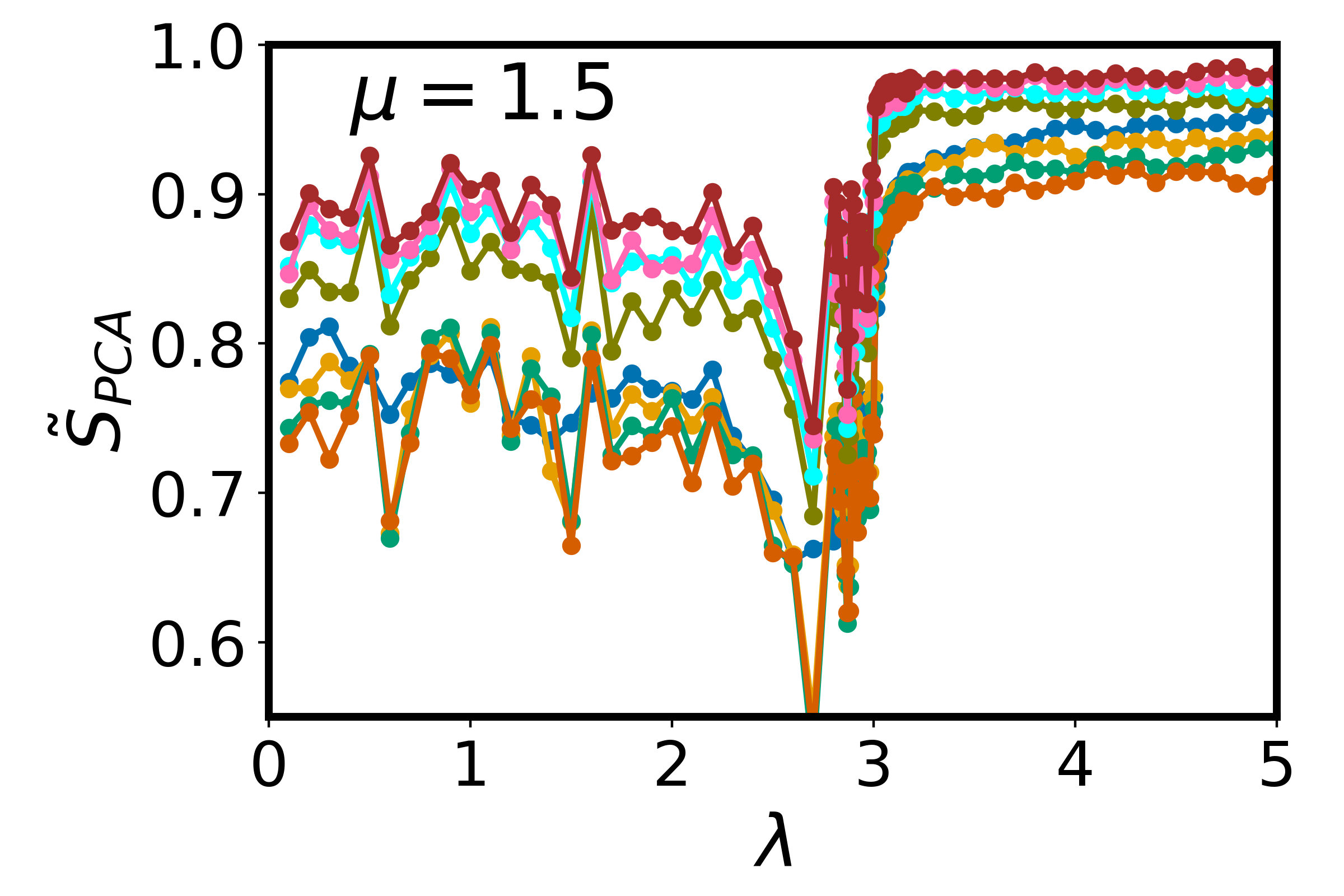}}
\stackunder{\hspace{-4cm}(d)}{\includegraphics[width=4.2cm,height=3.2cm]{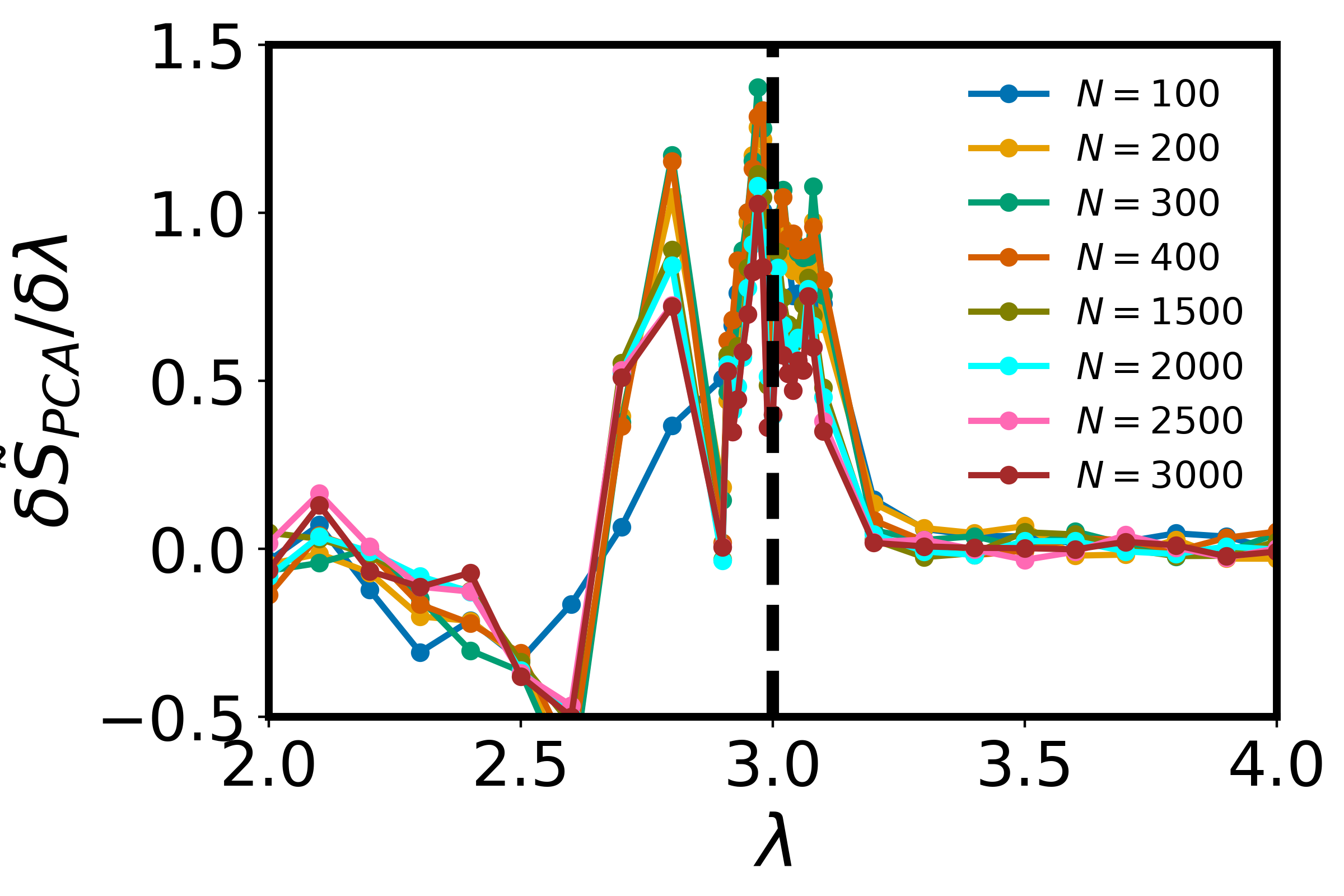}}
\caption{\label{fig6}For the extended Harper model, (a) PCA-entropy $\Tilde{S}_{PCA}$ as a function of hopping strength $\mu$ for different system sizes $N$ at fixed disorder strength $\lambda=0.5$.  
(b) The corresponding numerical derivative of PCA-entropy, $\Delta \Tilde{S}_{PCA}/\Delta \mu$, computed using $\delta \mu=0.01$ near the theoretically known transition point $\mu_c=1$, indicated by the black dashed line and $\delta \mu=0.1$ otherwise. (c) PCA-entropy $\Tilde{S}_{PCA}$ as a function of onsite disorder strength $\lambda$ for different system sizes $N$ at fixed hopping strength $\mu=1.5$. (d) The corresponding numerical derivative of PCA-entropy, $\Delta \Tilde{S}_{PCA}/\Delta \lambda$, computed using $\delta \lambda=0.01$ near the theoretically known transition point at $\lambda_c=3$ marked by the black dashed line and $\delta \lambda=0.1$ otherwise. In all cases, the number of disorder samples is fixed at $m=500$.}
\end{figure}

In contrast, in the localized phase, the probability density vectors are spatially isolated, with each eigenstate concentrated in small regions that decay rapidly outside these localized peaks. Minimal overlap between vectors results in weak correlations among the eigenstates. PCA, therefore, does not detect strong global patterns but instead captures a spread of variance across many components ($O(10^2)$), as each component reflects small, uncorrelated details of distinct localized states. This even distribution of variance leads to a high PCA entropy, analogous to a high configurational entropy system. At the multifractal point ($\lambda_c = 2$), PCA-entropy lies in the range $0.5 < \Tilde{S}_{PCA} < 1$, reflecting the existence of intermediary features.

To identify phase transitions, we calculate the numerical derivative of the PCA-entropy using Eq.~\ref{eq2}. The derivative highlights transitions near the delocalized-critical-localized boundaries. As expected, in Fig~\ref{fig5}(b) a sharp jump is observed near $\lambda_c = 2$, the known transition point for the AAH model, where states change from delocalized to multifractal and finally to localized. This confirms that PCA-entropy and its derivative effectively detect phase transitions in the system. This approach provides a quantitative and robust marker for detecting phase boundaries, complementing other established methods. 

We now extend this approach to the EAAH model described by Eq.~\ref{eq1} with no interaction ($U=0$). Two cases are considered~\cite{PhysRevLett.126.080602}: (i) $\lambda=1$, which exhibits a phase transition from the delocalized phase to the critical phase at $\mu_c=1$, and (ii) $\mu=1.5$, where a transition from the critical phase to the localized phase occurs at $\lambda_c=3$. 

In Fig.~\ref{fig6}, we plot the PCA-entropy $\Tilde{S}_{PCA}$ and its derivative with respect to $\lambda$ or $\mu$ to study these transitions. Fig.~\ref{fig6}(a) illustrates the variation of $\Tilde{S}_{PCA}$ as the system transitions from the delocalized to the critical phase. For larger system sizes (${L}$), the PCA-entropy in the delocalized phase is system size-independent, consistent with previous observations in the AAH model. However, in the critical phase, $\Tilde{S}_{PCA}$ shows a clear dependence on system size for both sets ${R}$ and ${L}$. At the theoretical phase transition point $\mu_c=1$, a distinct jump in $\Tilde{S}_{PCA}$ is observed, which is further accentuated in the numerical derivative plot shown in Fig.~\ref{fig6}(b). 

For the transition from the critical to the localized phase, we analyze $\Tilde{S}_{PCA}$ as a function of $\lambda$ at fixed $\mu=1.5$. A notable feature in the critical phase is the presence of fluctuations in $\Tilde{S}_{PCA}$, in contrast to the smoother behaviour observed in the delocalized and localized phases. The derivative plot in Fig.~\ref{fig6}(d) highlights a pronounced jump at $\lambda_c=3$, marking the phase transition point. These observations suggest that in the single-particle regime, PCA-entropy of the eigenvector PDs, along with its derivative, is an effective tool for detecting phase transitions. In the next section, we apply this methodology to the interacting version of the extended Harper model to assess its performance in the presence of interactions.

\begin{figure}[b]
\centering
\stackunder{\hspace{-3.5cm}(a)}{\includegraphics[width=4cm,height=3.1cm]{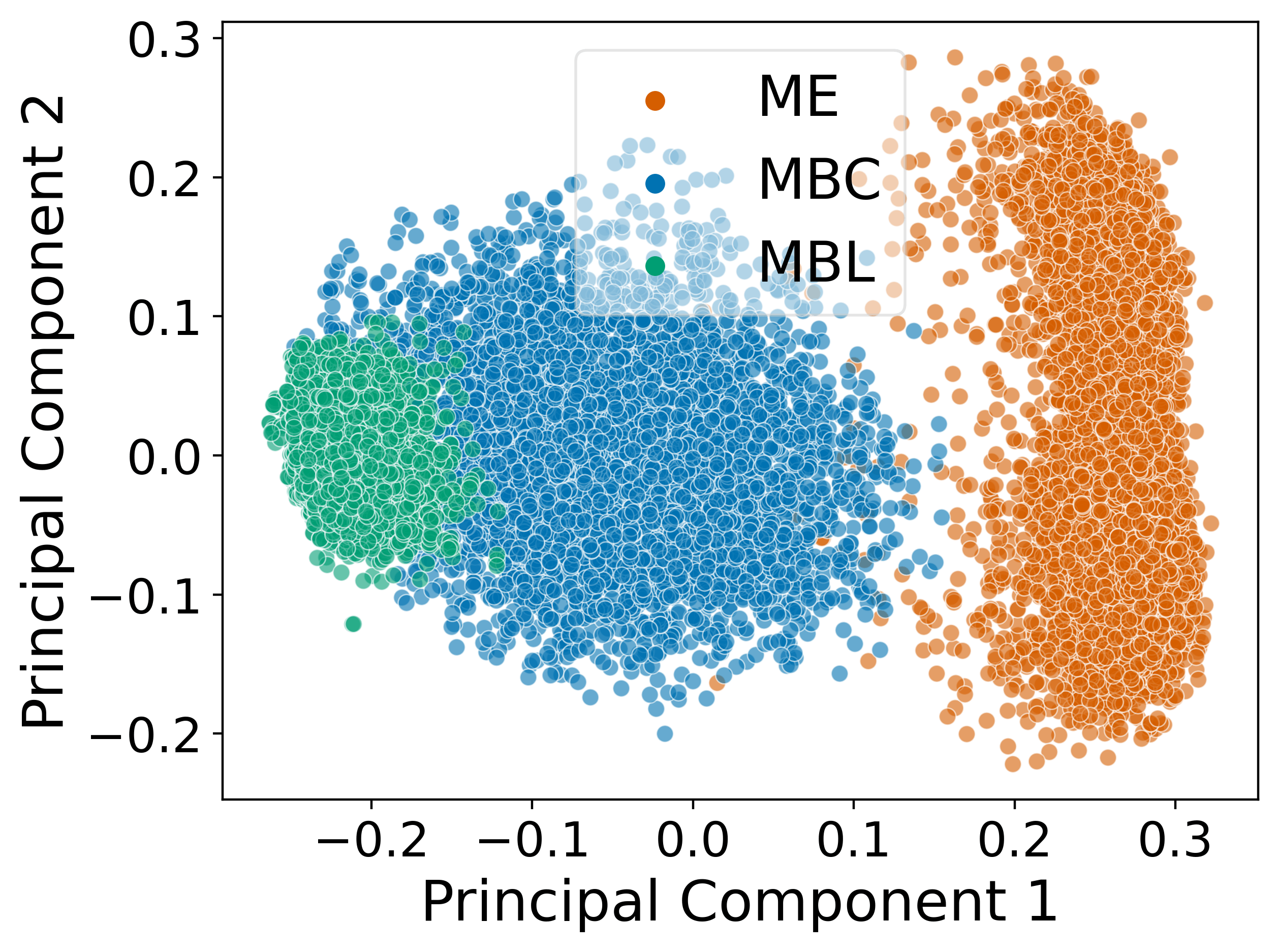}}
\stackunder{\hspace{-3.5cm}(b)}{\includegraphics[width=4cm,height=3.1cm]{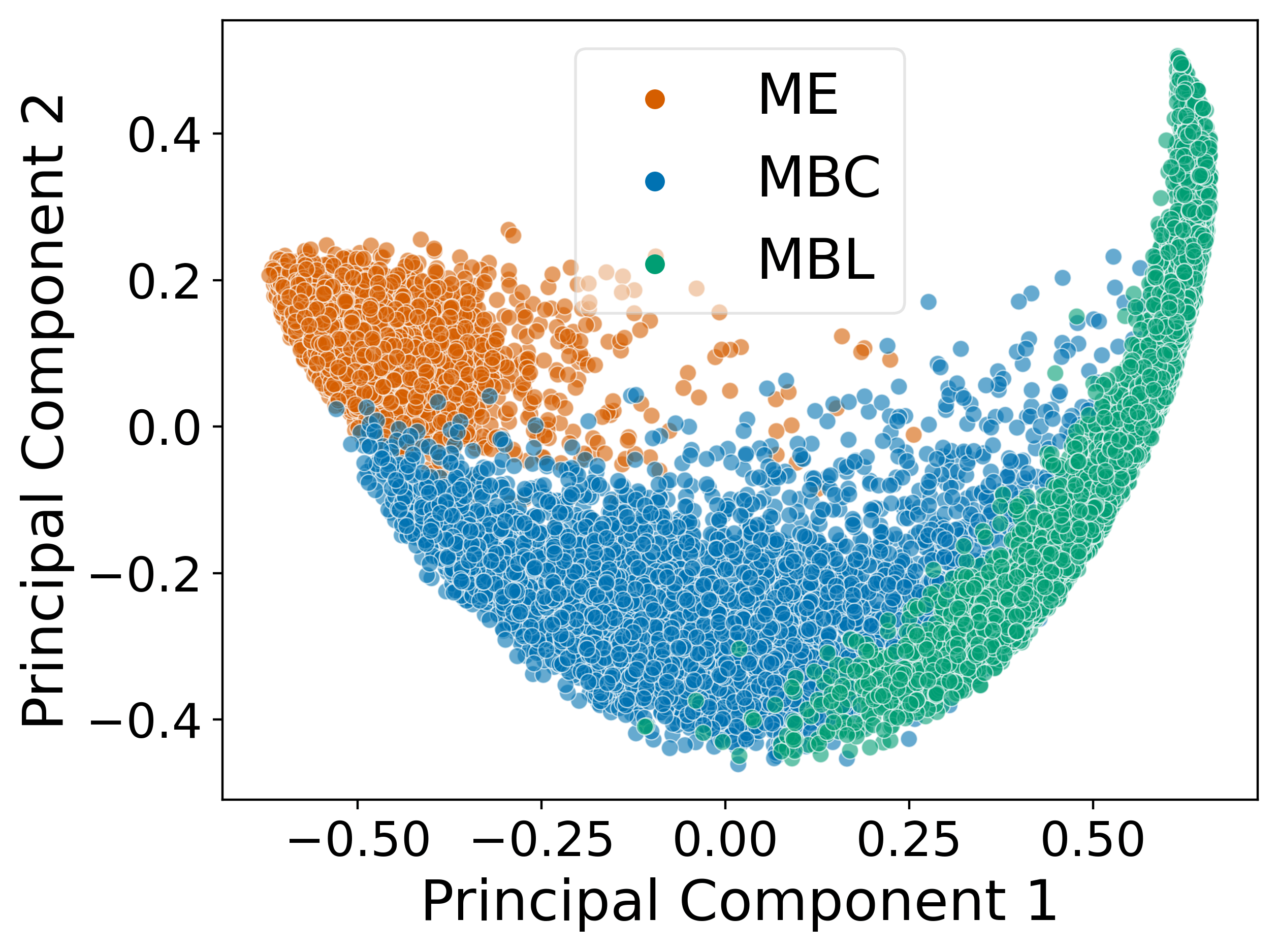}}
\caption{\label{fig7}The first principle component plotted against the second for (a) the normalized eigenvector PDs and (b) rearranged normalized eigenvector PDs. Here $5000$ datasets per class have been utilized for system size $N=14$ with half filling.}
\end{figure}

\subsection{Many-body system}
In this section, we analyze the many-body interacting system. We determine the principal components using the training data sets of eigenvector probability densities corresponding to various phases in Section~\ref{sec:level3}. In Fig.~\ref{fig7}(a), we plot the first principal component against the second principal component for the normalized PDs. We observe that data corresponding to the ergodic phase (ME) is distinguishable, while data from the MBL phase substantially overlaps with that from the MBC phase. Fig.~\ref{fig7}(b) shows the plot obtained by utilizing the normalized PDs with the elements of each vector rearranged in descending order. Here, we observe that while the ME class remains well-separated, the overlap between the MBC and MBL classes is reduced, indicating that the PCA provides better class separability. The normalization and rearrangement of the data serve as preprocessing steps that enhance the critical features for class distinction. This results in a more meaningful variance captured by the principal components, thereby improving phase classification. 

\begin{figure}
\centering
\stackunder{\hspace{-4cm}(a)}{\includegraphics[width=4.2cm,height=3.1cm]{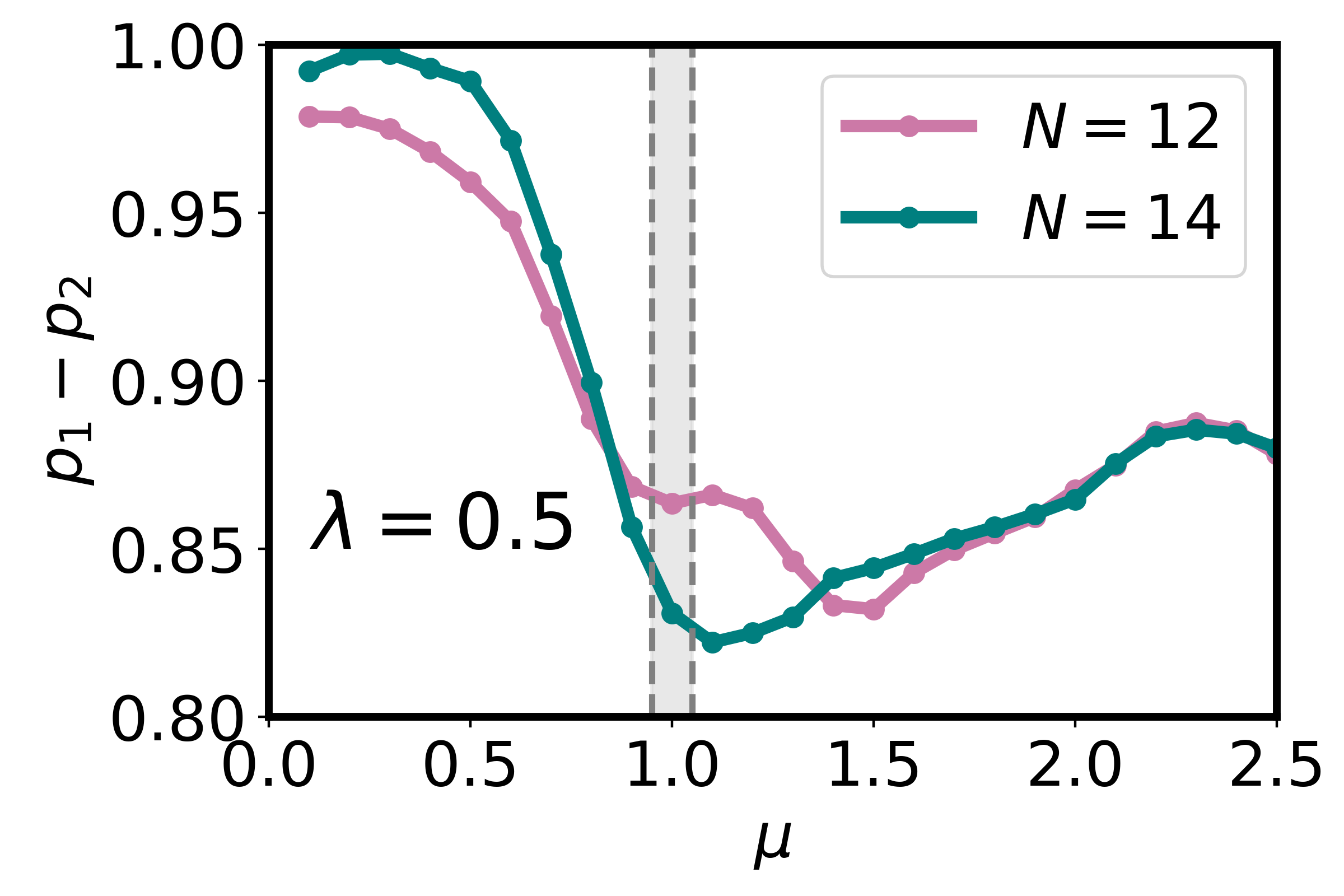}}
\stackunder{\hspace{-4cm}(b)}{\includegraphics[width=4.2cm,height=3.1cm]{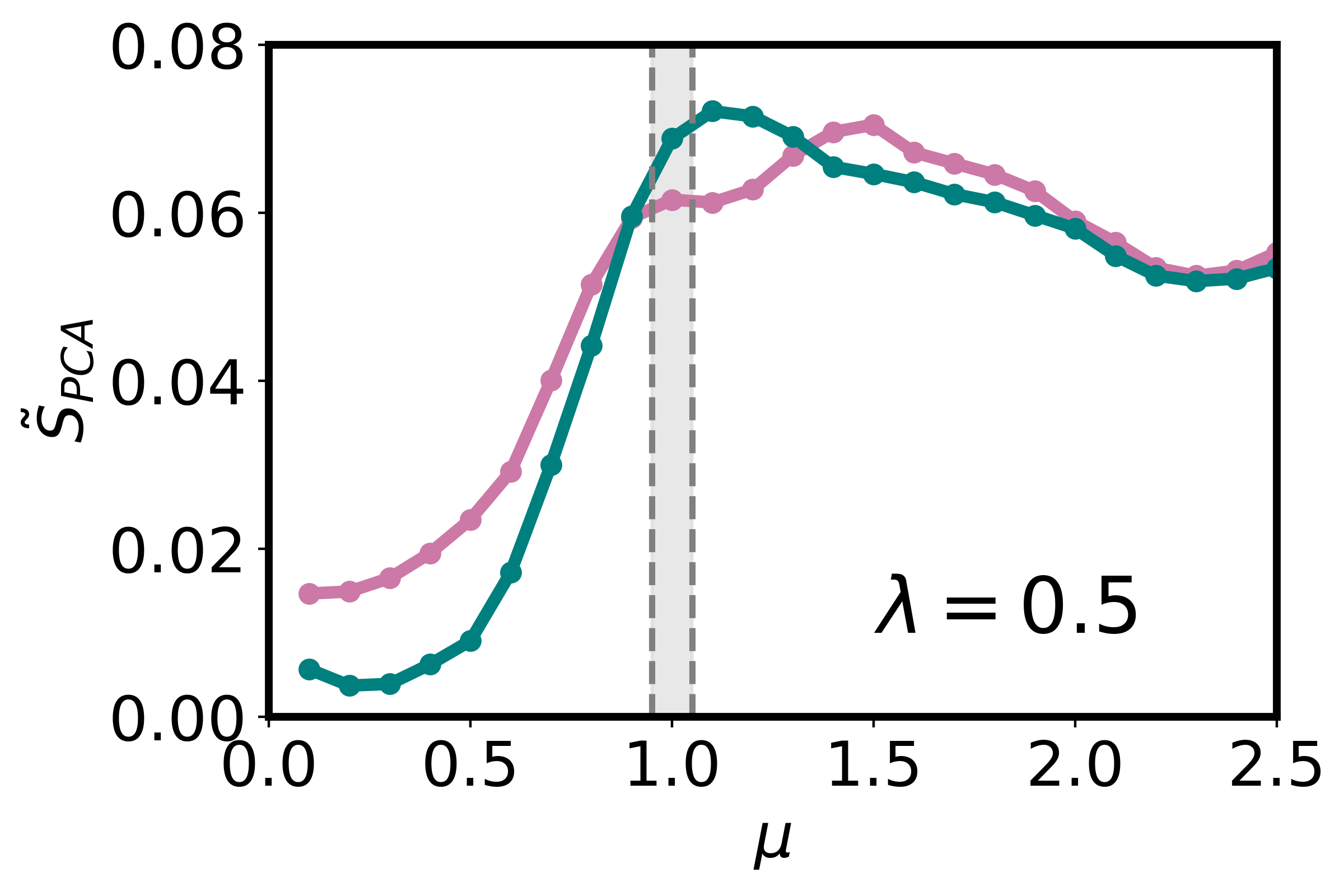}}
\vspace{-0.4cm}

\stackunder{\hspace{-4cm}(c)}{\includegraphics[width=4.2cm,height=3.1cm]{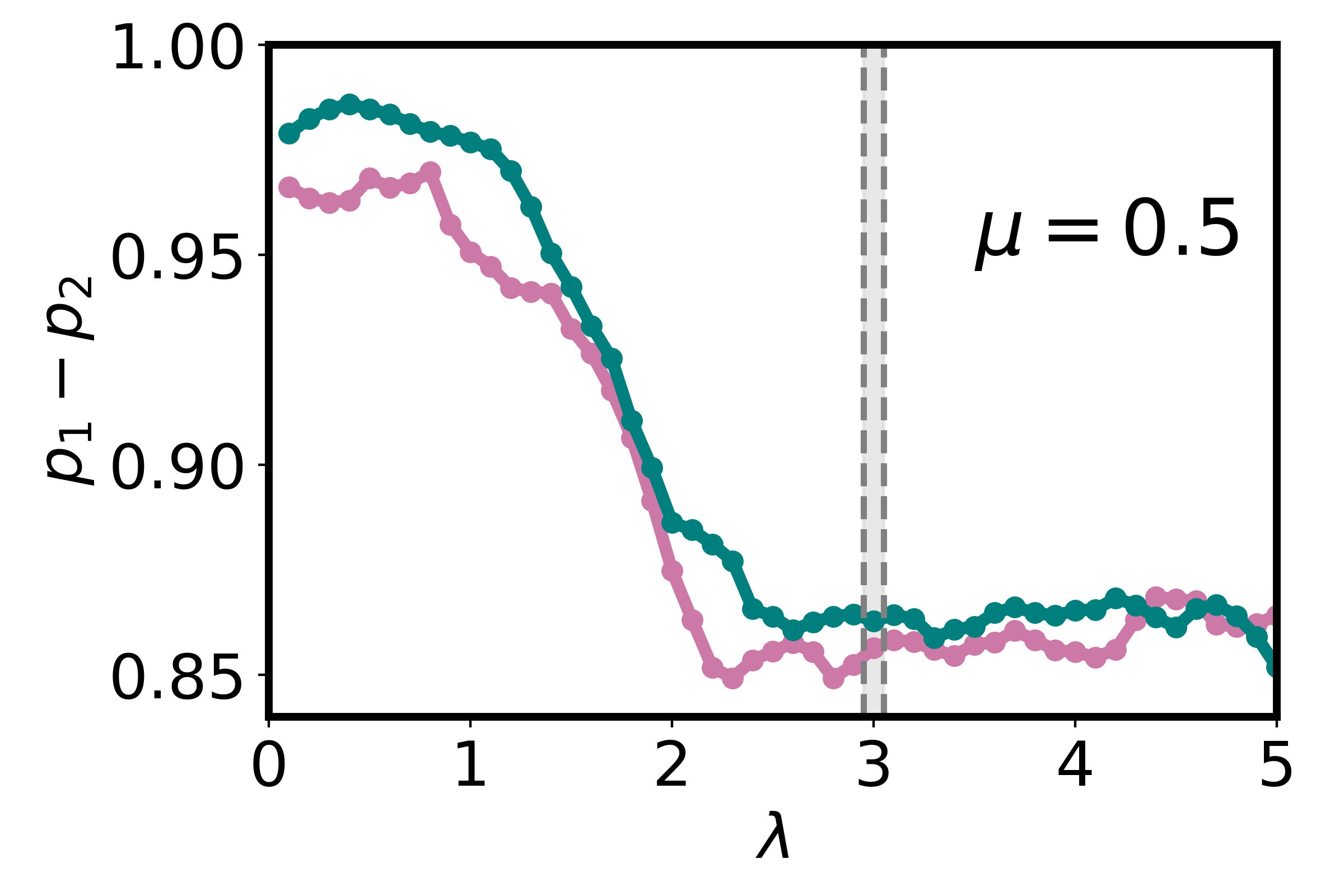}}
\stackunder{\hspace{-4cm}(d)}{\includegraphics[width=4.2cm,height=3.1cm]{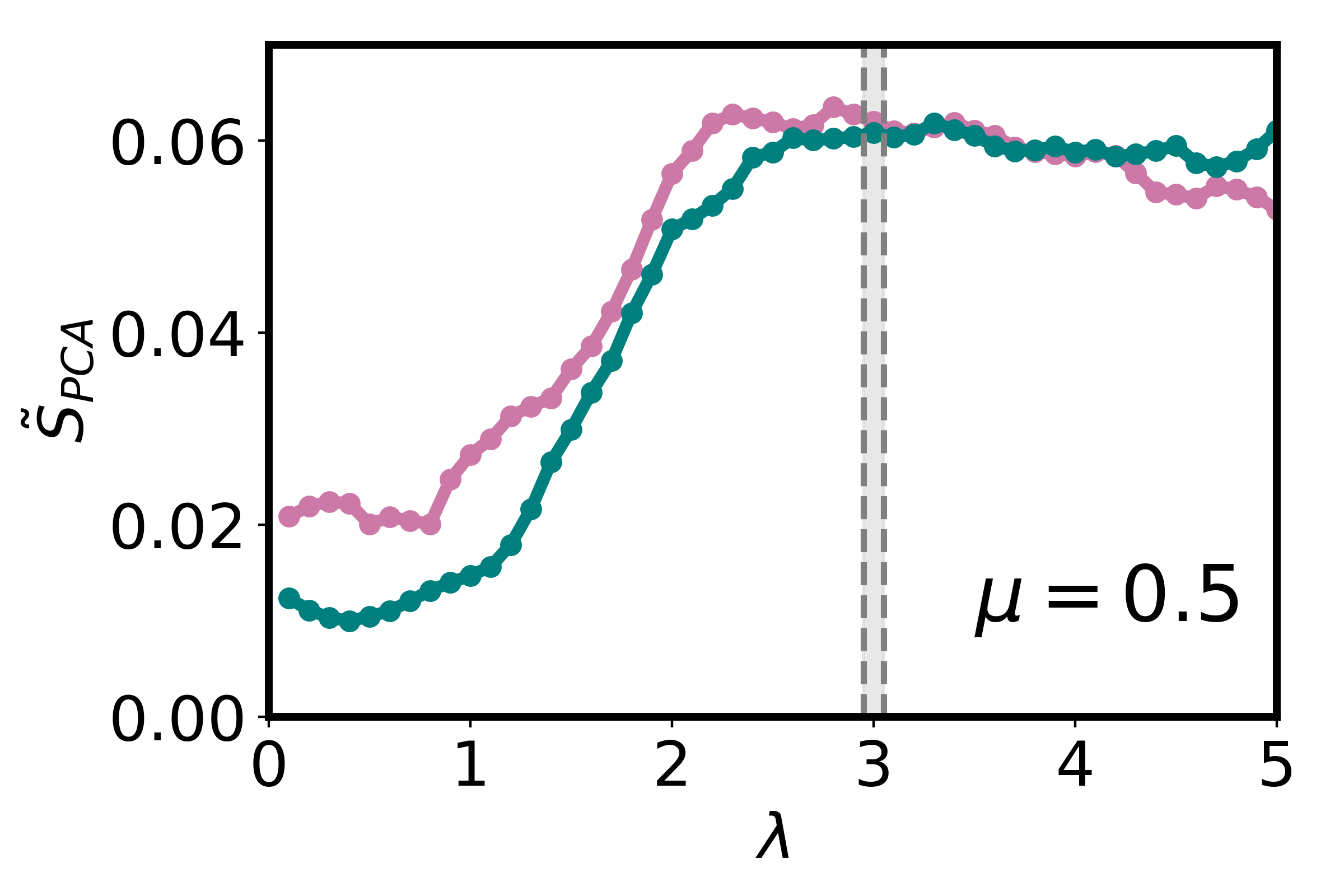}}
\caption{\label{fig8}The difference between the first explained variance ratio (EVR) $p_1$ and the second EVR $p_2$ is shown for (a) increasing hopping amplitude $\mu$ at fixed potential strength $\lambda=0.5$ (transition from the ME to MBC phase) and (c) increasing potential strength $\lambda$ at fixed hopping amplitude $\mu=0.5$ (transition from the ME to MBL phase). Panels (b) and (d) depict the corresponding Shannon entropy obtained from the PCA analysis. The theoretically established transition region in each figure, as outlined in~\cite{PhysRevLett.126.080602}, is shaded in grey for clarity. The datasets used for these analyses consist of rearranged normalized eigenvector probability densities (PDs), with $500$ samples considered for each system size.}
\end{figure}

Next, we compute the explained variance ratios and PCA-entropy using the testing datasets from Section~\ref{sec:level3}. For the normalized eigenvector PDs, we find that the number of principal components capturing $90\%$ of the variance ranges from $O(10)$ in the ME phase to $O(100)$ in the MBL phase, making the results from PCA less meaningful. On the other hand, by rearranging the elements of the normalized eigenvector PDs, we observe that $90\%$ of the variance is captured by the first principal component only. In Fig.~\ref{fig8}, for a given $\lambda$ and $\mu$, we use the testing dataset of rearranged normalized eigenvector PDs to compute the difference between the first two explained variance ratios ($p_1 - p_2$) and the normalized PCA entropy ($\tilde{S}_{PCA}$). We plot these quantities for (i) $\lambda=0.5$, which exhibits a phase transition from the ME to the MBC phase and (ii) $\mu=0.5$, which exhibits a transition from the ME to the MBL phase. 
%%%%%%%%%%%%%%%%%%%%%%%%%%%%%%%%%%%%%%%%%%%%%%%%%%%%%%%%

%%%%%%%%%%%%%%%%%%%%%%%%%%%%%%%%%%%%%%%%%%%%%%%%%%%%%%%%
In the ergodic phase, the first EVR $p_1$ approaches $1$ with increasing system sizes, with all remaining EVRs approaching $0$. Consequently, the PCA entropy $\tilde{S}_{PCA}$ approaches $0$. In both the MBC and MBL phases, the first EVR $p_1 \approx 0.9$, which is reflected in $\tilde{S}_{PCA} < 0.1$. From Figs.~\ref{fig8}(a) and (b), we observe that at $\lambda = 0.5$, the transition from the ME to the MBC phase occurs roughly around $\mu = 1$. Figs.~\ref{fig8}(c) and (d), for fixed $\mu = 0.5$, we observe that the transition from the ME to the MBL phase is roughly around $\lambda = 2.7$. Broadly, $p_1-p_2$ is larger and $\tilde{S}_{PCA}$ is smaller in the MBC phase compared to the MBL phase. However, the transition from the MBC to the MBL phase is not very clearly defined here, as the preprocessing of the eigenvector PDs may still not be sufficient to distinctly separate these classes, as seen in Fig.~\ref{fig7}(b). It would be helpful to explore additional preprocessing steps or other analysis methods that could further improve the distinction between these two phases. The fluctuations found in almost every plot are mostly intrinsic to the quasiperiodic model. This can be attributed to system sizes being non-Fibonacci numbers and the presence of quasiperiodic hopping~\cite{roy2025manybodycriticalphasequasiperiodic}.

Nevertheless, we demonstrate that with appropriate preprocessing, PCA analysis using rearranged normalized eigenvector probability densities (PDs) effectively identifies transitions from the ergodic to non-ergodic phases. Rearranging the eigenvector PDs in descending order emphasizes the most significant features, aligning the data structure with the variance-maximizing direction of PCA. This preprocessing highlights critical differences between classes and standardizes the data representation across samples. The improved class separability, with reduced overlap between the MBC and MBL phases, illustrates the enhanced efficacy of this approach. As a result, the variance captured by the principal components becomes more representative of phase-specific characteristics, enabling a robust identification of phase transitions.

\section{Finite-size scaling}\label{sec:level5}
In this section, we study the finite-size scaling of the output probability obtained from a binary classifier trained using the components of inverse participation ratios (IPR) as inputs. We compare the critical exponents obtained from this analysis with those derived directly from the finite-size scaling of the IPR. The IPR for any single-particle normalized state \(\ket{\psi}\) is mathematically defined as: 
\begin{equation}
    \text{IPR} = \sum_i^N |\psi_i|^4,
\end{equation}
where \(i\) denotes the site index and $|\psi_i|^4$'s are the components of IPR. For a perfectly localized state, the IPR is \(1\), while for a perfectly delocalized state, the IPR approaches \(1/N\). For multifractal states, the IPR lies between these extremes, \(1/N < \text{IPR} < 1\). 

We utilize the scaling ansatz~\cite{PhysRevB.111.024205}
\begin{equation}
    \text{IPR} = N^{-w} f(|\tau-\tau_c|N^{1/\nu}),
    \label{eq5}
\end{equation}
where \(|\tau - \tau_c|\) represents the distance from the critical value in parameter space, \(\nu\) is the critical exponent associated with the correlation length, and \(w\) is an observable-dependent scaling exponent e.g. fractal dimension in case of IPR. To determine these exponents, we employ a cost function defined as~\cite{PhysRevB.102.064207}:
\begin{equation}
    C_Q = \frac{\sum_i^{n_q-1} |Q_{i+1} - Q_i| }{\max\{Q_i\} - \min\{Q_i\}}-1,
\end{equation}
where \(\{Q_i\}\) is the dataset with total \(n_q\) values of IPR$\times N^{w}$, corresponding to different values of \(|\tau - \tau_c|N^{1/\nu}\) and for all system sizes \(N\). The values in \(\{Q_i\}\) are sorted in increasing order corresponding to \(|\tau - \tau_c|N^{1/\nu}\). The best estimates of critical exponents correspond to the minimum value of the cost function $C_Q$. For a perfect scaling collapse, \(C_Q = 0\). 
\begin{figure}[b]
\centering
\stackunder{\hspace{-4cm}(a)}{\includegraphics[width=4.2cm,height=3.1cm]{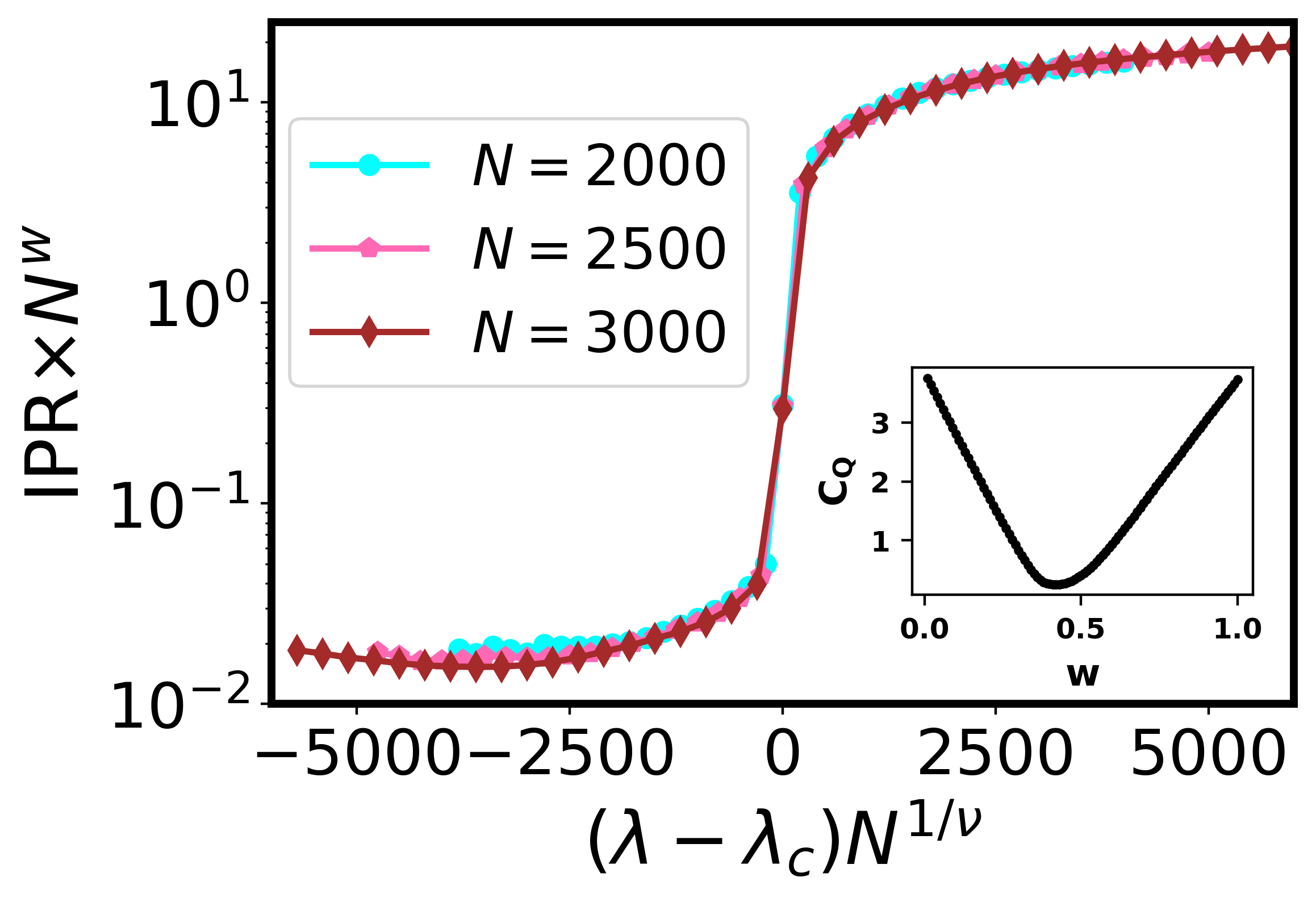}}
\stackunder{\hspace{-4cm}(b)}{\includegraphics[width=4.2cm,height=3.1cm]{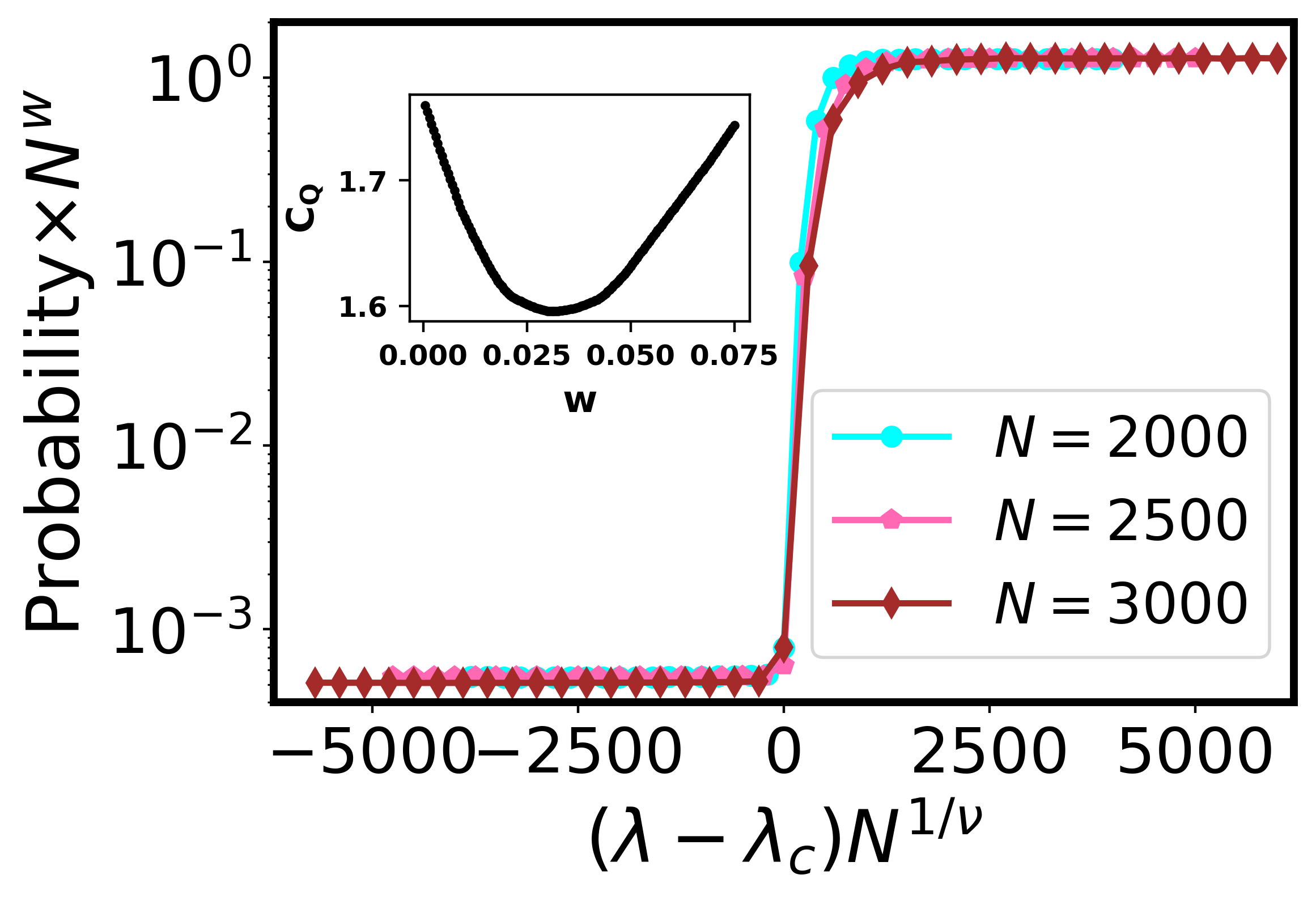}}
\vspace{-0.4cm}

\stackunder{\hspace{-4cm}(c)}{\includegraphics[width=4.2cm,height=3.1cm]{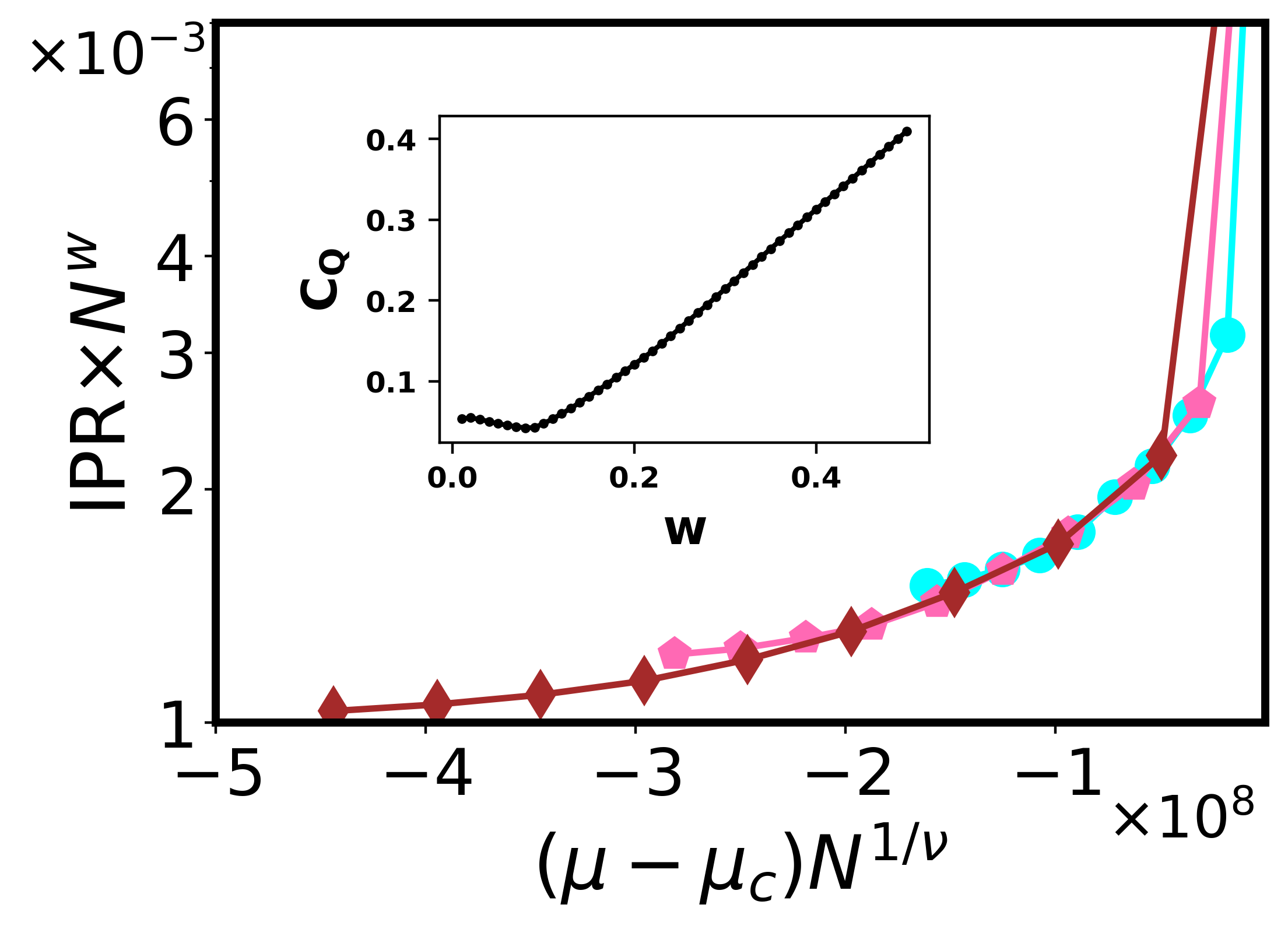}}
\stackunder{\hspace{-4cm}(d)}{\includegraphics[width=4.2cm,height=3.1cm]{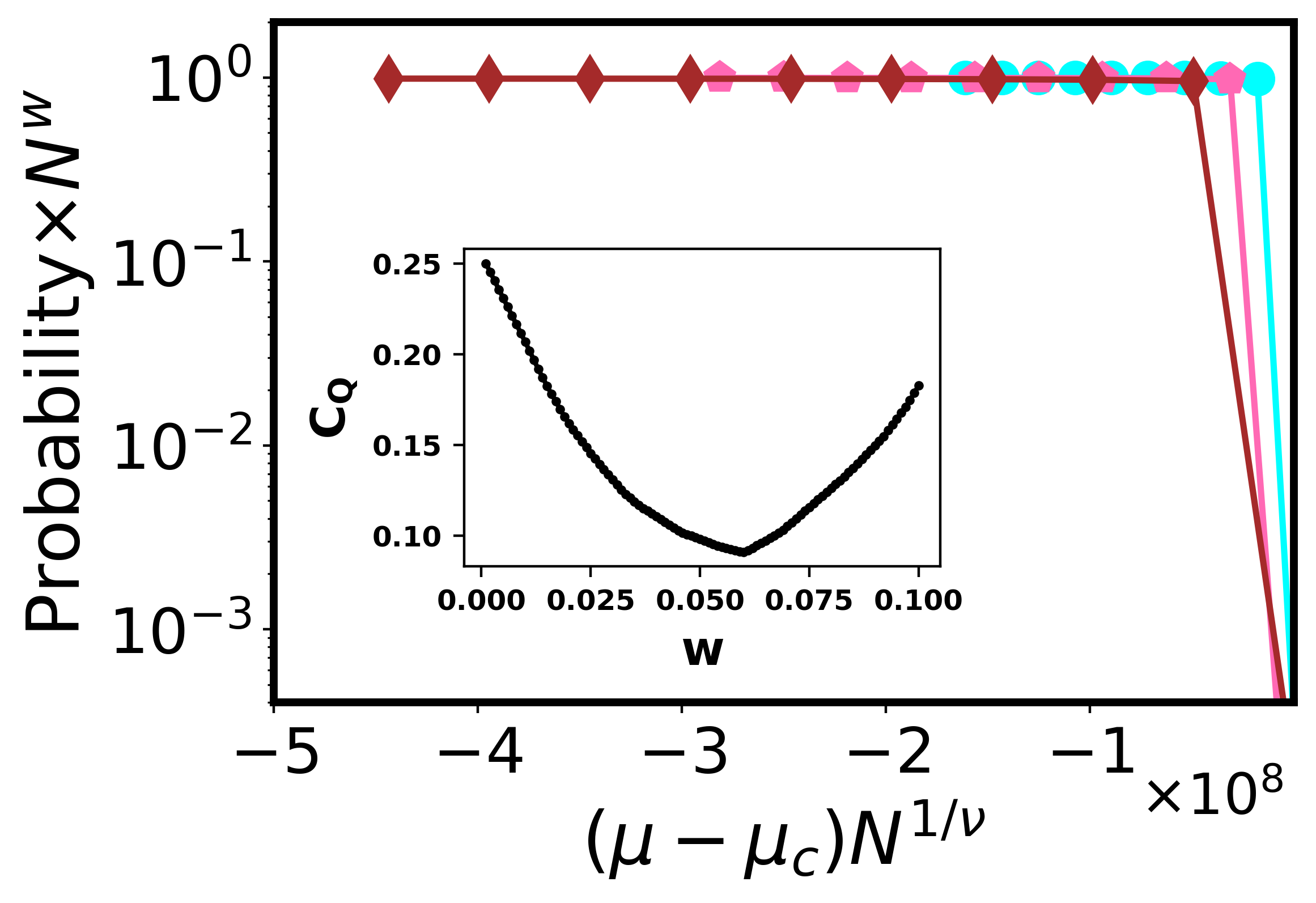}}
\vspace{-0.4cm}

\stackunder{\hspace{-4cm}(e)}{\includegraphics[width=4.2cm,height=3.1cm]{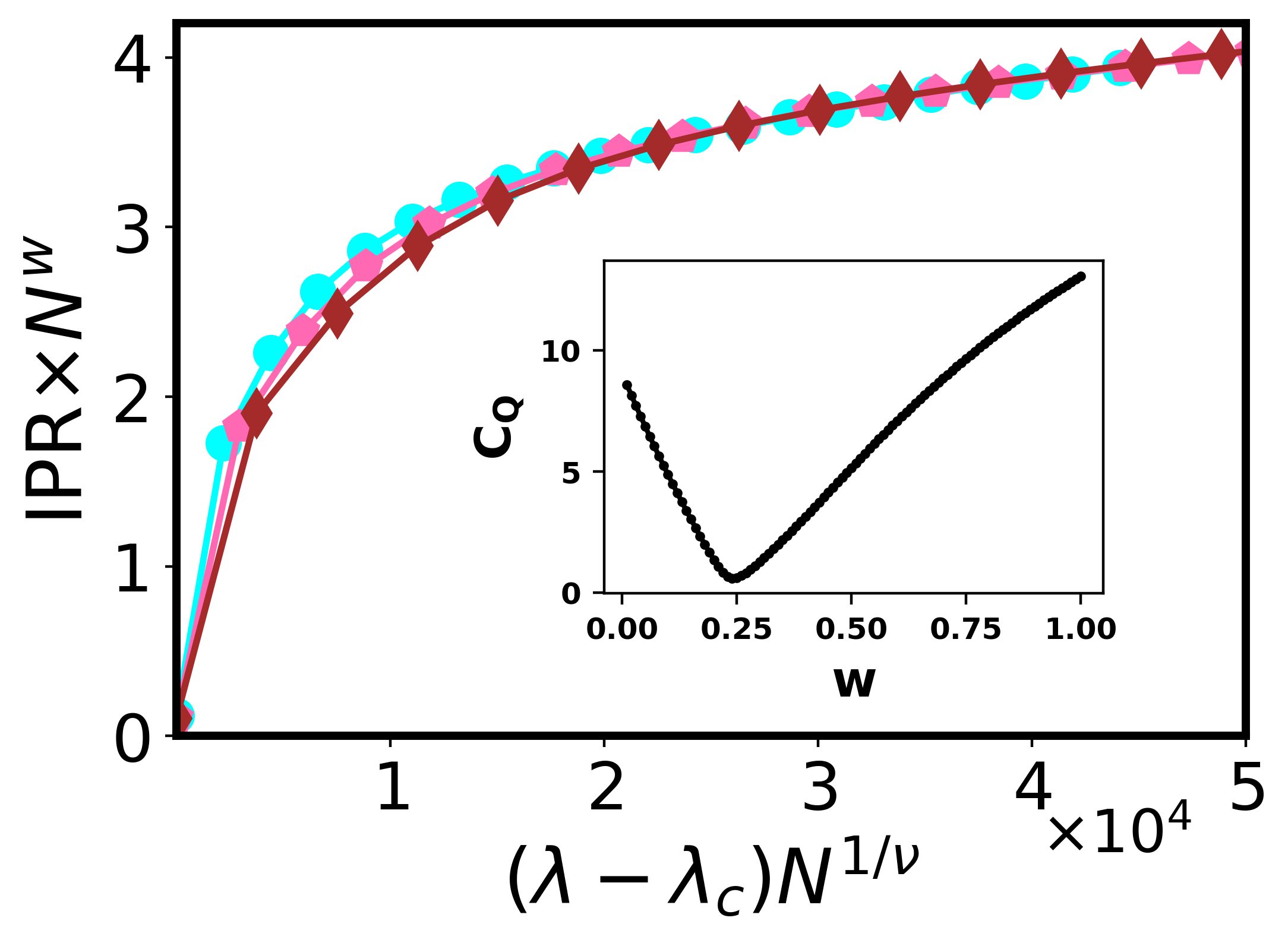}}
\stackunder{\hspace{-4cm}(f)}{\includegraphics[width=4.2cm,height=3.1cm]{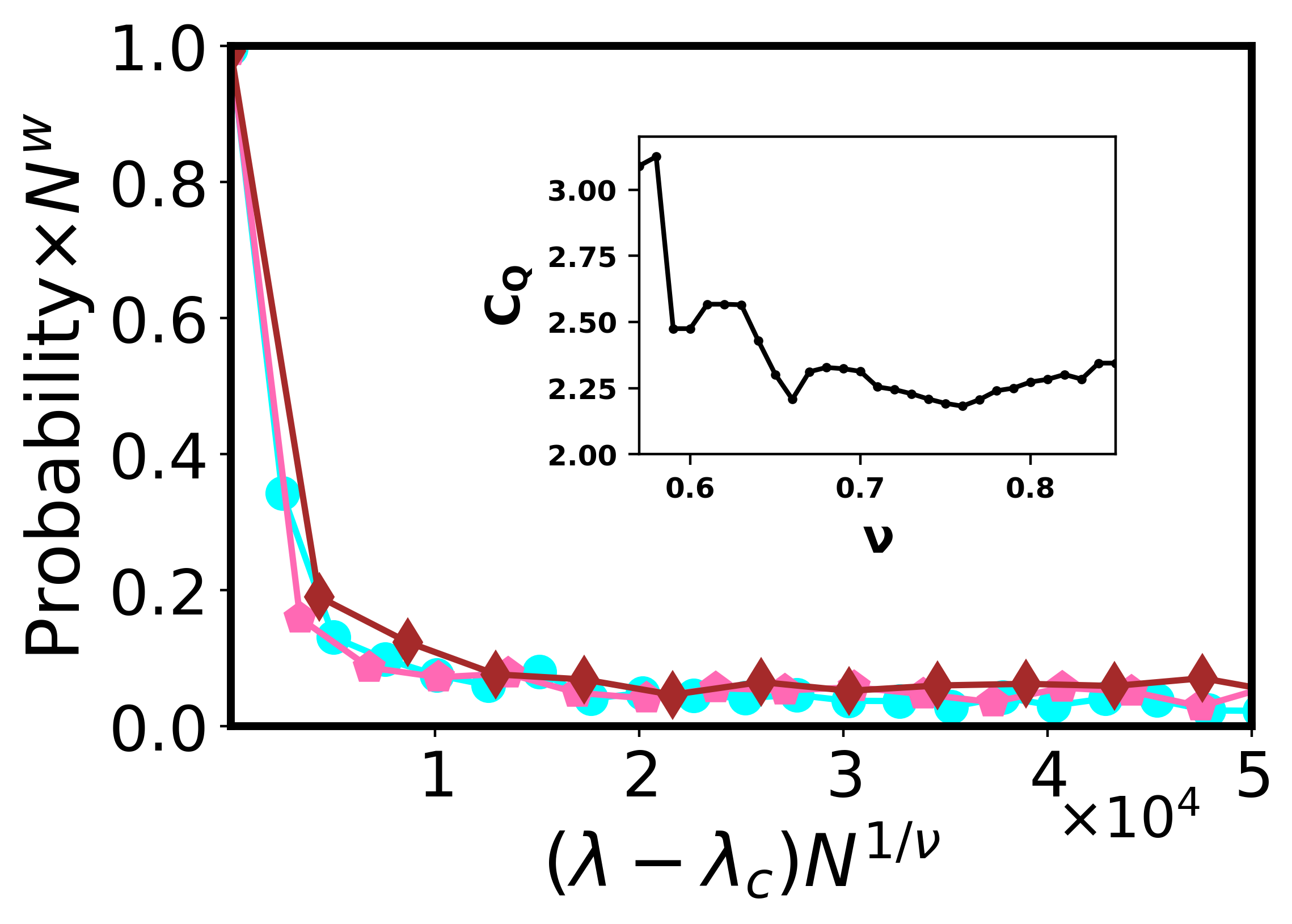}}
\caption{\label{fig9}Finite-size scaling of the IPR and binary classifier probability using IPR-components as inputs. (a) Scaling collapse for the delocalized-localized transition $\lambda_c = 2,\mu = 0$ using the IPR. (b) Corresponding scaling collapse using the classifier probability. (c) Scaling collapse for the delocalized-critical transition $\mu_c = 1, \lambda = 0.5$, approaching from the delocalized side of the phase diagram. (d) Corresponding scaling collapse using the classifier probability. (e) Scaling collapse for the critical-localized transition $\lambda_c = 3,\mu = 1.5$, approaching from the localized side of the phase diagram. (f) Corresponding scaling collapse using the classifier probability. The insets correspond to cost function $C_Q$ in all cases. $500$ realizations for each system size have been considered for IPR and as testing data sets for binary classifiers.}
\end{figure}

In our analysis, we consider \(500\) disorder realizations of mid-spectrum states across various phase transitions. The IPR is calculated for each realization, and a scaling collapse is performed to extract the critical exponents. Concurrently, we train a binary classifier (same as described in Table~\ref{tab1} for the many-body case) using \(5000\) samples per class of IPR data. For delocalized and localized classes, the samples are extracted at $\mu=0$, $\lambda=0.5$ and $\mu=0$, $\lambda=3.5$ respectively. For the multifractal class, we obtained samples from $\mu=2$ and $\lambda=0.3,1,2.5,3$. Once trained, the network predicts the probabilities for the \(500\) test samples from the disorder realizations discussed earlier. The training process requires only \(2\) epochs to achieve \(99\%\) accuracy across all cases.

 For the single-particle system, we analyze three distinct transitions: (i) delocalized-localized transition (\(\mu = 0\)), (ii) delocalized-critical transition (\(\lambda = 0.5\)), and (iii) critical-localized transition (\(\mu = 1.5\)). For the well-known AAH model (\(\mu = 0\)), the delocalization-localization transition occurs at the critical disorder strength \(\lambda_c = 2\). Figure~\ref{fig9}(a) shows the finite-size scaling of the IPR for this transition, while Fig.~\ref{fig9}(b) illustrates the scaling of the probability obtained from a binary classifier trained on IPR values deep in the delocalized and localized phases. The classifier assigns a probability \(P\) to the input being localized and \(1-P\) to being delocalized. In both cases, the cost function is minimized for the critical exponent \(\nu = 1\).

For the delocalized-critical transition along \(\lambda = 0.5\), we analyze the scaling behaviour as the transition point $\mu_c=1$ is approached from the delocalized side. Figures~\ref{fig9}(c) and~\ref{fig9}(d) compare the IPR scaling and the scaling of the binary classifier probability. The classifier is trained on IPR values from the delocalized and critical phases. The resulting network gives probability $P$ corresponding to the input belonging to the delocalized phase and $1-P$ to the input belonging to the critical phase. We observe that the cost function is minimized in both cases for the critical exponents $\nu=0.4$. For the critical-localized transition at $\lambda_c=3$ \(\mu = 1.5\), the IPR scaling yields a critical exponent \(\nu = 0.76\) , as shown in Fig.~\ref{fig9}(e). The same value of \(\nu\) is obtained from the classifier trained on IPR values from the critical and localized phases, as shown in Fig.~\ref{fig9}(f). The classifier assigns a probability \(P\) to the critical phase and \(1-P\) to the localized phase. 
%However, we again observe that the scaling exponent \(w\) is not consistent between the two methods.

\begin{table}[b]
\begin{tabular}{ | m{8em} | m{8em} | m{8em} | }
  \hline
  Phase Transition & Critical exponents for scaling collapse of IPR & Critical exponents for scaling collapse of classifier probability \\
  \hline
      & &  \\

  Delocalized to Localized (\(\mu = 0\)) 
  & \(\lambda_c = 2\), \(\nu = 1\), \(w = 0.42\) 
  & \(\lambda_c = 2\), \(\nu = 1\), \(w = 0.03\) \\
  \hline
        & &  \\
  Delocalized to Critical (\(\lambda = 0.5\)) 
  & \(\mu_c = 1\), \(\nu = 0.4\), \(w = 0.08\) 
  & \(\mu_c = 1\), \(\nu = 0.4\), \(w = 0.06\) \\
  \hline
      & &  \\
  Critical to Localized (\(\mu = 1.5\)) 
  & \(\lambda_c = 3\), \(\nu = 0.76\), \(w = 0.24\) 
  & \(\lambda_c = 3\), \(\nu = 0.76\), \(w = 0\) \\
  \hline
\end{tabular}
\caption{\label{tab3} Critical exponents obtained from the scaling analysis of the IPR and binary classifier probability using IPR as inputs for single-particle phase transitions (see Fig.~\ref{fig9}).}
\end{table}

The critical exponents are summarized in detail in Table~\ref{tab3}. The values of critical correlation length exponent $\nu$ found from scaling collapse of the probability obtained from a binary classifier trained on the IPR are consistent with the same found from the scaling derived directly from the IPR for all three types of transitions. However, apparently, values of the other exponent $w$ differ between classifier probability and IPR. This discrepancy arises because \(w\) captures the system size dependence of the IPR or classifier probability. To be precise $w$ is the fractal dimension of eigenstates corresponding to localization transitions~\cite{PhysRevB.42.8121}. Hence, $w$ is a fraction when the transition involves a localized phase.
On the other hand, the classifier probability \(w  \approx 0\) for all three types of transitions, indicates that it is almost independent of the system size \(N\) once scaled by the correlation length exponent \(\nu\). This suggests that the classifier effectively identifies phase boundaries without depending much on \(N\) due to the training method. Minimal system size dependence could be attributed to the presence of critical states at the transition point which was excluded in the training process. These states may subtly influence the classifier's sensitivity to system size.

Overall, we demonstrate the robust capability of the binary classifier to facilitate consistent scaling collapse using IPR components as inputs. Extending this analysis to many-body systems, we employ normalized eigenvector PDs and normalized IPR components as inputs to the classifier. Using the scaling ansatz \(\text{P} = f(|\tau-\tau_c|N^{1/\nu})\), for the machine learning probability $P$ analogous to Eq.~\ref{eq5}, we observe a satisfactory collapse in the MBC and MBL phases (not shown here). However, at the minimum of the cost function \(C_Q\), the critical disorder parameter \(\tau_c\) and critical exponent \(\nu\) deviate from the standard values reported in Ref.~\cite{PhysRevLett.126.080602}. Given that this analysis was conducted with system sizes \(N = 12, 14\), we conclude that incorporating larger system sizes might improve the results. Alternatively, the inability to achieve a consistent scaling collapse may reflect the inherent limitations of machine-learning approaches in many-body systems, as previously noted in Ref.~\cite{PhysRevB.100.224202}. 

Additionally, we perform scaling collapse analyses for the PCA entropy \(\tilde{S}_{\text{PCA}}\) in single-particle systems for two sets of data: \(R\) (where the system size is smaller than the number of samples, \(n < m\)) and \(L\) (where the system size is larger than the number of samples, \(n > m\)) (see  Section~\ref{sec:level5}B). In both cases, \(\tilde{S}_{\text{PCA}}\) does not collapse well under the ansatz \(\tilde{S}_{\text{PCA}} = f(|\tau-\tau_c|N^{1/\nu})\). Furthermore, although the critical disorder strengths \(\tau_c\) obtained are accurate, the critical exponent \(\nu\) significantly deviates from the expected range. In conclusion, a consistent scaling collapse for \(\tilde{S}_{\text{PCA}}\) was not achieved, underscoring the need for further investigation, e.g. as shown in Ref.~\cite{PhysRevB.110.024204}.

\section{Conclusion and Outlook}\label{sec:level6}
In this work, we investigate critical phases alongside delocalized and localized phases both in interacting and non-interacting quasiperiodic disordered systems using simplified yet effective inputs, such as eigenvalue spacings and eigenvector probability densities (PDs). These features, derived from fundamental system properties, provide a practical and efficient means of analyzing phase transitions in such systems. By leveraging fully connected neural networks and adopting a phase detection via confusion approach within the multi-class classifiers, we demonstrate their ability to classify diverse phases, including the ergodic, MBL, and intermediate MBC phases, while revealing indications of sub-phases within it. This approach highlights the feasibility of systematically probing phase transitions in disordered systems using minimal yet informative inputs.

In the single-particle regime, unsupervised learning techniques such as PCA prove invaluable for identifying phase transitions. By analyzing eigenvector PDs, we find that the PCA entropy and its numerical derivative are effective tools for detecting transitions. While these methods excel in single-particle systems, their extension to many-body systems presents greater challenges. To improve class separability between ergodic and non-ergodic phases, we apply data preprocessing that enables PCA to more effectively capture phase distinctions. Notably, the first principal component retains high variance after these adjustments, which remains consistent with increasing system size and underscores PCA's robustness in capturing the correlation structure of eigenvector PDs. However, further refinements in data preparation are necessary to fully distinguish the MBC and MBL phases, as the current approach only partially mitigates the overlap between these phases.

Furthermore, we perform a systematic comparison of critical exponents obtained through finite-size scaling analysis in the single-particle regime, using both traditional IPR-based methods and neural network outputs. The strong agreement between these approaches validates the reliability of our methods, particularly in the single-particle case, where the extracted critical exponents align well with theoretical predictions. However, deviations observed in the scaling of PCA entropy in the single-particle regime and in network outputs based on eigenvector derivatives in the many-body case indicate areas that warrant further investigation. These findings highlight the need for larger system sizes and more refined scaling strategies to achieve a deeper understanding of critical behaviour in complex quantum systems.

Overall, our study demonstrates the potential of machine learning to bridge traditional techniques with data-driven approaches, providing a robust framework for investigating phase separability, critical exponents, and structural correlations in disordered quantum systems, even where any prior information about phase transitions is not available. By integrating simplified inputs and advanced methodologies, this work lays a foundation for exploring complex interacting regimes, critical phases, and topological transitions, showcasing the versatility of machine learning in understanding quantum systems. Future work could extend these methods to address finite-size effects and enhance the precision of critical exponent estimation. Additionally, advanced preprocessing techniques, such as the auto encoders, and feature engineering hold promise for improving phase classification in challenging many-body regimes. These efforts, combined with the exploration of critical phases, could further expand the applicability of machine learning in unravelling the complexities of quantum systems.

\begin{acknowledgments}
We are grateful to Daniel Leykam for his valuable comments on the manuscript. NR thanks Sumilan Banerjee for fruitful and interesting discussions about the phase diagram of the interacting EAAH model. All neural networks used in this work were implemented in Python using TENSORFLOW~\cite{45166}.
\end{acknowledgments}

%%%%%%%%%%%%%%%%%%%%%%%%%%%%%%%%%%%%%%%%%%%%%%%%%%%%%%%%
%%%%%%%%%%%%%%%%%%%%%%%%%%%%%%%%%%%%%%%%%%%%%%%%%%%%%%%%
%\section*{APPENDIX}
\appendix
\section{Gap Ratio and MIPR}\label{appA}
In Section~\ref{sec:level3}C of the main text, we presented a detailed analysis of the three-class classifier, which employed supervised learning techniques on eigenvalue spacings and eigenvector probability densities to classify ME, MBC, and MBL phases. To complement this machine learning-based approach, we now utilize the same testing dataset to compute the established measures, namely, the gap ratio and many-particle inverse participation ratio (MIPR). 

The gap ratio serves as a robust diagnostic for phase transitions in disordered systems by characterizing the statistical properties of energy level spacings. Mathematically, $r_{av}$~\cite{PhysRevLett.110.084101} is defined as:
\begin{eqnarray}
r_{av} = \bigg\langle \frac{1}{N-2} \sum\limits_{i=1}^{N-2} \frac{\min[s_i, s_{i+1}]}{\max[s_i, s_{i+1}]} \bigg\rangle,
\label{eq4}
\end{eqnarray}
where the energies $E_i$ are arranged in ascending order, and the energy level spacings are given by $s_i = E_{i+1} - E_i$. The angular brackets in Eq.~\ref{eq4} denote averaging over disorder realizations. In the ergodic phase, energy levels follow the Wigner-Dyson statistics, leading to $r_{av}\approx0.528$, indicative of strong level repulsion. Conversely, in the MBL phase, energy levels obey Poisson statistics, resulting in $r_{av}\approx0.386$, reflecting the lack of level correlations~\cite{PhysRevLett.110.084101}. These distinct values provide a clear signature of phase separation and the intermediate behaviour observed in the MBC phase highlights its hybrid characteristics, bridging ergodic and localized regimes.
%\label{sec:level7}

\begin{figure}
\centering
\stackunder{\hspace{-4cm}(a)}{\includegraphics[width=4.2cm]{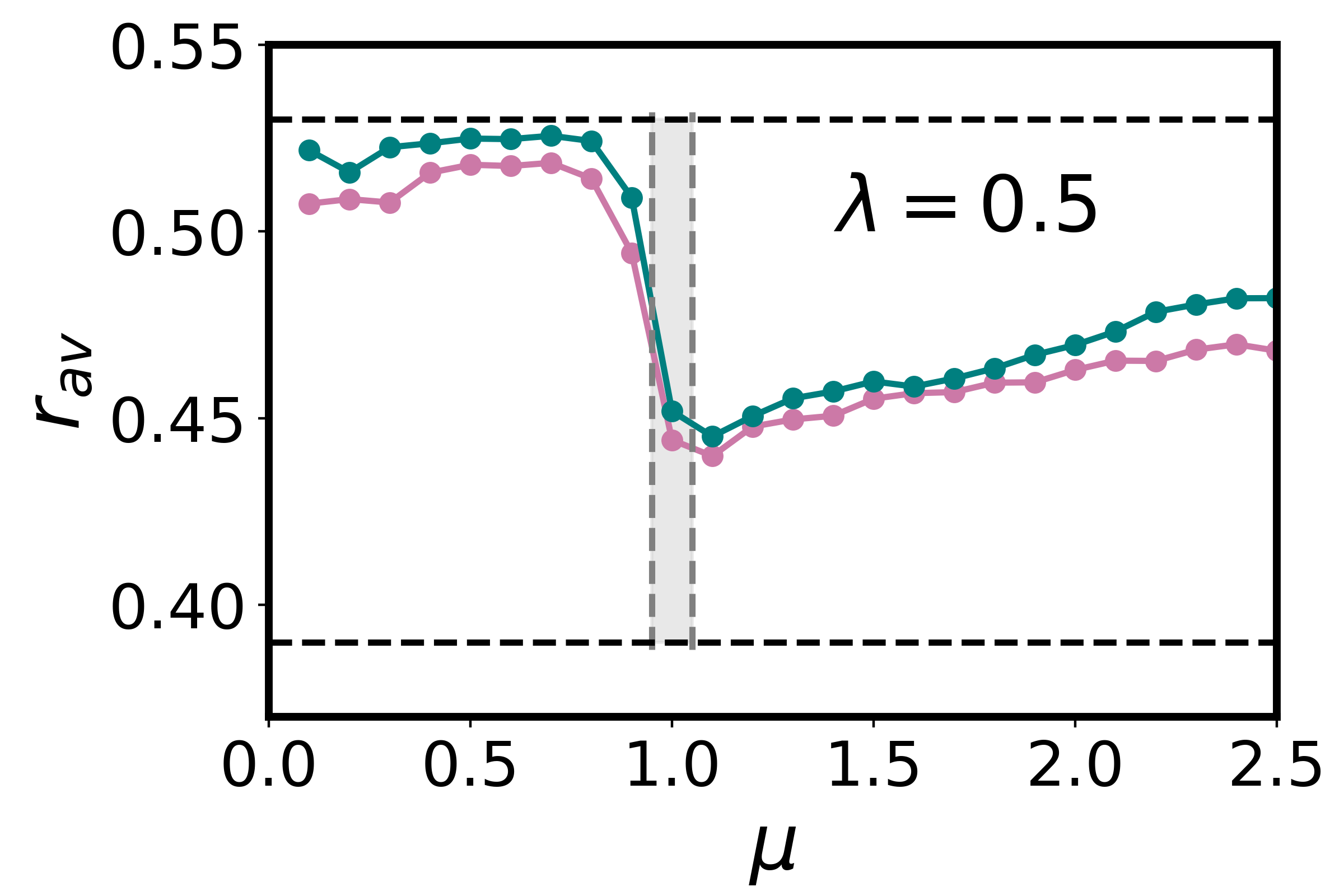}}
\stackunder{\hspace{-4cm}(b)}{\includegraphics[width=4.2cm]{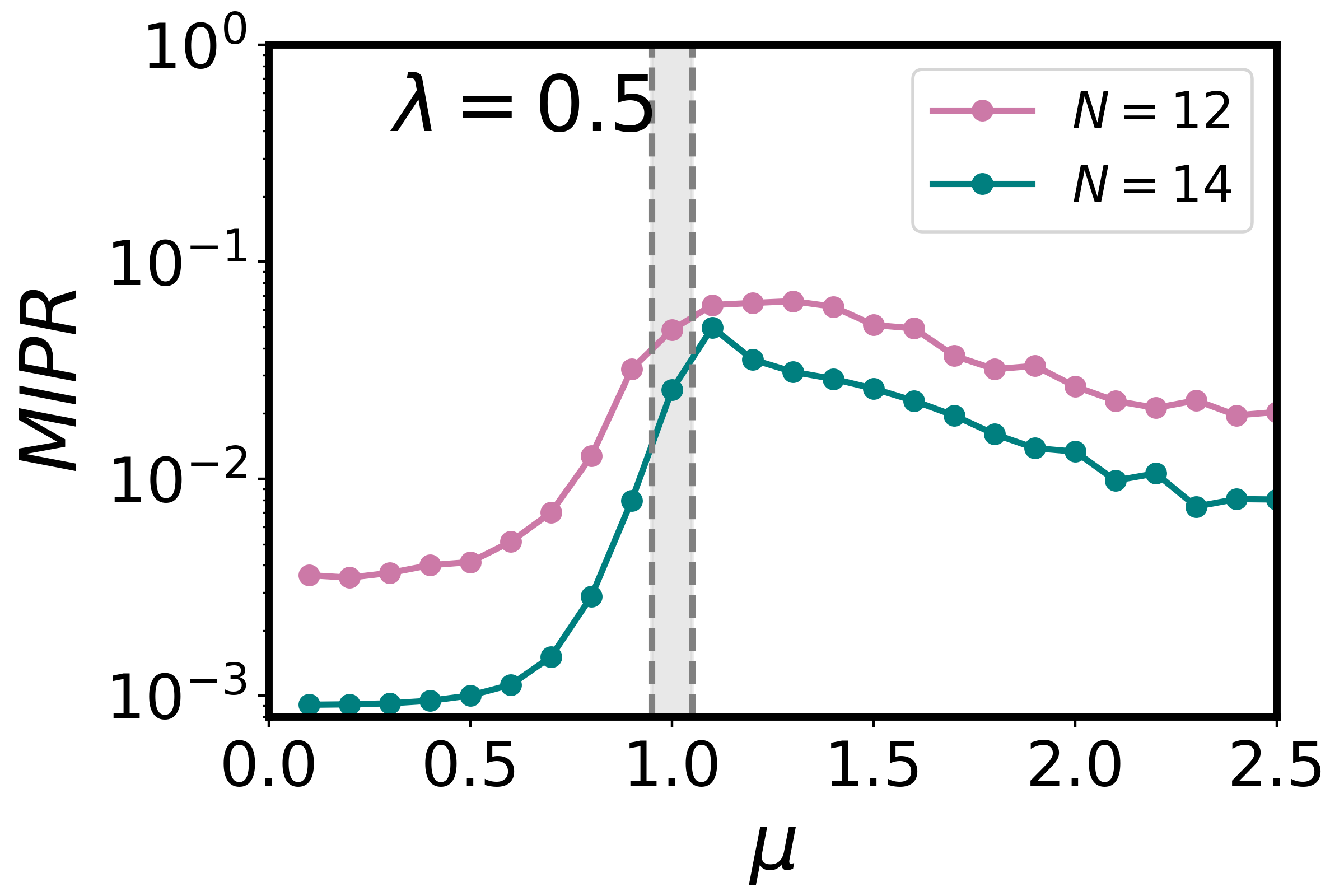}}
\vspace{-0.4cm}

\stackunder{\hspace{-4cm}(c)}{\includegraphics[width=4.2cm]{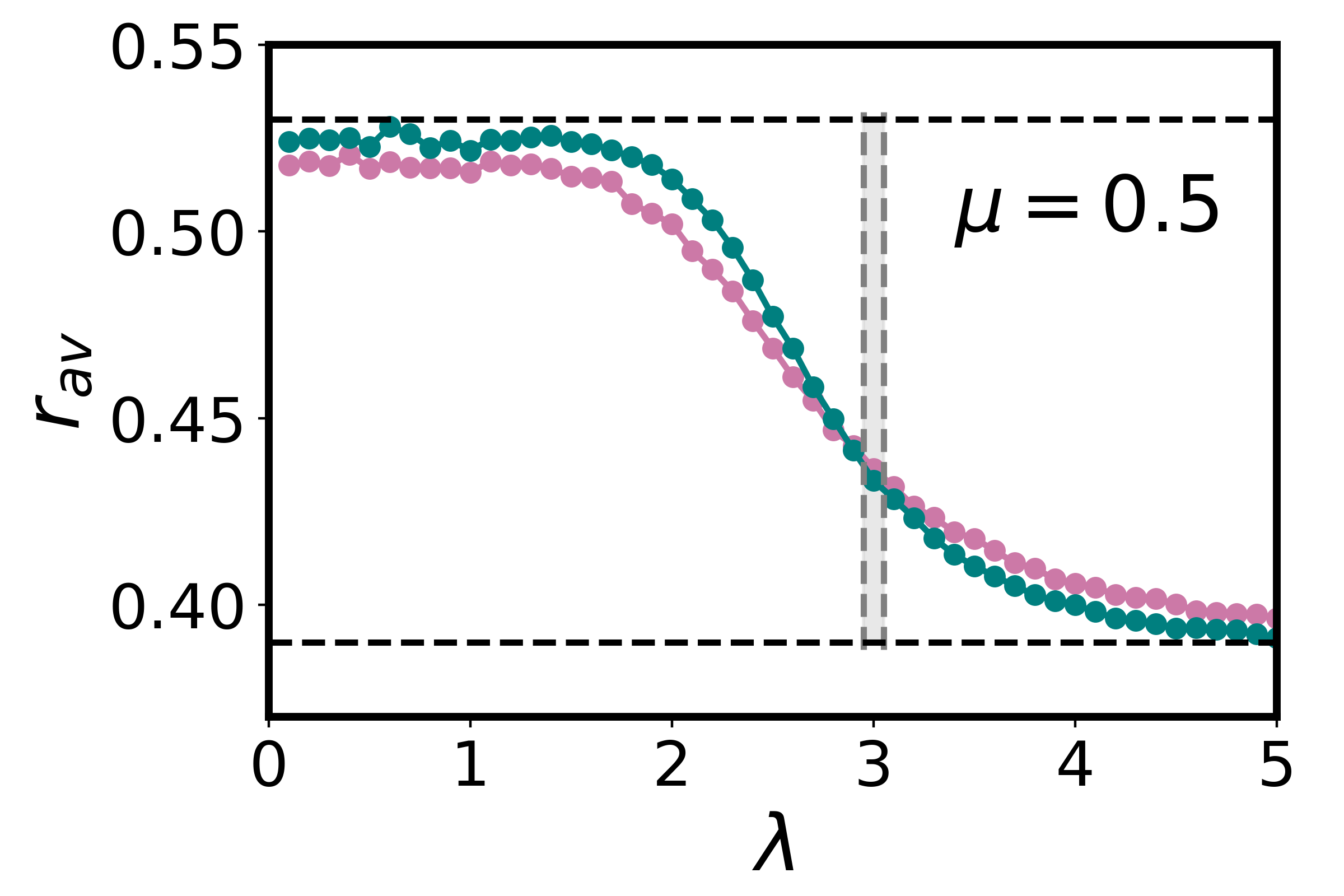}}
\stackunder{\hspace{-4cm}(d)}{\includegraphics[width=4.2cm]{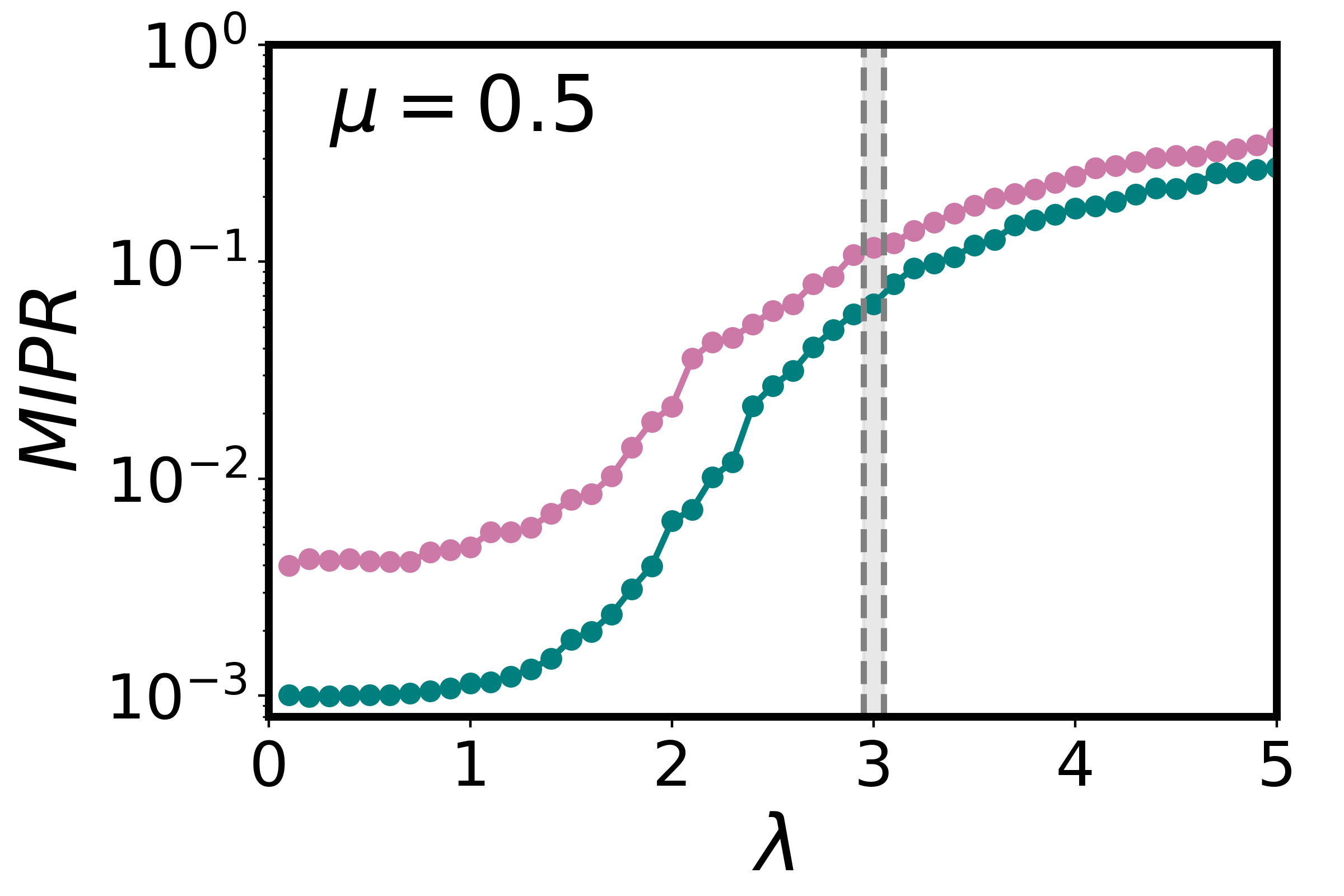}}
\vspace{-0.4cm}

\stackunder{\hspace{-4cm}(e)}{\includegraphics[width=4.2cm]{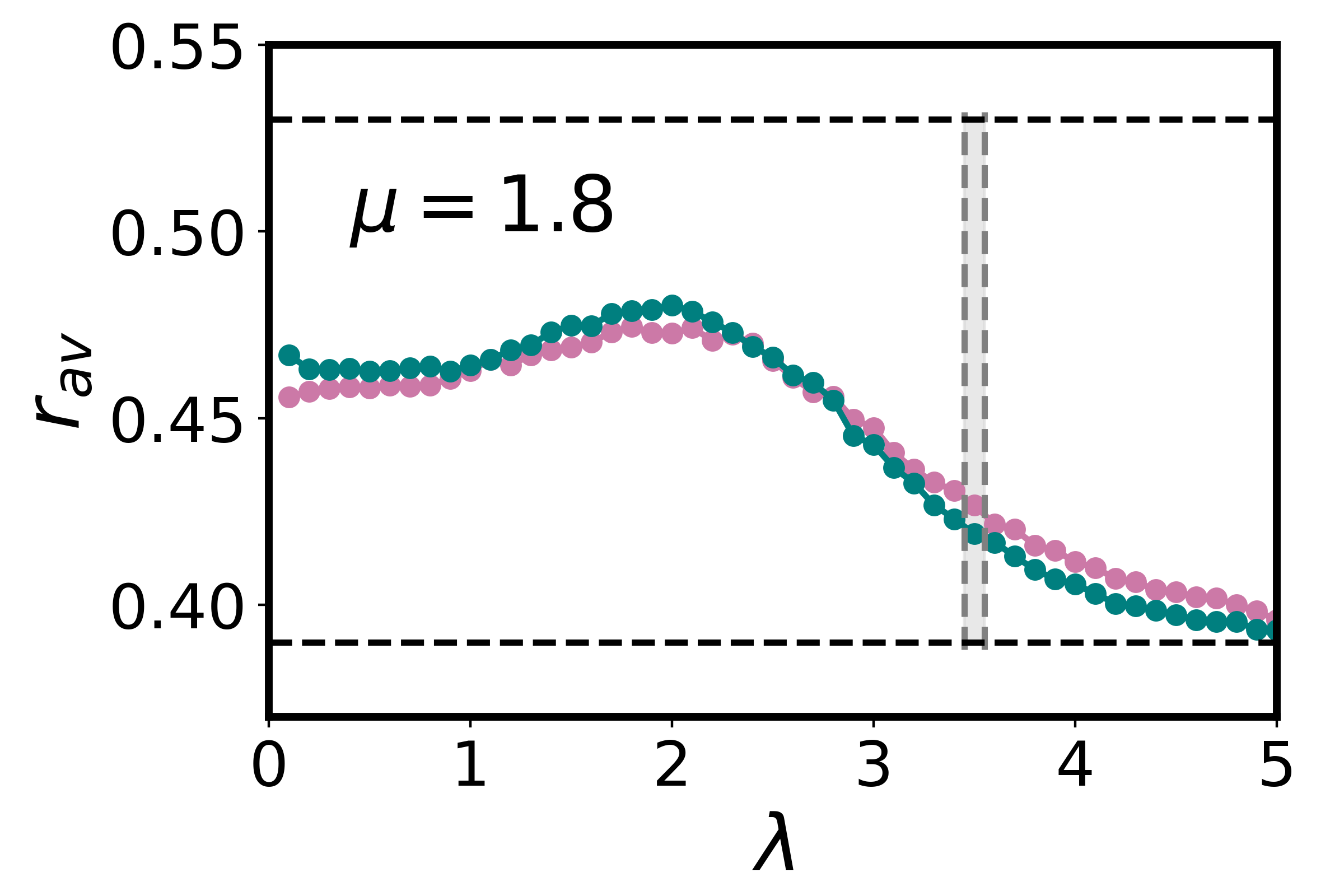}}
\stackunder{\hspace{-4cm}(f)}{\includegraphics[width=4.2cm]{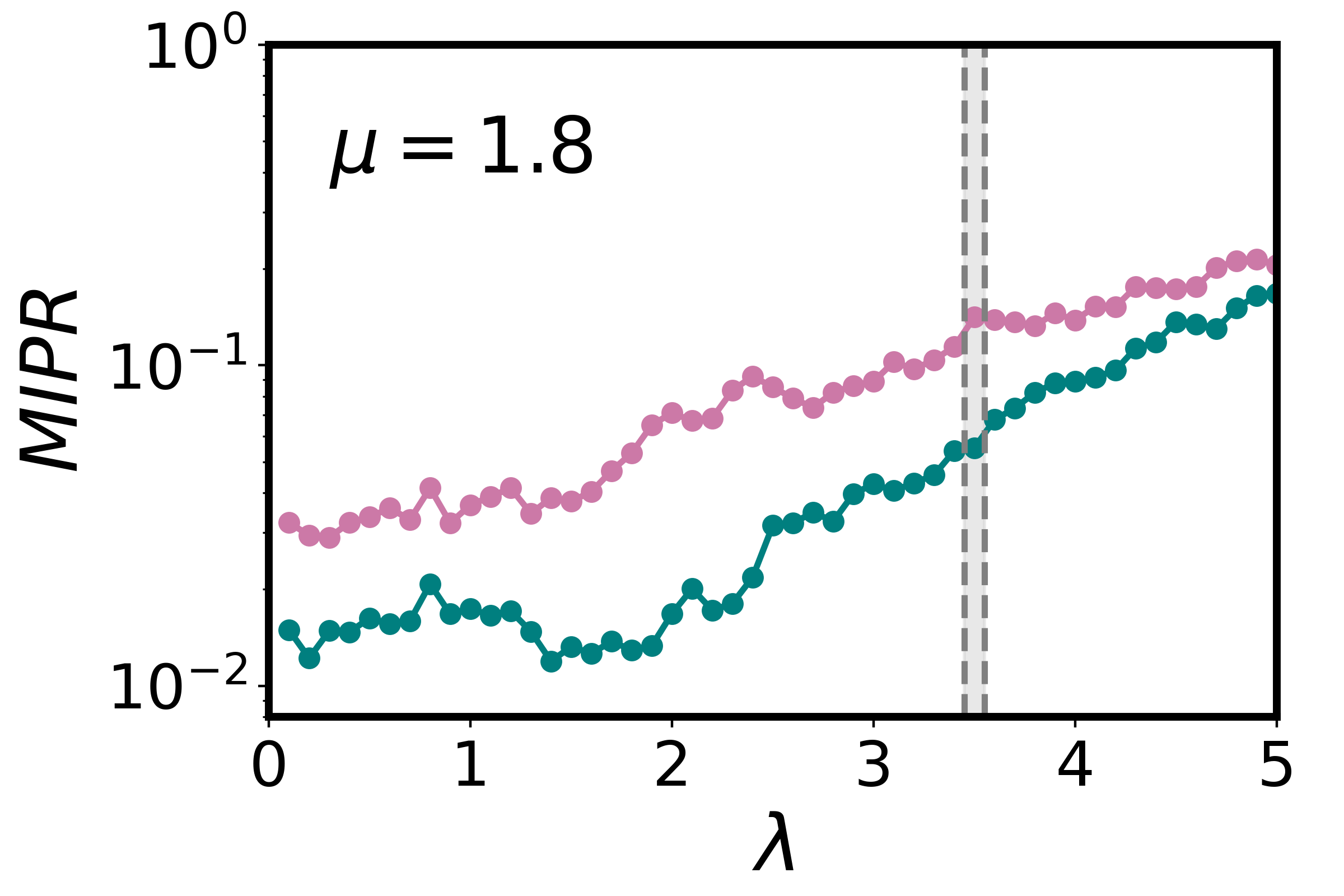}}
\caption{\label{fig10} Gap ratio (a), (c), (e) and MIPR (b), (d), (f) for varying parameters: (a-b) $\lambda=0.5$, (c-d) $\mu=0.5$, and (e-f) $\mu=1.8$. For $N=12$ and $14$, $1000$ and $300$ eigenvalue disorder realizations, and $1500$ and $500$ eigenvector disorder realizations have been used, respectively. The standard reference values are indicated by dashed lines in the $r_{\text{av}}$ plots. The theoretically established transition region in each figure, as outlined in~\cite{PhysRevLett.126.080602}, is shaded in grey for clarity. }
\end{figure}

\begin{figure*}
\centering
\stackunder{\hspace{-5cm}(a)}{\includegraphics[width=4.7cm]{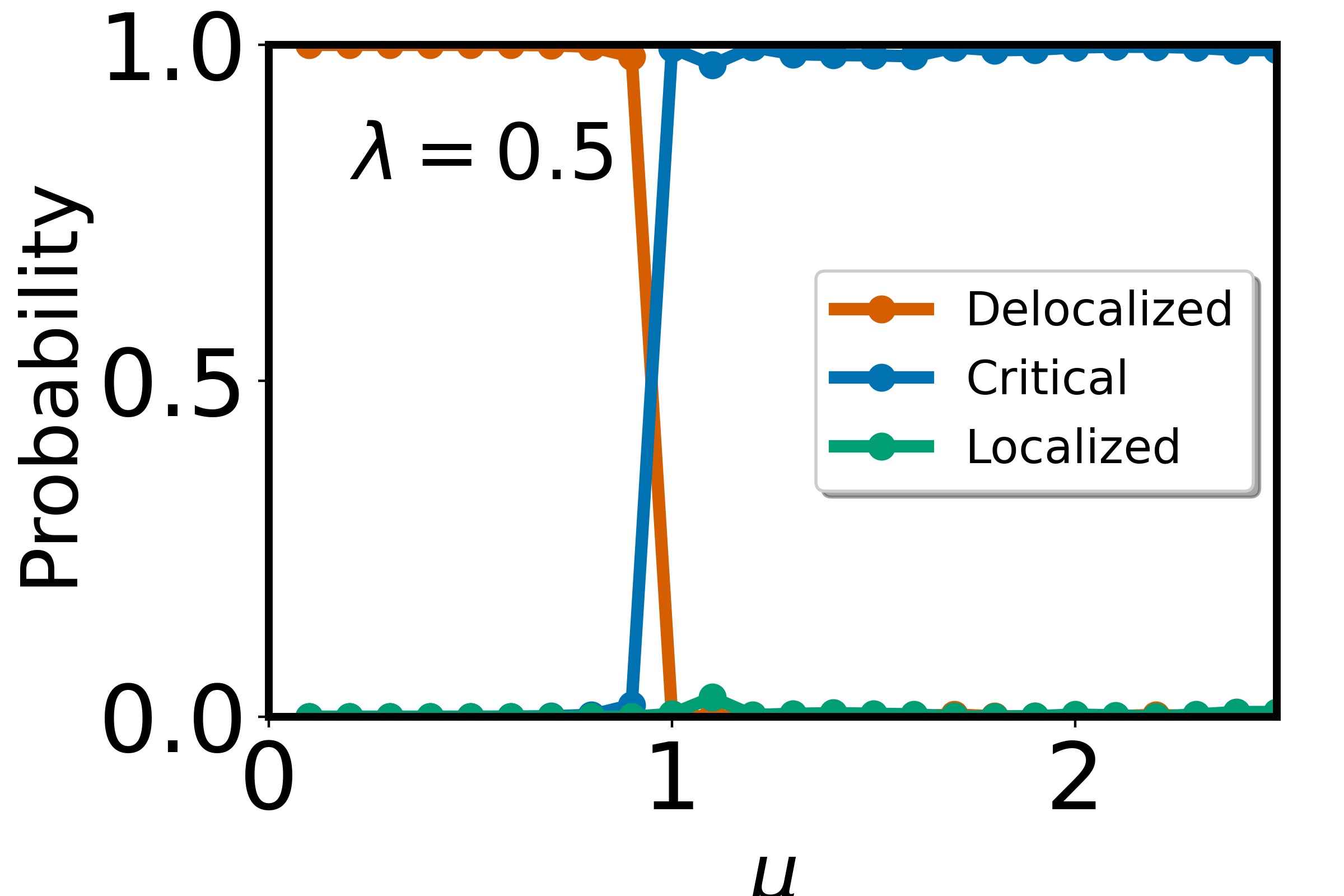}}
\stackunder{\hspace{-5cm}(b)}{\includegraphics[width=4.7cm]{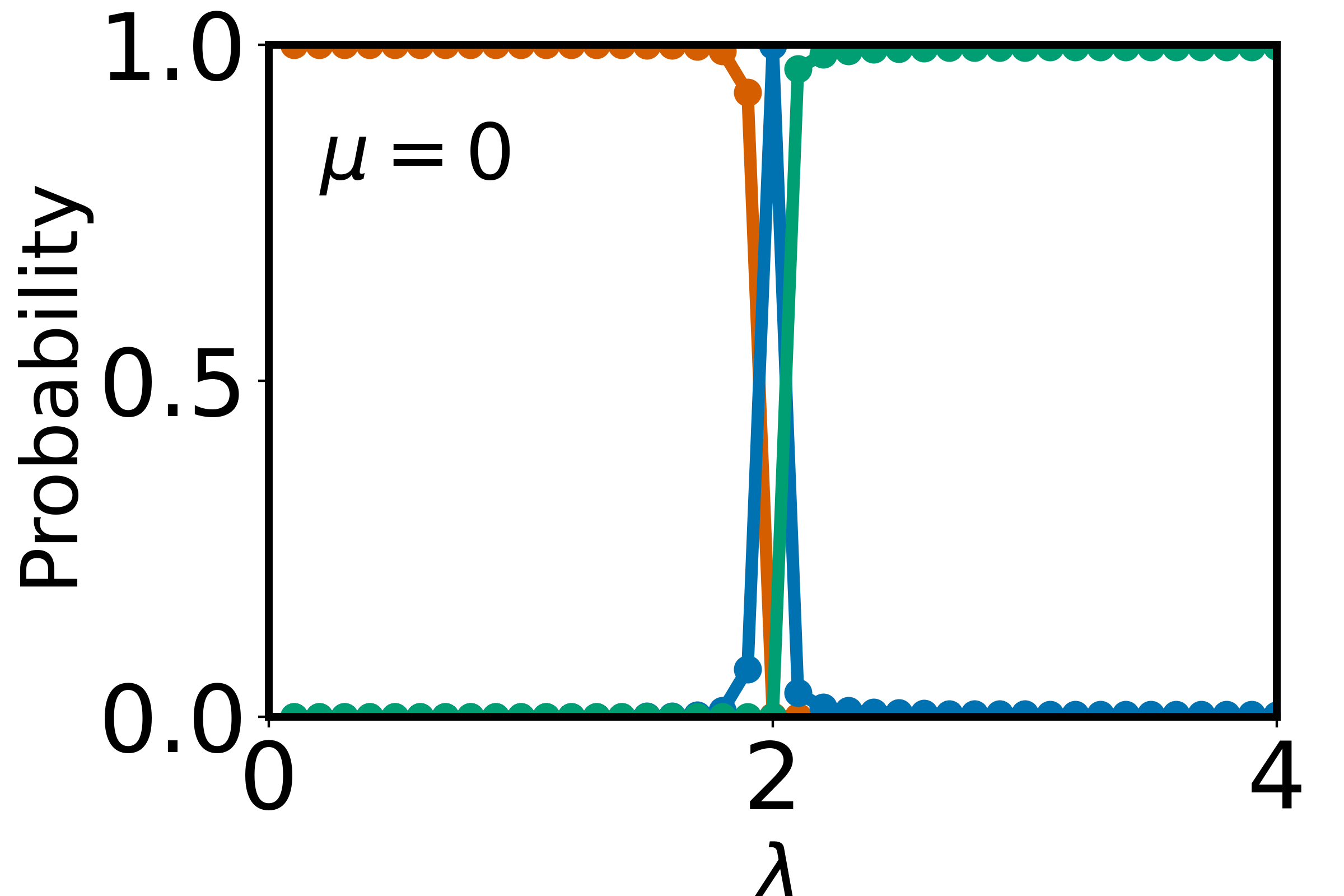}}
\stackunder{\hspace{-5cm}(c)}{\includegraphics[width=4.7cm]{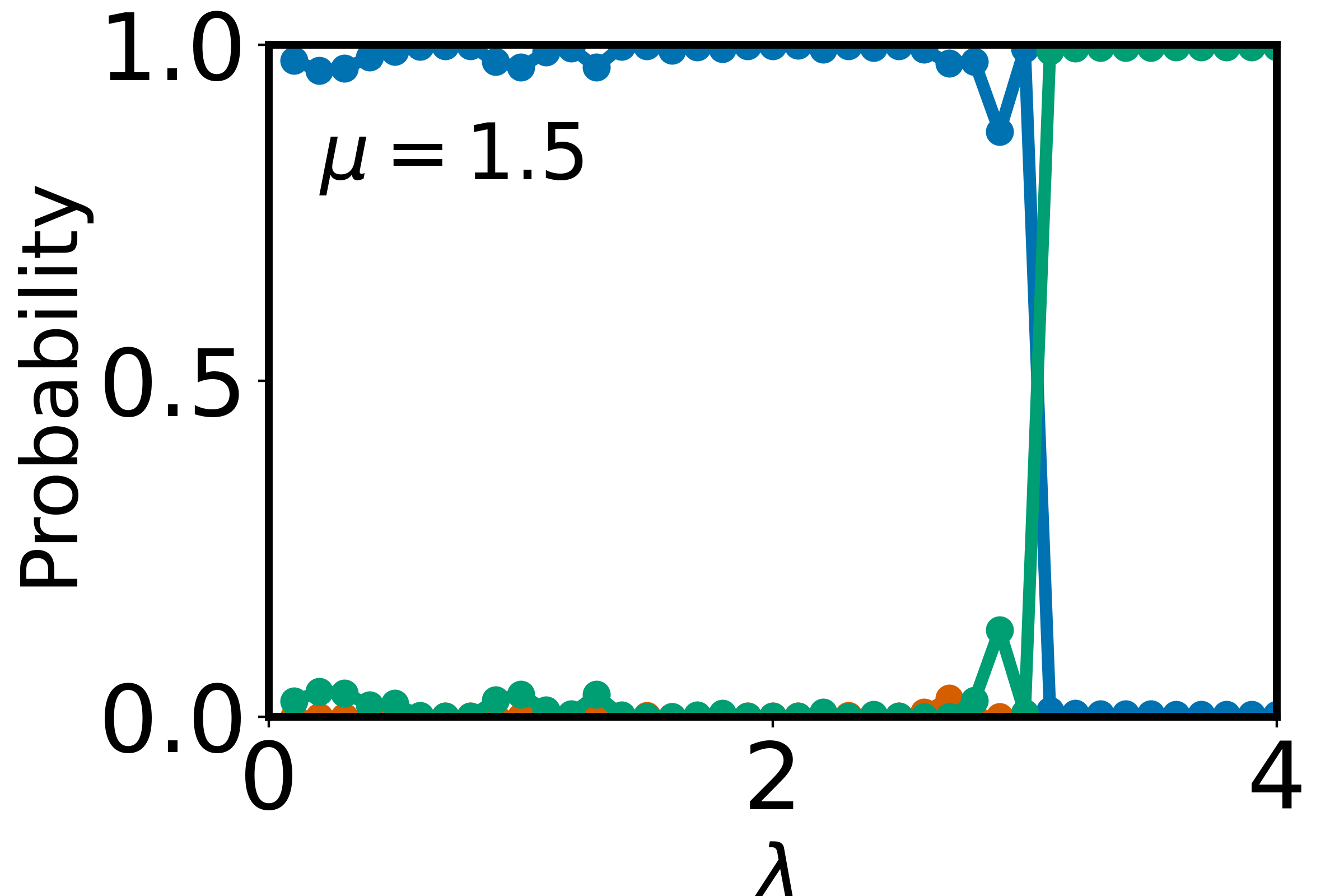}}
\caption{\label{fig11}Neural network probability for the $3$-class classifier: (a) as a function of hopping amplitude $\mu$ for fixed $\lambda=0.5$, (b) as a function of disorder strength $\lambda$ for fixed $\mu=0$, and (c) as a function of $\lambda$ for fixed $\mu=1.5$. Input data comprise normalized eigenvector probability densities corresponding to the infinite-temperature single-particle state, with $5000$ samples for training and averaged over $500$ test datasets. The system size considered is $N=3000$.}
\end{figure*}

Additionally, we compute the MIPR. For a normalized many-body eigenstate $|\Psi\rangle$ in the particle-number-constrained space, expressed as $|\Psi\rangle = \sum_{i=1}^D C_i |i\rangle$, the MIPR is defined as:
\begin{equation}
\mathrm{MIPR} = \sum_{i=1}^D \left|C_i\right|^4.
\end{equation}
Typically $\mathrm{MIPR} \sim D^{-\eta}$.
For a perfectly ergodic eigenstate $\eta=1$, reflecting the uniform distribution of the wavefunction across all basis states, while for a highly localized eigenstate, $\eta \gtrsim 0$, a small fraction, indicating that the wavefunction is fractal but essentially concentrated on a small subset of basis states. For MBC or NEE states, in general, $0<\eta<1$, a larger fraction which implies NEE states are fractal and distributed non-uniformly on the basis states.  

Figure~\ref{fig10}(a) shows $r_{av}$ as a function of $\mu$ across an ME-to-MBC transition for $\lambda = 0.5$. As the system size increases, $r_{av}$ approaches $0.53$ in the ergodic phase ($\mu < 1$), whereas in the MBC phase ($\mu > 1$), it attains a value intermediate between ME and MBL. Similarly, Fig.~\ref{fig10}(b) depicts the behaviour of MIPR, where it approaches $\sim 1/D$ in the ME phase and reaches much higher values than $1/D$ in the MBC phase.

Next, we examine the ME-MBL transition at $\mu = 0.5$. In the ergodic phase, $r_{av}$ converges to $0.53$, while in the MBL phase, it approaches $0.38$, as shown in Fig.~\ref{fig10}(c). Correspondingly, Fig.~\ref{fig10}(d) demonstrates that MIPR approaches $\mathcal{O}(1)/D$ in the ergodic phase (approximately for $\lambda < 2.5$), while in the MBL phase, it deviates hugely from the ergodic value.

Finally, we study the MBC-MBL transition along $\lambda$ at $\mu = 1.8$. In the MBC phase, $r_{av}$ remains between the standard ergodic and MBL values (see Fig.~\ref{fig10}(e)), but in the MBL phase, it approaches $0.38$. Similarly, Fig.~\ref{fig10}(f) reveals that MIPR increases very little in the MBL phase with increasing system sizes, while in the MBC phase, it shows substantial system size dependence.

This analysis corroborates the distinct characteristics of ME, MBC, and MBL phases and demonstrates the use of $r_{av}$ and MIPR in distinguishing these phases. Additionally, the entanglement entropy of eigenstates in the ME, MBC and MBL phases shows the thermal volume law, subthermal volume law and area law, respectively~\cite{PhysRevLett.126.080602}.  However, it is evident from the standard metrics such as the gap ratio and MIPR, as described here, that one requires data for a number of larger system sizes in order to define the phase boundaries clearly for most of the cases. For a single (smaller) system size, it is even more confusing to detect the phase boundaries from the behaviour of  $r_{av}$ and MIPR. This is where our data-driven approach to detect phase boundaries comes to help as discussed in the main text.

\section{Single Particle Supervised Learning}
\label{appB}

In this section, we present the three-class classifier for the single-particle system (\(U=0\)) and analyze its output probabilities across various phases. The training dataset comprises $5000$ realizations of infinite-temperature state PDs sampled from distinct regions of the phase diagram. Specifically, PDs at \( \mu=0,\lambda=0.5\) correspond to the delocalized phase, \(\mu=0, \lambda=3.5\) to the localized phase, and \(\mu=2\) with \(\lambda=0.3, 1, 2.5, 3\) to the critical (multifractal) phase. For training, we employ the neural network architecture described in Table~\ref{tab2}. Since this analysis focuses on the single-particle system, the number of neurons in the input layer equals the number of sites \(N\). For the normalized PDs the training process spans $50$ epochs, achieving a network accuracy of $95\%$. Testing datasets comprise $500$ realizations of normalized infinite-temperature eigenstate PDs sampled along various phase transitions (data utilized for PCA in Section~\ref{sec:level4}).

The classifier outputs three probabilities, \(p_1\), \(p_2\), and \(p_3\), corresponding to the delocalized, critical, and localized phases, respectively. These probabilities satisfy \(p_1 + p_2 + p_3 = 1\), such that if any \(p_i \to 1\), the network identifies the corresponding phase with high confidence. In Fig.~\ref{fig11}(a), we plot the output probabilities of the multiclass classifier for the delocalized-to-critical transition at \(\mu=0.5 \) for fixed $\lambda=0.5$. Similarly, in Fig.~\ref{fig11}(b), the classifier probabilities for the delocalized-to-localized transition at $\lambda_c=2$ for fixed \(\mu=0\) are presented, showing that the multifractal states are observed only at \(\lambda=2\). Finally, Fig.~\ref{fig11}(c) illustrates the probabilities along the criitcal-to-localized transition at $\lambda_c=3$ for fixed \(\mu=1.5\). The phase transition points identified by the neural network align well with previously established results~\cite{PhysRevB.50.11365,chang1997multifractal}, demonstrating the effectiveness of the supervised learning approach in detecting phase transitions in single-particle systems. Notably, the network predictions are robust with increasing system size. 

While it is possible to perform this analysis using bare eigenvector PDs as in the many-body case, this approach requires a larger number of training epochs. In contrast to the many-body case, where normalized PDs lead to discrepancies in predictions, the results for single-particle systems with normalized PDs remain consistent and accurate. This could be attributed to the simpler structure of single-particle states.

\section{Testing the trained network on a model with inconclusive level statistics}
\label{appC}

In this section, we demonstrate how ML techniques outperform conventional measures in identifying phases. We apply a neural network trained on eigenvalue spacings of the EAAH model~\eqref{eq1} to a different Hamiltonian, where level statistics yield inconclusive results.

The system under investigation is the interacting all-bands-flat (ABF) diamond lattice with Aubry-Andr\'e potential applied antisymmetrically~\cite{ahmed2023interplay}.  The Hamiltonian is given by:
\begin{equation}
    \hat{H}_{ABF}= \hat{H}_{\text{hop}}+ \hat{H}_{\text{os}}+ \hat{H}_{\text{int}},
\end{equation}
where
\begin{align}
\nonumber \hat{H}_{\text{hop}}=&-J \sum_{k=1}^{N / 3}\left(-\hat{u}_{k}^{\dagger} \hat{c}_{k}+\hat{d}_{k}^{\dagger} \hat{c}_{k}+\hat{c}_{k}^{\dagger} \hat{u}_{k+1}+\hat{c}_{k}^{\dagger} \hat{d}_{k+1}+\text{H.c.}\right),\\
\nonumber \hat{H}_{\text{os}}=&W \sum_{k=1}^{N / 3}\cos(2\pi kb+\theta_p) \left( \hat{u}_{k}^{\dagger} \hat{u}_{k}- \hat{d}_{k}^{\dagger} \hat{d}_{k}\right),\\
\nonumber \hat{H}_{\text{int}}=& V \sum_{k=1}^{N / 3}\left( \hat{u}_{k}^{\dagger} \hat{u}_{k} \hat{c}_{k}^{\dagger} \hat{c}_{k}+\hat{d}_{k}^{\dagger}\hat{d}_{k} \hat{c}_{k}^{\dagger}\hat{c}_{k}+\hat{c}_{k}^{\dagger} \hat{c}_{k}\hat{u}_{k+1}^{\dagger}\hat{u}_{k+1} \right.\\
& +\left. \hat{c}_{k}^{\dagger}\hat{c}_{k}\hat{d}_{k+1}^{\dagger} \hat{d}_{k+1}\right).
\label{eq5}
\end{align}

Each unit cell consists of three sites, $\alpha_k=\left\lbrace u_k, d_k, c_k\right\rbrace$, with a total of $N$ lattice sites. The hopping amplitude $J$ and nearest-neighbor interaction strength $V$ are both set to $1$. The disorder strength is denoted by $W$, and the quasiperiodicity parameter is the golden mean $(\sqrt{5}-1)/2$. The phase shift $\theta_p$ is randomly drawn from a uniform distribution $[0,2\pi]$.  

\begin{figure}
\centering
\stackunder{\hspace{-4cm}(a)}{\includegraphics[width=4.4cm]{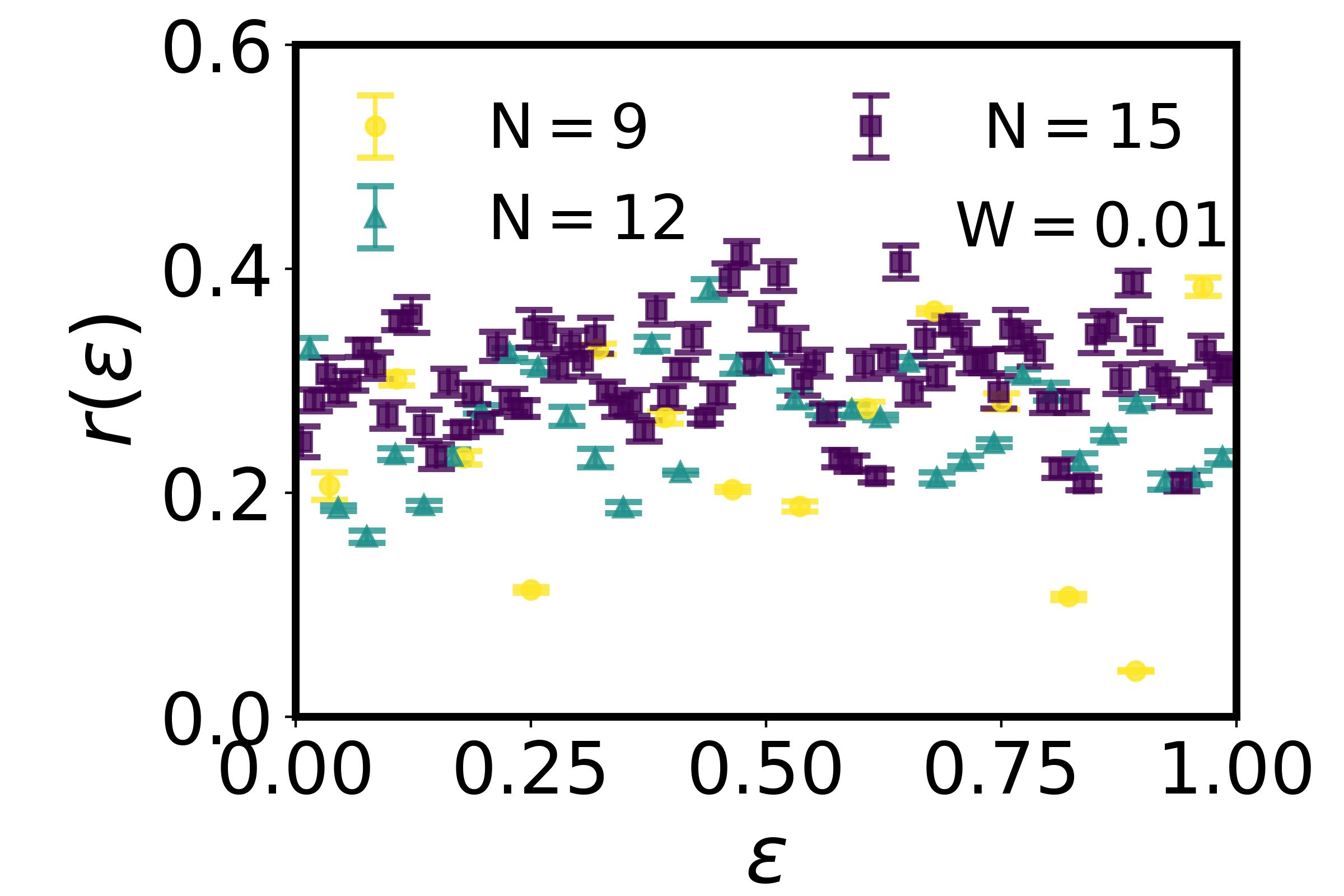}}\hspace{-0.35cm}
\stackunder{\hspace{-4cm}(b)}{\includegraphics[width=4.4cm]{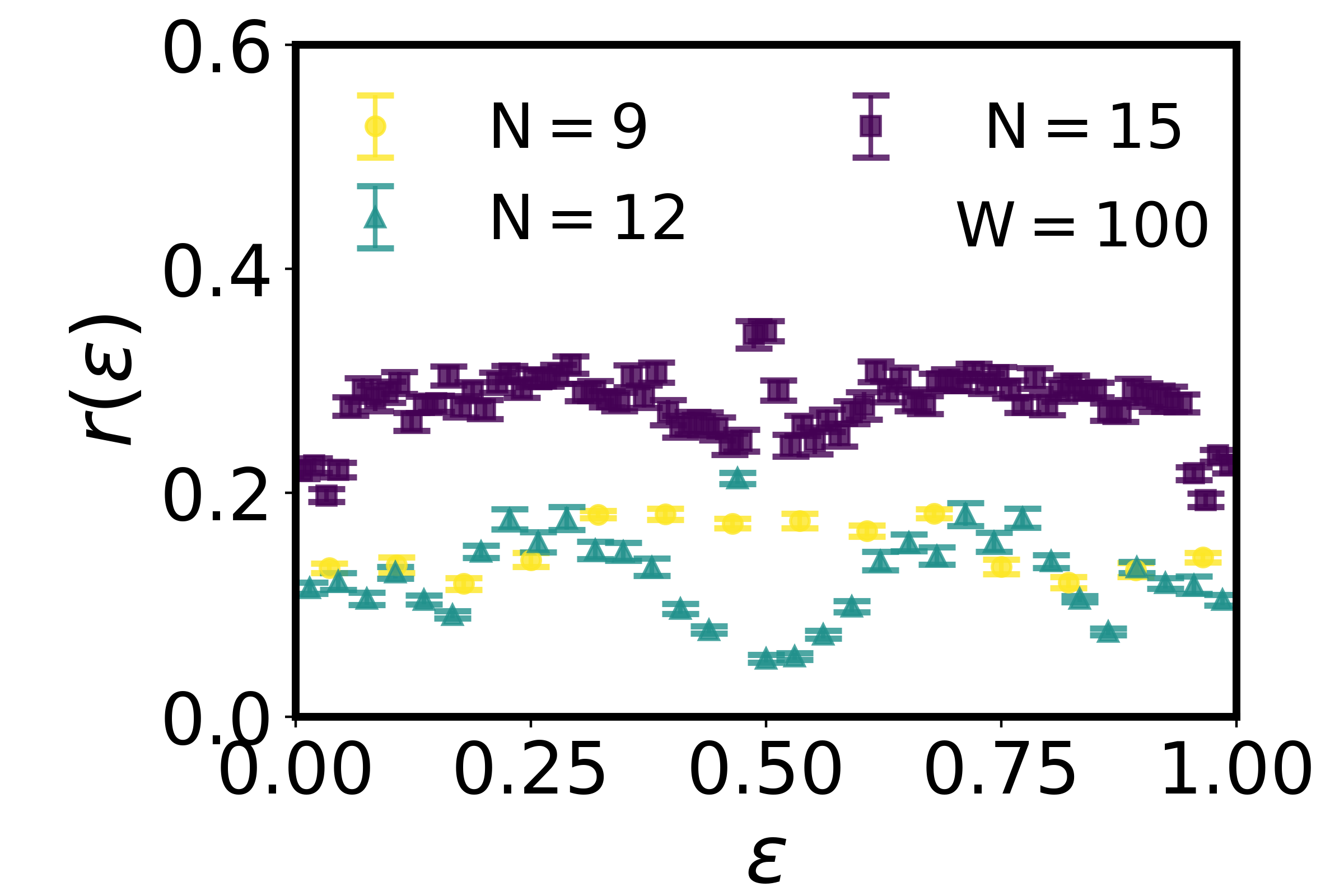}}
\vspace{-0.2cm}

\stackunder{\hspace{-8cm}(c)}{\includegraphics[width=8cm]{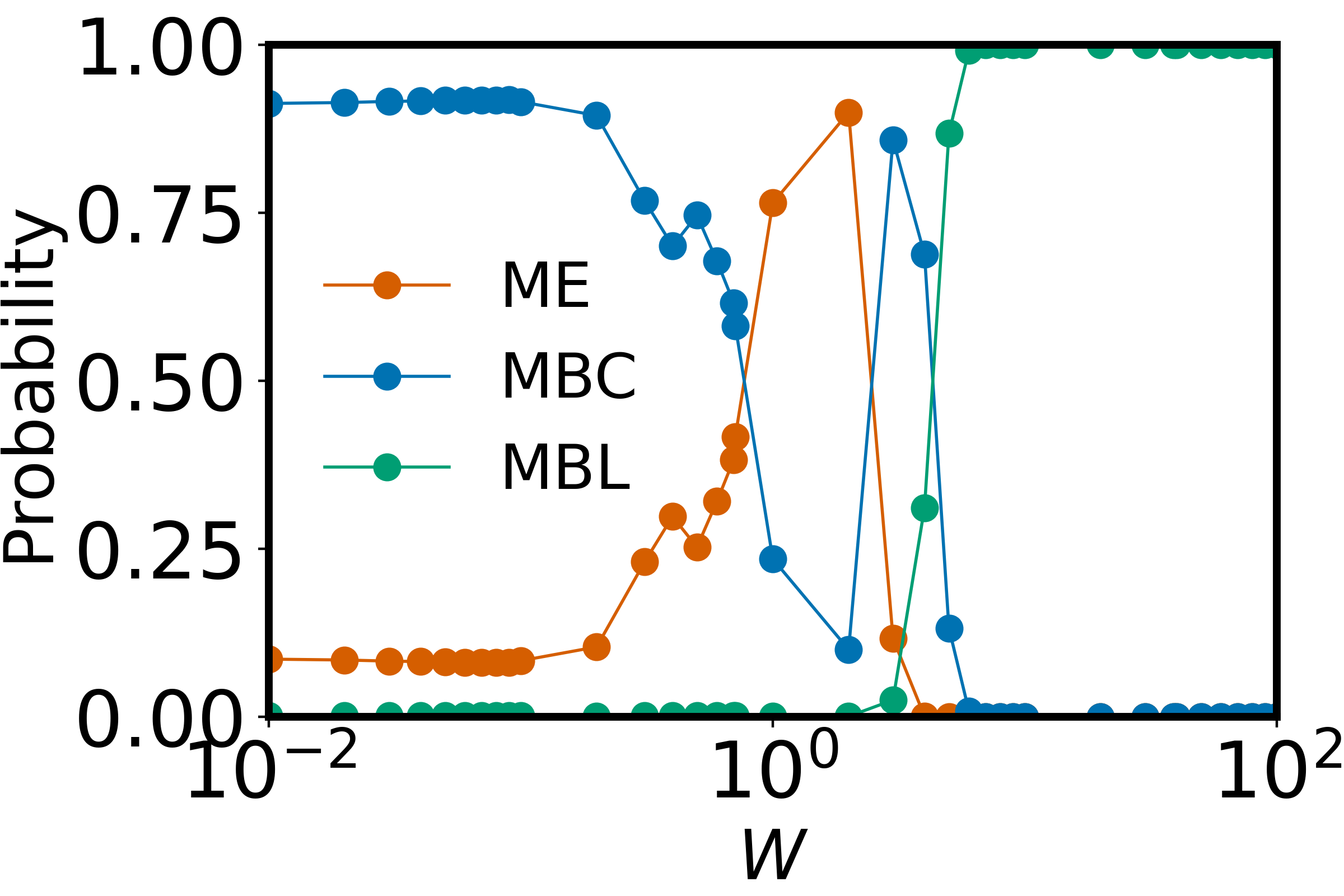}}
\caption{\label{fig12}For the system described by the ABF Hamiltonian in Eq.~\ref{eq5} , energy-resolved gap-ratio $r$ as a function of the fractional eigenstate index $\epsilon$ for disorder strength (a) $W = 0.01$ (low) and (b) $W = 100$ (high). The number of disorder realizations is
$500$, $400$, and $50$ for system sizes $N = 9, 12$, and $15$, respectively. (c) Neural network predictions for the $3-$ class classifier with increasing disorder strength $W$. Input data comprises eigenvalue spacings, with $3000$ samples per class for training and averaged over $500$ test datasets for system size $N=12$. Here filling fraction is fixed to $\nu=N/3$.}
\end{figure}

In our earlier work~\cite{ahmed2023interplay} using measures such as the MIPR and the one-particle density matrix (OPDM)~\cite{PhysRevB.96.060202}, we have established the existence of three distinct phases with increasing disorder strength $W$: a nonergodic (MBC) phase at low disorder, a thermal phase near around $W = 2$, and an MBL-like phase at higher disorder.
However, level statistics failed to reliably identify these phases. Specifically, we computed the energy-resolved level spacing ratio $r$ (see Eq.~\ref{eq4}), by segmenting the many-body energy spectrum and averaging $r$ within each segment. Figures~\ref{fig12}(a) and \ref{fig12}(b) show $r$ for the low and high disorder cases, respectively. In the low disorder (MBC) regime, $r$ is expected to assume an intermediate value between those of the thermal and MBL phases (between $0.53$ and $0.386$). However, this behavior is not observed here. Even at high disorder, the level statistics fail to approach the MBL value ($r \approx 0.386$). This discrepancy arises due to quasi-degeneracies and gaps in the spectrum, making level spacing an unreliable indicator of localization in such systems.

To overcome this limitation, we apply our multi-class neural network trained on the EAAH model, to the level-spacing data of the system discussed here. We consider a system size $N=12$ and filling fraction $\nu=N/3$. The training dataset consists of $3000$ disorder realizations of eigenvalue spacings for each class, with the following parameter choices: for the ME phase, we use $\mu=0.3$ and $\lambda=1, 2.5$; for the MBC phase, $\mu=2$ and $\lambda=0.3,1,2.5,3$; and for the MBL phase, $\mu=0.5, 2$ and $\lambda=4.5$ of the EAAH model. The neural network architecture is described in Table~\ref{tab2}. Training is performed over $20$ epochs, achieving $99\%$ accuracy. The testing dataset comprises of $500$ realizations of $\theta_p$ from Eq.~\eqref{eq5}. Figure~\ref{fig12}(c) shows the classifier output probabilities as a function of disorder strength $W$. The network correctly identifies the MBC phase at low disorder, followed by the ME phase near $W\approx2$ and a transition back to MBC at the critical point where ME to MBL transition happens, followed by a clear MBL phase at the higher disorder. 

These results highlight the advantage of ML in phase identification, even when traditional measures fail. Unlike level statistics, which become inconclusive due to spectral complexities, the neural network successfully classifies phases by recognizing subtle patterns in eigenvalue spacings. Thus, ML is not merely an alternative but a powerful tool for identifying localization behavior in complex systems.

%%%%%%%%%%%%%%%%%%%%%%%%%%%%%%%%%%%%%%%%%%%%%%%%%%%%%%%%

%%%%%%%%%%%%%%%%%%%%%%%%%%%%%%%%%%%%%%%%%%%%%%%%%%%%%%%%
\section{Signature of crossover within MBL phase from supervised learning}
\label{sec:MBL_subphases}
In this section, we extend our analysis of exploring subphases within the established phase diagram in Section~\ref{4-class}. Specifically, we investigate whether the MBL phase can be meaningfully partitioned into distinct subphases by dividing the phase diagram at $\mu = 1$. To this end, we employ a neural network trained to classify input data into four phases, with output probabilities corresponding to the confidence of the data belonging to each phase.  

The neural network architecture follows the specifications in Table~\ref{tab2}, using the SGD optimizer with its default learning rate. For eigenvalue spacing data, the network achieves an accuracy of $\approx 99\%$ within $30$ epochs, while for eigenvector probability densities (PDs), it attains $\approx 95\%$ accuracy after $1200$ epochs. We divide the MBL phase into two subphases: MBL-I ($\mu < 1$) and MBL-II ($\mu > 1$). The training data sets consists of $10000$ and $20000$ samples per class for the eigenvalue spacing and eigenvector PDs, respectively, deep within these phases. For the ME phase, we use $\mu = 0.3$ with $\lambda = 1.0$ and $2.5$; for the MBC phase, $\mu = 2.0$ with $\lambda = 0.3$, $1.0$, $2.5$, and $3.0$; for MBL-I, $\mu = 0.5$ and $\lambda = 4.5$; and for MBL-II, $\mu = 1.5$ and $\lambda = 4.5$.  

\begin{figure}
\centering
\stackunder{\hspace{-4cm}(a)}{\includegraphics[width=4.2cm]{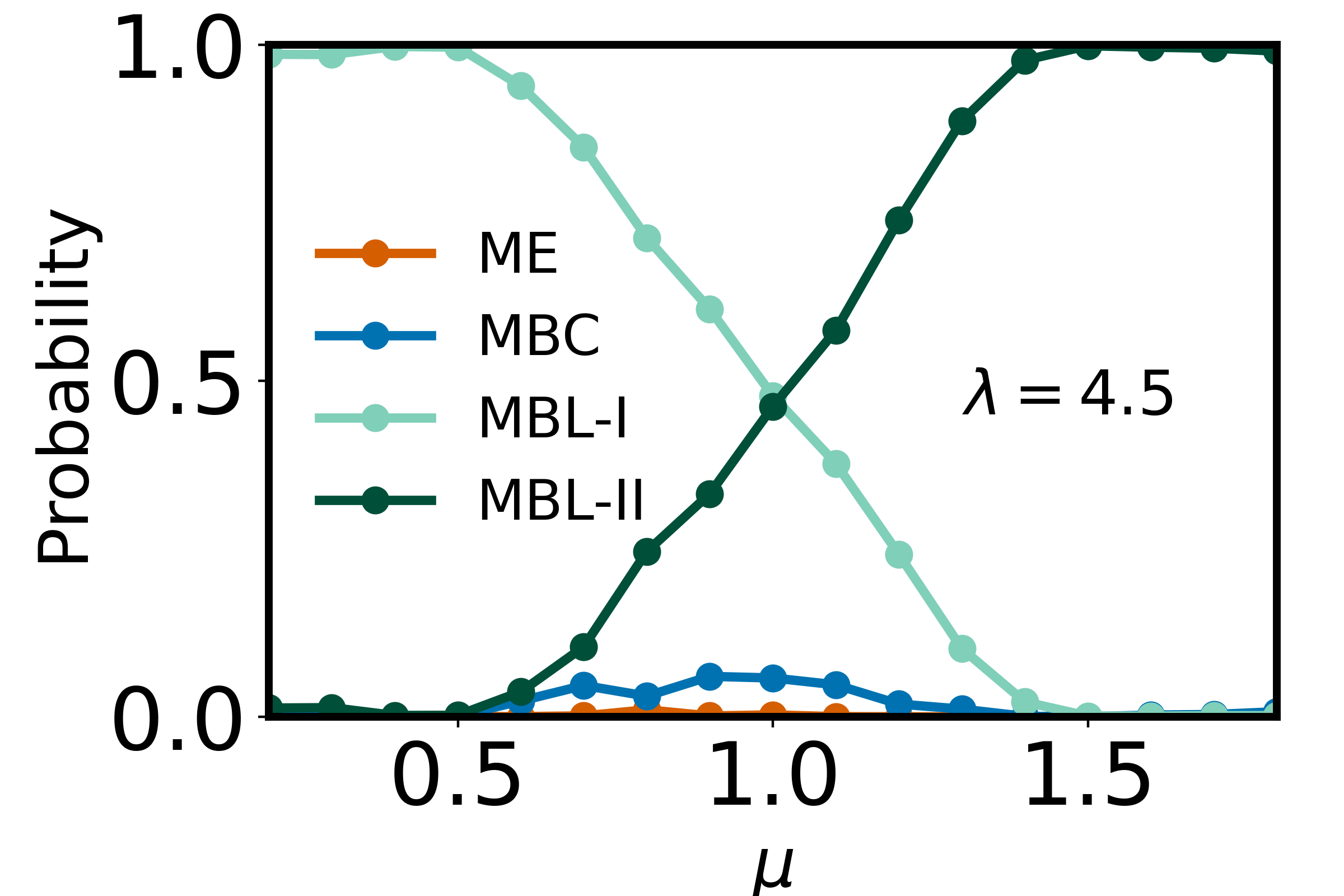}}
\stackunder{\hspace{-4cm}(b)}{\includegraphics[width=4.2cm]{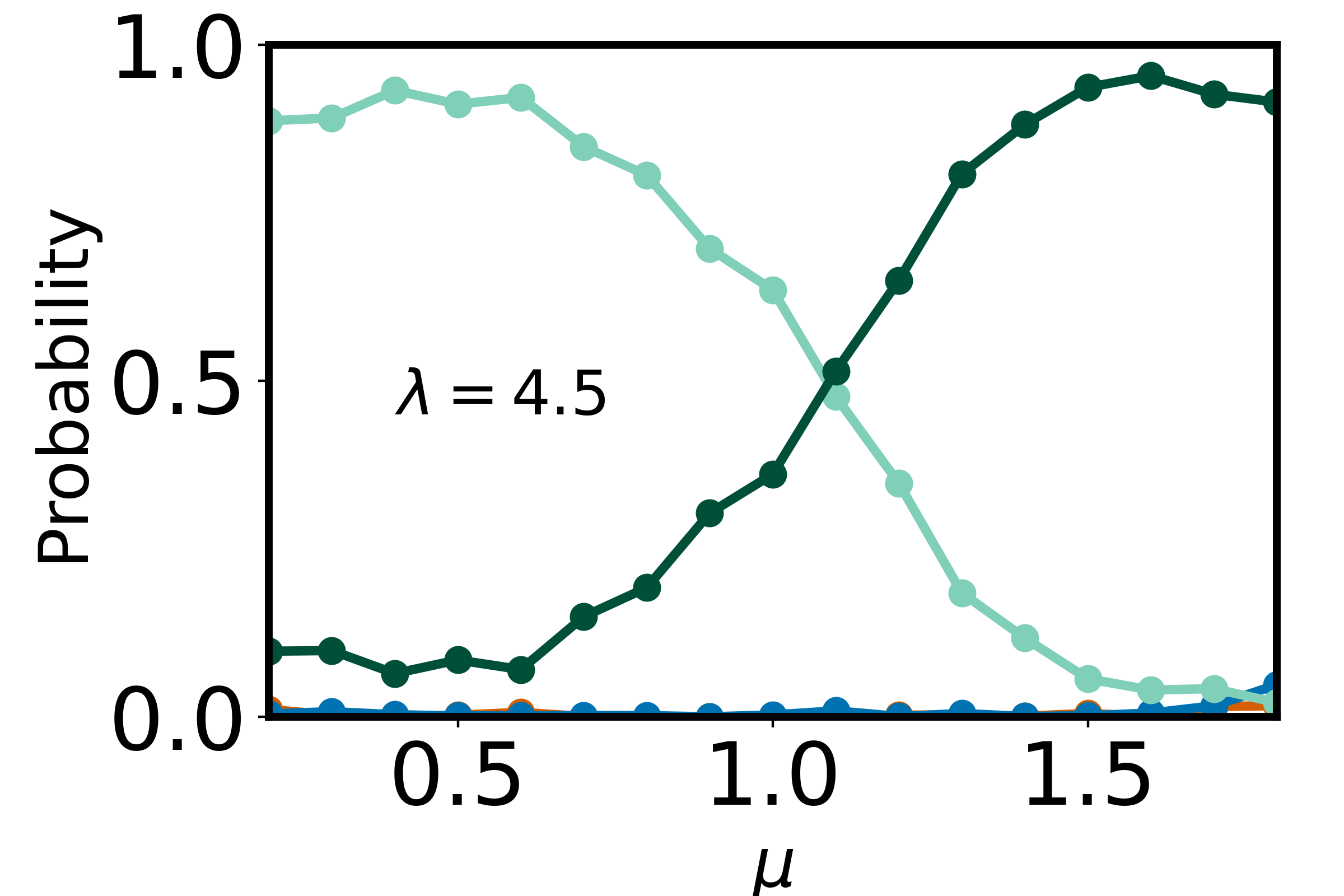}}
\caption{\label{fig13} Neural network probabilities for the $4$-class classifier, considering two MBL phases, are shown for increasing hopping amplitude $\mu$ at fixed disorder strength $\lambda = 4.5$. Here data considered are (a) eigenvalue spacings, and (b) probability densities corresponding to eigenstates, averaged over $400$ test datasets. The system size is $N=14$, and half-filling is assumed. Both the figures indicate a crossover within the MBL phase at $\mu=1$. }
\end{figure}

Figure~\ref{fig13}(a) presents the results of the four-class classifier trained on eigenvalue spacings, while Fig.~\ref{fig13}(b) shows the corresponding results for eigenvector PDs. Notably, both MBL-I and MBL-II exhibit broad probability distributions, with a clear crossover at $\mu \approx 1$, confirming their existence. This analysis robustly demonstrates that the MBL phase can indeed be partitioned into two subphases separated at $\mu = 1$.  

Our machine learning approach thus reveals that the MBL phase comprises two distinct subphases, MBL-I and MBL-II, suggesting a possible transition or crossover in the localization properties of many-body eigenstates near $\mu \approx 1$~\cite{roy2025manybodycriticalphasequasiperiodic}. This highlights the power of ML analysis in uncovering subtle changes in phases, even with smaller system sizes, unlike the conventional measures which typically require investigation with much larger system sizes for a clear detection of phase transitions.

%%%%%%%%%%%%%%%%%%%%%%%%%%%%%%%%%%%%%%%%%%%%%%%%%%%%%%%%
\color{black}
\bibliography{ml_new}
\end{document}